\documentclass[article]{aa} 

\usepackage{longtable,lscape}
\usepackage{multicol}
\usepackage{natbib}
\bibpunct{(}{)}{;}{a}{}{,} 

\usepackage{txfonts}
\usepackage{color}
\usepackage{hyperref}
\usepackage{cleveref}
\usepackage{graphicx}
\usepackage[labelformat=simple]{caption,subcaption}
\usepackage{gensymb}

\renewcommand\thesubfigure{\Alph{subfigure}}

\def\m2s2{\,m$^{2}$\,s$^{-2}$} 

\begin{document}

   \title{Orbiting a binary: \\
   SPHERE characterisation of the HD~284149 system}
 

\titlerunning{SPHERE characterisation of the HD~284149 system}

  \author{M. Bonavita\inst{\ref{inst1},\ref{inst2}}
    \and V. D'Orazi \inst{\ref{inst1}} 
    \and D. Mesa \inst{\ref{inst1}}
    \and C. Fontanive \inst{\ref{inst2}}
    \and S. Desidera \inst{\ref{inst1}}
    \and S. Messina \inst{\ref{inst4}}
    \and S. Daemgen \inst{\ref{inst3}} 
    \and R. Gratton \inst{\ref{inst1}}
    \and A. Vigan \inst{\ref{inst5}}
    \and M. Bonnefoy \inst{\ref{inst7}} 
    \and A. Zurlo \inst{\ref{inst5}, \ref{inst20}}
    \and J. Antichi \inst{\ref{inst13}, \ref{inst1}}
    \and H. Avenhaus \inst{\ref{inst3}, \ref{inst6}, \ref{inst16}}
    \and A. Baruffolo \inst{\ref{inst1}} 
    \and J.~L. Baudino\inst{\ref{inst19}}
    \and J.~L. Beuzit \inst{\ref{inst7}}
    \and A. Boccaletti \inst{\ref{inst10}}
    \and P. Bruno \inst{\ref{inst4}}
    \and T. Buey \inst{\ref{inst10}}
    \and M. Carbillet \inst{\ref{inst20}}
    \and E. Cascone\inst{\ref{inst14}}
    \and G. Chauvin \inst{\ref{inst7}}  
    \and R.~U. Claudi\inst{\ref{inst1}}
    \and V. De Caprio \inst{\ref{inst14}}
    \and D. Fantinel \inst{\ref{inst1}}
    \and G. Farisato \inst{\ref{inst1}}
    \and M. Feldt \inst{\ref{inst6}}
    \and R. Galicher \inst{\ref{inst10}}
    \and E. Giro \inst{\ref{inst1}}
    \and C. Gry \inst{\ref{inst5}}
    \and J. Hagelberg \inst{\ref{inst7}}
    \and S. Incorvaia \inst{\ref{inst15}} 
    \and M. Janson \inst{\ref{inst6}, \ref{inst9}}
    \and M. Jaquet \inst{\ref{inst5}}
    \and A.~M. Lagrange \inst{\ref{inst7}}
    \and M. Langlois \inst{\ref{inst5}}
    \and J. Lannier \inst{\ref{inst7}}
    \and H. Le Coroller \inst{\ref{inst5}}
    \and L. Lessio \inst{\ref{inst1}}
    \and R. Ligi \inst{\ref{inst5}}
    \and A.~L. Maire \inst{\ref{inst6}}
    \and F. Menard \inst{\ref{inst7}, \ref{inst8}}
    \and C. Perrot \inst{\ref{inst10}} 
    \and S. Peretti \inst{\ref{inst12}} 
    \and C. Petit \inst{\ref{inst22}}
    \and J. Ramos \inst{\ref{inst6}}
    \and A. Roux \inst{\ref{inst7}}
    \and B. Salasnich \inst{\ref{inst1}} 
    \and G. Salter \inst{\ref{inst5}}
    \and M. Samland \inst{\ref{inst6}} 
    \and S. Scuderi \inst{\ref{inst4}}
    \and J. Schlieder \inst{\ref{inst6}, \ref{inst17}}
    \and M. Surez \inst{\ref{inst22}}
    \and M. Turatto \inst{\ref{inst1}} 
    \and L. Weber \inst{\ref{inst12}}
    }

   \authorrunning{M. Bonavita et al.}


\institute{INAF -- Osservatorio Astronomico di Padova,  
    Vicolo dell'Osservatorio 5, I-35122, Padova, Italy \label{inst1}
    \and Institute for Astronomy, The University of Edinburgh, 
        Royal Observatory, Blackford Hill, Edinburgh, EH9 3HJ, U.K. \label{inst2}
    \and Institute for Astronomy, ETH Zurich, Wolfgang-Pauli Strasse 27, 
        8093 Zurich, Switzerland \label{inst3}
    \and INAF -- Osservatorio Astronomico di Catania,  
        Via Santa Sofia, 78,Catania, Italy  \label{inst4}
    \and Aix Marseille Univ, CNRS, LAM, Laboratoire d'Astrophysique de 
        Marseille, Marseille, France \label{inst5}
    \and Max-Planck Institute for Astronomy, K\"onigstuhl 17, 
        69117 Heidelberg, Germany \label{inst6}
    \and Univ. Grenoble Alpes, CNRS, IPAG, F-38000 Grenoble, France \label{inst7}
    \and European Southern Observatory (ESO), 
        Karl-Schwarzschild-Str. 2,85748 Garching, Germany \label{inst8}
    \and Department of Astronomy, Stockholm University, 
        AlbaNova University Center, 106 91 Stockholm, Sweden \label{inst9}
    \and LESIA, Observatoire de Paris, PSL Research University, CNRS, 
         Sorbonne Universités, UPMC Univ. Paris 06, 
         Univ. Paris Diderot, Sorbonne Paris Cité\label{inst10}
    \and Observatoire de Haute-Provence, CNRS, Universit\'e d'Aix-Marseille, 
        04870 Saint-Michel-l'Observatoire, France \label{inst11}
    \and Observatoire Astronomique de l'Universit\'e de Gen\`eve, 
        Chemin des Maillettes 51, 1290 Sauverny, Switzerland \label{inst12}
    \and INAF - Osservatorio Astrofisico di Arcetri, 
        L.go E. Fermi 5, 50125 Firenze, Italy \label{inst13}
    \and INAF - Osservatorio Astronomico di Capodimonte, 
        Salita Moiariello 16, 80131 Napoli, Italy \label{inst14}
    \and INAF-Istituto di Astrofisica Spaziale e Fisica Cosmica di Milano, 
        via E. Bassini 15, 20133 Milano, Italy \label{inst15}
    \and Universidad de Chile, Camino el Observatorio, 1515 Santiago, Chile \label{inst16}
    \and NASA Goddard Space Flight Center, Greenbelt, MD 20771, USA \label{inst17}
    \and Department of Physics, University of Oxford, Parks Rd, 
        Oxford OX1 3PU, UK\label{inst19}
    \and N\'ucleo de Astronom\'ia, Facultad de Ingenier\'ia, Universidad Diego Portales, Av. Ejercito 441, Santiago, Chile \label{inst20}
    \and Universite Cote d’Azur, OCA, CNRS, Lagrange, France \label{inst21}
    \and ONERA (Office National d’Etudes et de Recherches Aérospatiales),B.P.72, F-92322 Chatillon, France \label{inst22}    
}

\date{Received  / Accepted }

\abstract{}{In this paper we present the results of the SPHERE observation of the \object{HD~284149} system, aimed at a more detailed characterisation of both the primary and its brown dwarf companion.}
{We observed \object{HD~284149} in the near-infrared with SPHERE, using the imaging mode (IRDIS+IFS) and the long-slit spectroscopy mode (IRDIS-LSS). The data were reduced using the dedicated SPHERE pipeline, and algorithms such as PCA and TLOCI were applied to reduce the speckle pattern. }
{The IFS images revealed a previously unknown low-mass ($\sim 0.16~M_{\odot}$) stellar companion (\object{HD~294149}~B) at $\sim 0.1~^{\prime\prime}$, compatible with previously observed radial velocity differences, as well as proper motion differences between Gaia and Tycho-2 measurements. The known brown dwarf companion (\object{HD 284149}~b) is clearly visible in the IRDIS images. This allowed us to refine both its photometry and astrometry. The analysis of the medium resolution IRDIS long slit spectra also allowed a refinement of temperature and spectral type estimates. A full reassessment of the age and distance of the system was also performed, leading to more precise values of both mass and semi-major axis.}
{As a result of this study, \object{HD~284149}~ABb therefore becomes the latest addition to the (short) list of brown dwarfs on wide circumbinary orbits, providing new evidence to support recent claims that object in such configuration occur with a similar frequency to wide companions to single stars.}


\maketitle

\clearpage 
\section{Introduction}

Several new sub-stellar companions to nearby young stars have been directly imaged in the last decade, spanning a wide range of masses and separations 
\citep[see e.g.][]{chauvin2005a,chauvin2005b,marois2008,marois2010,lagrange2010,biller2010,carson2013,delorme2013,rameau2013,bailey2014,bonavita2014,gauza2015,stone2016,wagner2016}.
These objects, some of which have masses near and below $15 M_{Jup}$ \citep[see e.g.][]{marois2008,marois2010,lagrange2010,rameau2013,bailey2014,kraus2014,51Eri,naud2014, artigau2015, bowler2017}, may represent the bottom end of the stellar companion mass function as well as the top end of the planet population, though both scenarios pose challenges to conventional formation models. 

On one hand, the binary star formation process \citep[e.g.][]{bate2003} rarely predicts such low mass ratios.
On the other hand, the standard core accretion model would struggle to form super-Jupiter planets at $>50$~AU. 
Even though the alternative gravitational instability (GI) model might become more plausible at large separations \citep[see][]{meru2010}, especially around more massive stars with larger disks,  it is still unclear whether or not such hypothesis is correct \citep[see e.g.][]{janson2012}.

Some of the most recent GI models \citep[see e.g.][]{forgan2013, forgan2015} seem to suggest that the most likely outcome of such formation process is a large fraction of relatively massive objects at large semi-major axis. Such a population would be easily detectable with direct imaging, and the lack of detections can be used to place strong constraints on how frequently disk fragmentation occurs \citep[see e.g.][]{vigan2017}.

An in depth characterisation of the few known members of this population is therefore highly desirable, but the lack of multi-wavelength spectroscopy and photometry make precise characterisation of these systems quite challenging, leading to poorly constrained values of some fundamental properties such mass, radius, effective temperature etc. 

\noindent New dedicated instruments, which allow both precise multi band photometry and low and medium resolution spectroscopy, are now becoming available and allow a better characterisation of low mass companions. SPHERE \citep{beuzit2008}, the new planet finder mounted at the VLT, is one of those. 

\noindent SPHERE includes three scientific modules: IFS \citep{claudi2008} and IRDIS \citep{dohlen2008}, both operating in the NIR, and ZIMPOL \citep{Thalmann2008} which uses visible light instead. 
The main IRDIFS imaging mode uses the IFS and IRDIS channels simultaneously: low resolution (R=50) spectra are obtained with IFS while dual band images  \citep[DBI, see][]{Vigan2010} are taken using the IRDIS H2-H3 filter pair at 1.593 $\mu m$ and 1.667 $\mu m$. 
Lower resolution but wider spectra can be obtained using the IRDIFS\_EXT mode. In this case the IFS spectra have R=30 and the K1-K2 filter pair is used for IRDIS, taking images at 2.110 $\mu m$ and 2.251 $\mu m$. 
Finally, IRDIS can be used in long slit spectroscopy \citep[LSS][]{vigan2008} mode, supplying medium resolution (R=350) and low resolution (R=50) spectra. 

\noindent Since the start of operation in December 2014, SPHERE has been used to characterise several directly imaged companions \citep[see e.g.][]{vigan2016, maire2016,bonnefoy2016,zurlo2016,mesa2016}.\\

The brown dwarf companion at $\sim 3.6$ $^{\prime\prime}$ from the F8 star \object{HD~284149} was discovered by \citet{bonavita2014} as part of a direct imaging survey of 74 members the Taurus star forming region \citep{daemgen2015}. 
 Together with several other targets of the same survey, \object{HD~284149} has been proposed to be part of the so-called Taurus-Ext association \citep{luhman2017, kraus2017, daemgen2015}, a group of stars with similar space position and kinematics of the Taurus star forming region but distinctly older ages.
A dedicated age estimate was then performed for \object{HD~284149} by \cite{bonavita2014}, using several youth indicators, leading to 
an adopted age of $25^{+25}_{-10}$~Myrs. They therefore finally estimated the mass of \object{HD~284149~b} to be $32^{+18}_{-14}~M_{Jup}$. From the available photometry they were able to infer a spectral type between M8 and L1 and an effective temperature of $2537^{+95}_{-182}$~K, but no spectroscopic characterisation has been performed up to now, leaving these estimates highly uncertain. 

\noindent Here we present the result of the observations of \object{HD~284149} performed with SPHERE, aimed at a precise characterisation of the whole system. 

\noindent Section~2 contains a detailed description of the observations and data reduction while an update to the stellar characteristics given the new information available, including those coming from the Gaia mission, is given in Section~3. 
The results are described and discussed in Section~4 and include the description of a newly discovered close stellar companion, an update of the astrometry of the known wide brown dwarf companion and the comparison with the models of the medium resolution spectra obtained with the IRDIS LSS mode. 

\section{Observations and data reduction}
\object{HD~284149} was observed with SPHERE in IRDIFS mode on 2015-10-25 and with IRDIFS\_EXT mode on 2015-11-27 as part of the SHINE (SpHere INfrared survey for Exoplanet) GTO campaign.
The second epoch observations were taken without coronagraph to confirm a faint candidate that was imaged for the first time very close to the edge of the coronagraphic mask in the first observation. 
In order to take full advantage of the angular differential imaging technique \citep[ADI, see][]{marois2006a}, for both epochs the target was observed in pupil-stabilised mode to allow the rotation of the field of view.
The reduction of all the IRDIS and IFS data sets was performed through the SPHERE Data Center (DC) using the SPHERE Data Reduction and Handling (DRH) automated pipeline \citep{pavlov08}. In the case of the IFS data, the DRH pipeline is complemented with additional routines that allow for an improved wavelength calibration as well as for a correction of both coherent and incoherent cross-talk effects \citep[see][for a detailed description on method and theory, respectively]{mesa2015, antichi2009}. The reduced images are then finally processed using the SHINE Specal pipeline (R. Galicher, private comm.), that provides anamorphism and flux normalisation, as well as speckle correction, using various flavors of angular and spectral differential imaging algorithms. A detailed description of the various tools is given in the following sections in case dedicated routines are applied for the reduction.

\noindent Observations of \object{HD~284149} and its brown dwarf companion were also acquired on 2015-02-03 employing the long slit spectroscopic (LSS) mode of the IRDIS instrument.
Contrary to the imaging ones, the LSS observations are always performed in field-stabilised mode. This ensures that the object is kept within the slit during the complete integration. 
Each LSS observing sequence included images taken with the coronagraph, an off-axis reference PSF (that is an image of the target star offset from the coronagraph), and the spectrum of an early-type star used for telluric calibration. Finally, a series of sky backgrounds was acquired.

Table~\ref{tab:obs} summarises the main observational setup for all the epochs. The reduced non-coronagraphic images for both IRDIS and IFS are shown in Figure~\ref{fig:fig1}.

\begin{table*}[ht]
\caption{Main characteristics of the  setup of the SPHERE observations of the \object{HD~284149} system}
    \centering
    \begin{tabular}{c|ccc|cc|c}
    \hline \hline 
    Date  & Mode & Filter & Coronograph    & \multicolumn{2}{|c|}{Total Integration Time (s)} &  Total Field of View  \\
                &      &        &                      &  IFS & IRDIS      &  Rotation (deg) \\
             \hline
        2015-02-03    & IRDIS\_LSS    & YJH  & Y             & --   & 180        & --    \\    
        2015-10-25    & IRDIFS        & H23  & Y             & 256  & 256        & 22.13 \\
        2015-11-27    & IRDIFS\_EXT   & K12  & N             & 82.5 & 56.08      & 29.71 \\
        \hline \hline 
    \end{tabular}
    \label{tab:obs}
\end{table*}

\subsection{IFS data}
\label{sec:2.1}
The IFS calibration data (dark, detector flat, spectral position frames, wavelength calibration and instrument flat) were treated using the data reduction and handling (DRH) software \citep{pavlov08}.\\
\noindent For the coronagraphic data of 2015-10-25 we obtained a calibrated data cube for each of the 64 frames which were then used to apply the principal component analysis procedure \citep[PCA, see][for details]{soummer2012, amara2012pynpoint}. They were registered using images with satellite spots symmetrical with respect to the central star \citep[waffles, see e. g.][]{2006ApJ...647..620S,2006ApJ...647..612M,2013aoel.confE..63L} and flux calibrated exploiting images taken with the star outside the coronagraph. 
The PCA routine that was used applies both the Angular Differential Imaging and the Spectral Differential Imaging techniques \citep[ADI and SDI, respectively. See e.g.][]{marois2006a}.
A more detailed description of the reduction procedures can be found in \citet{mesa2015} and \citet{mesa2016}.

\noindent The non-coronagraphic data of 2015-11-27 were instead binned and stacked in sets of 10, to obtain 160 frames rotated of $\sim$ $0.2^{\circ}$ from each other. This final data cube was reduced in a similar way as the coronagraphic data described above, with the exception of the flux calibration which was not needed, and the centering which was performed using the CNTRD IDL procedure.  

Figure~\ref{f:ifs_limits} shows the limits achieved with the non-coronagraphic observations, both in terms of magnitude and minimum mass of detectable companions. 

\subsection{IRDIS imaging data}\label{sec:irdis}
Data reduction for the IRDIS data was performed following the procedures described in \citet{zurlo2014}.
The IRDIS raw images were pre-reduced performing background subtraction, bad-pixels correction, and flat fielding. 
In a similar way to what done for the IFS coronographic images (see Section~\ref{sec:2.1}), waffle images were used to obtain a precise measurement of the position of the star center for each frame, also taking into account the detector's dithering positions.

In the case of the non-coronographic images no waffle or PSF reference images were taken. 
We therefore used one of the images in the sequence as reference in order to be able to apply the DC data reduction procedure. 
The resulting data cube had to be then manually registered to ensure an accurate centering.
For both epochs, after the preprocessing of each frame the speckle pattern subtraction was finally performed using both the PCA \citep{soummer2012, amara2012pynpoint} and the TLOCI \citep{Marois2014} algorithms, combined with the the ADI technique. 
Figure~\ref{f:limits} shows the achieved performances, in terms of magnitude and minimum companion mass, for both epochs.

\begin{figure*}[t]
\includegraphics[width=9cm]{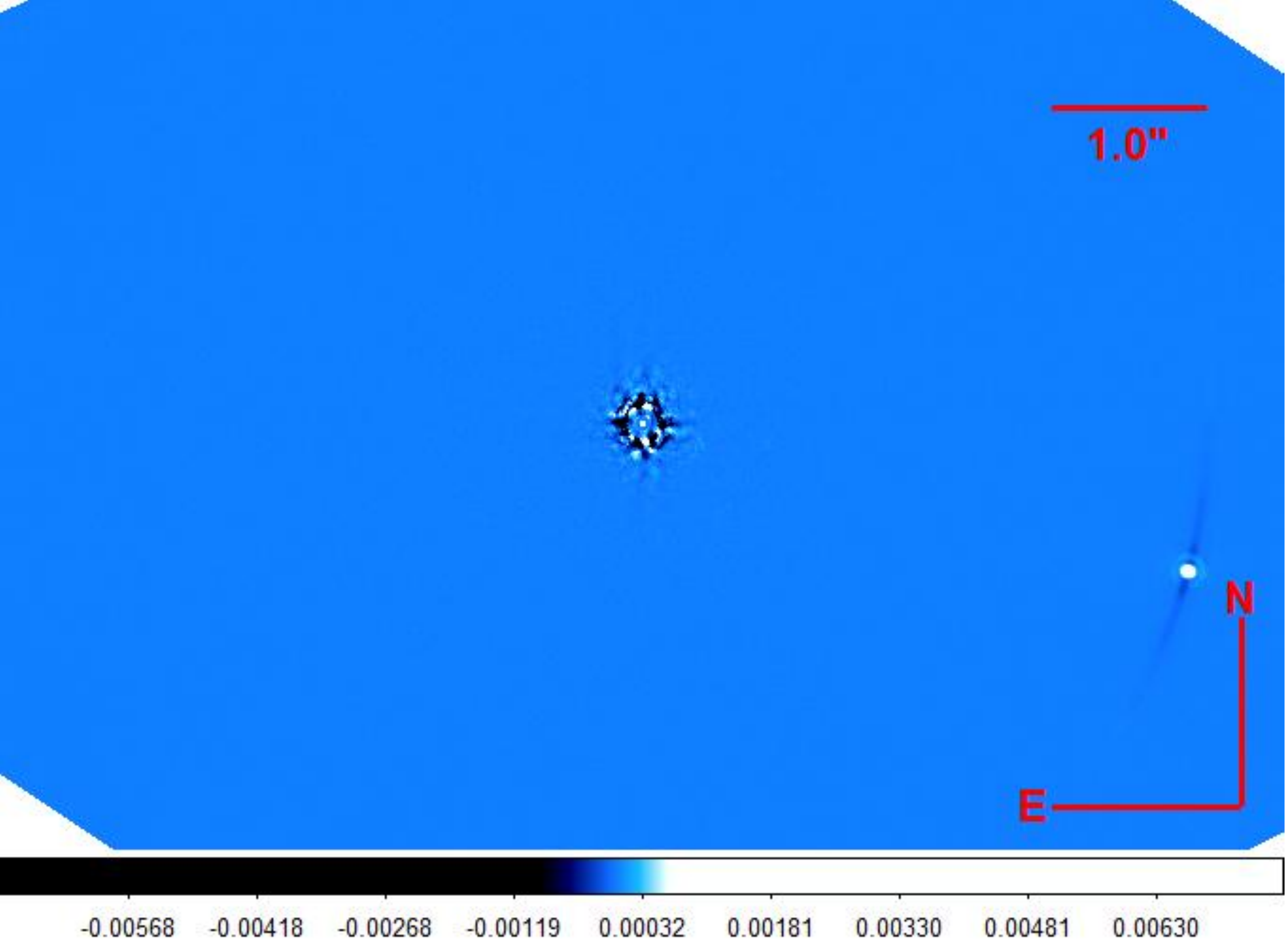}
\includegraphics[width=9cm]{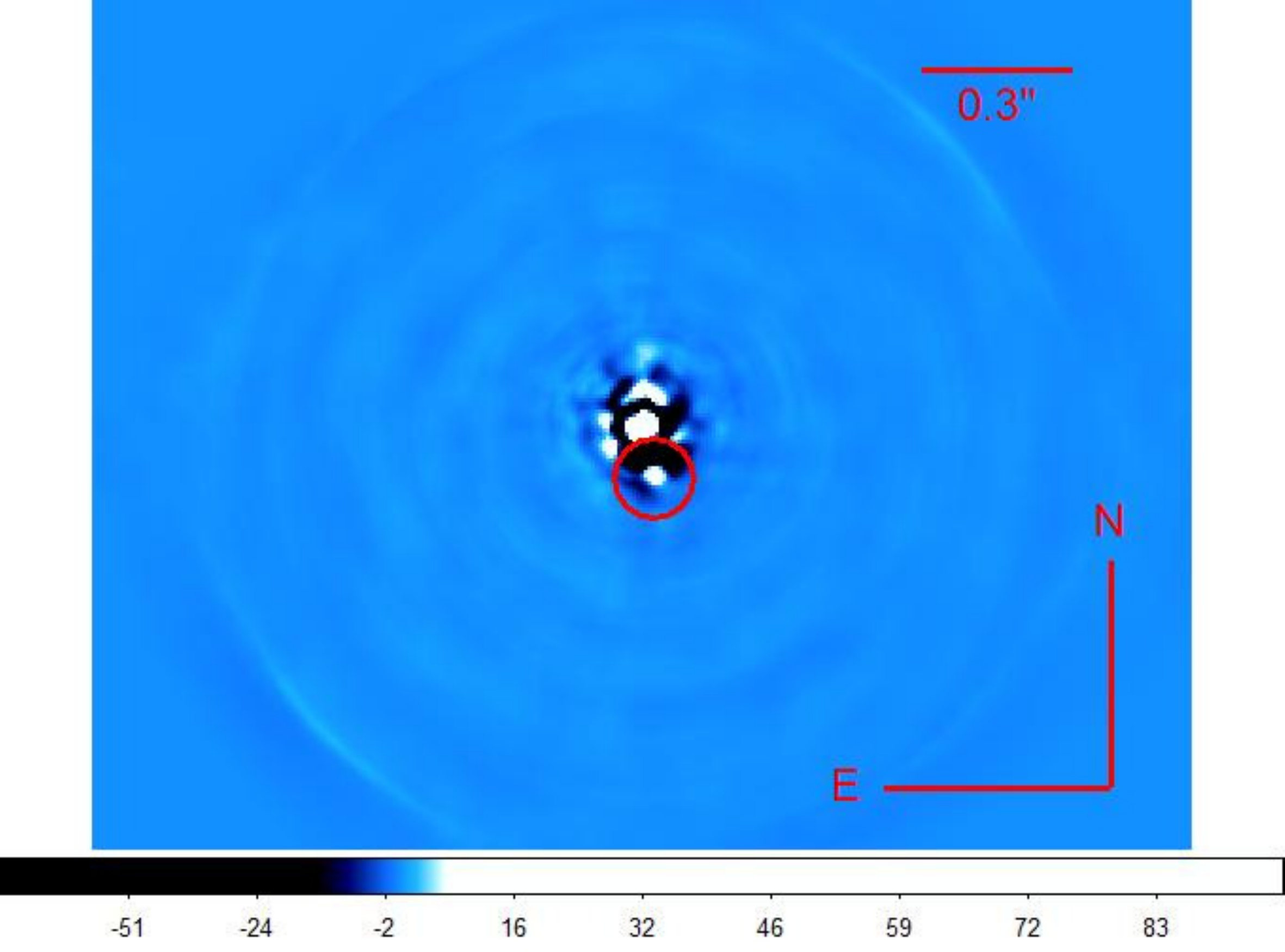}
\caption{Non coronagraphic IRDIFS\_EXT images of the \object{HD~284149} system acquired on November 27th 2015. 
{\bf Left:} Full-frame, TLOCI post-processed IRDIS image. The newly discovered close stellar companion \object{HD~284149~B} (see Section~\ref{sec:close}) is only barely visible in the IRDIS images. The known brown dwarf companion \object{HD~284149~b} (see e.g. Section~\ref{sec:4.2}) is clearly visible in the lower right corner.
{\bf Right:} PCA post-processed IFS image. The red circle marks the position of \object{HD~284149~B}.}
\label{fig:fig1}
\end{figure*} 

\begin{figure*}[htbp]
\begin{centering}
\includegraphics[width=0.4\textwidth]{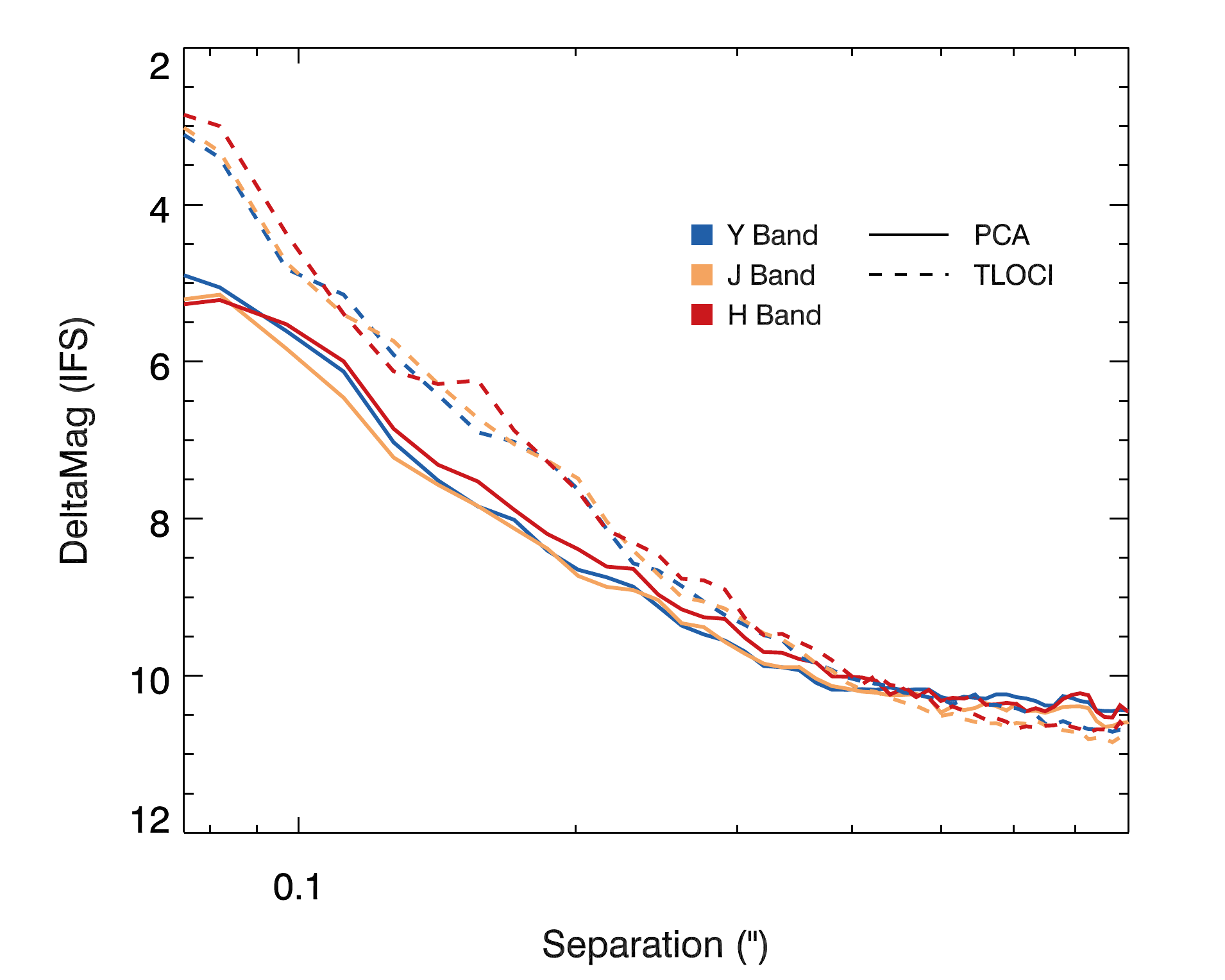}
\includegraphics[width=0.4\textwidth]{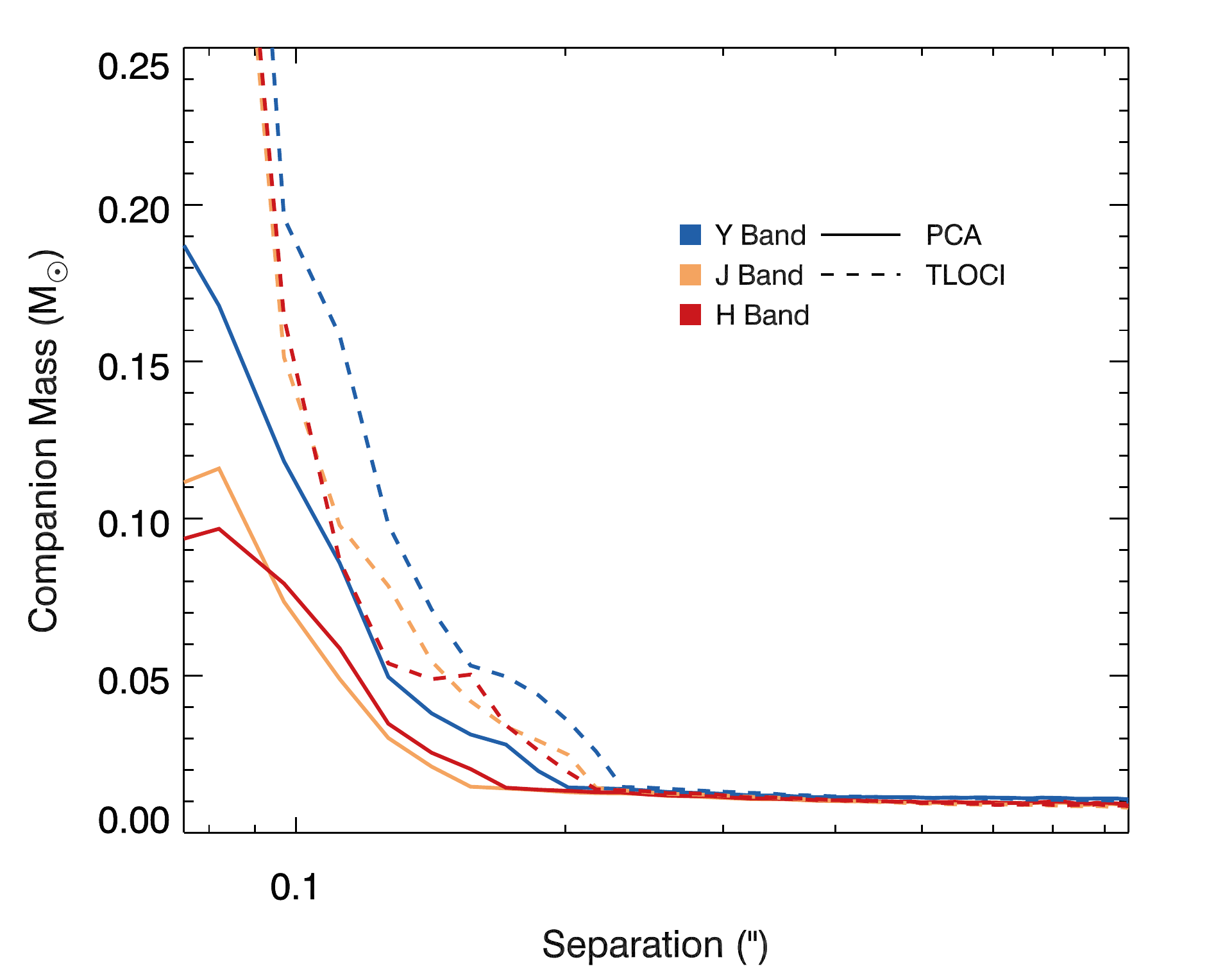}
\caption{IFS detection limits achieved during the non-coronagraphic observations of \object{HD~284149} in terms of $\Delta$mag (left panel) and minimum companion mass (right panel) vs separation.
The mass limits are evaluated using the COND models by \cite{baraffe03} and assuming an age of 35 Myrs. The different colours show the limits for the {\it Y}, {\it J} and {\it H} IFS filters respectively. The dashed line shows the limits evaluated using the images processed using the TLOCI method for speckle suppression \citep{Marois2014}, while the solid lines show the ones obtained using the images reduced using the PCA algorithm \citep{soummer2012, amara2012pynpoint}. }\label{f:ifs_limits}
\end{centering}
\end{figure*}

\begin{figure*}[htbp]
\begin{centering}
\includegraphics[width=0.4\textwidth]{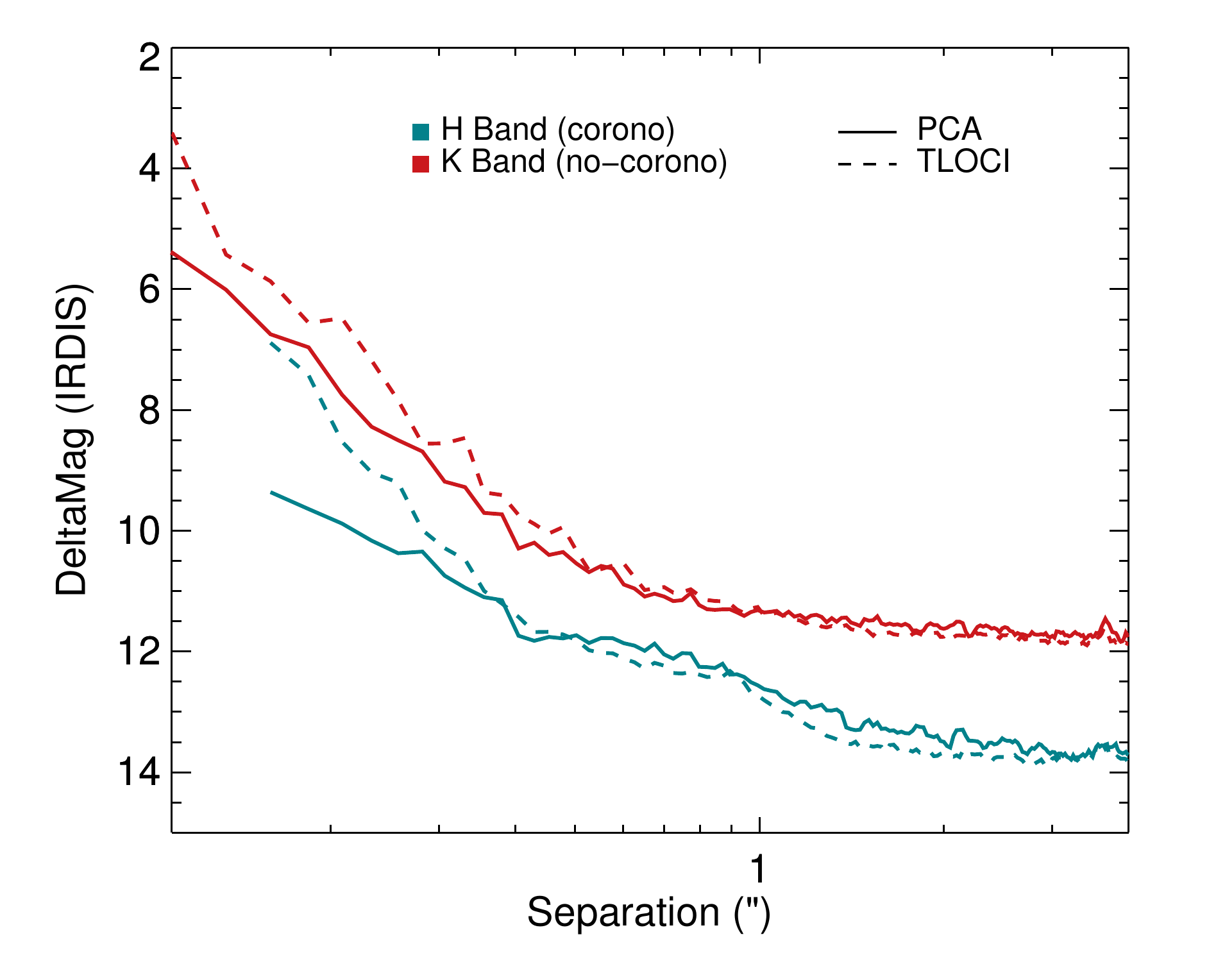}
\includegraphics[width=0.4\textwidth]{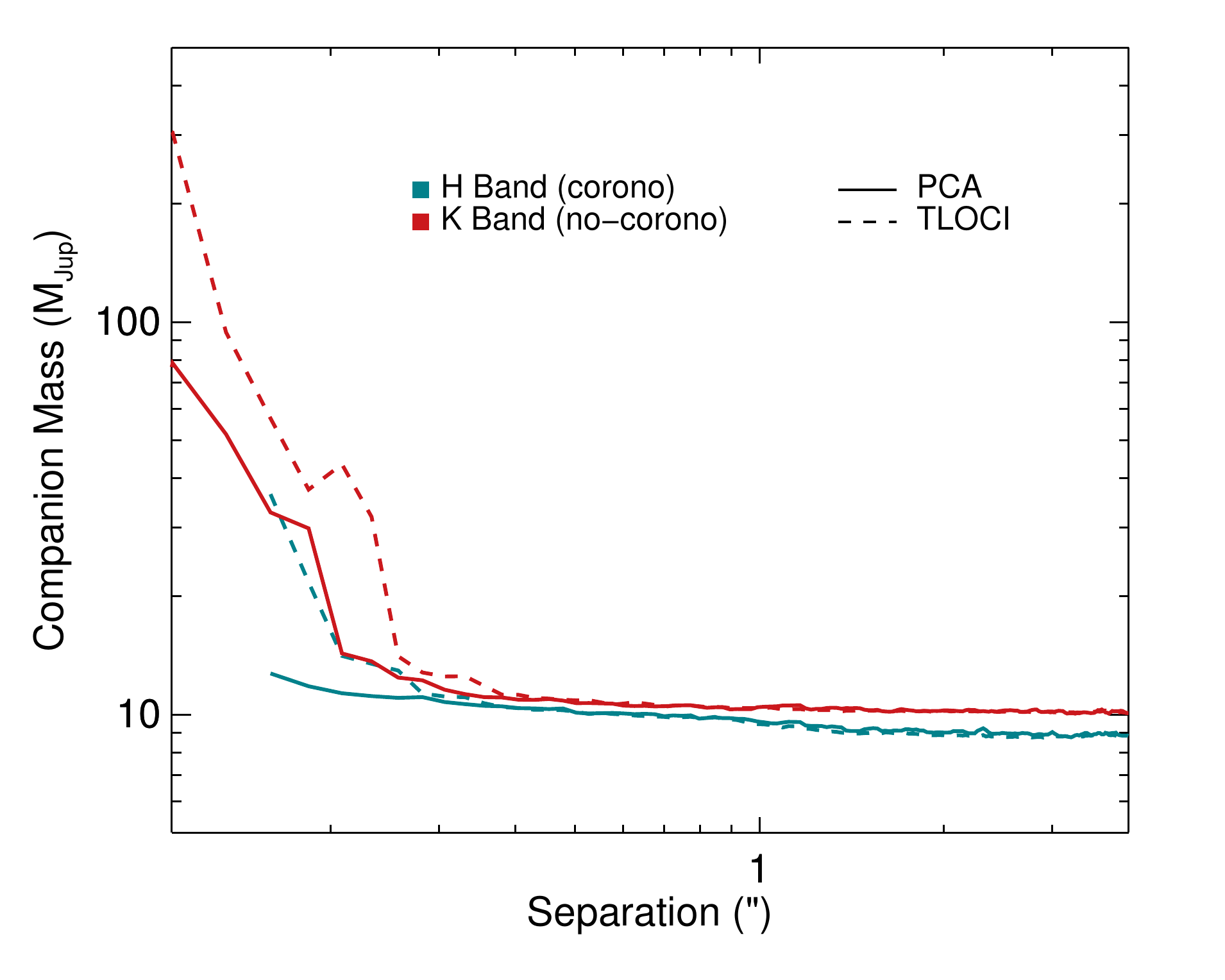}
\caption{IRDIS detection limits achieved during the observations of \object{HD~284149} in terms of $\Delta$mag (left panel) and minimum companion mass (right panel) vs separation. The mass limits are evaluated using the BT-Settl models by \cite{btsettl} and assuming an age of 35 Myrs. The blue and red curves show the limits achieved for the coronagraphic and non coronagraphic images, respectively. The dashed line shows the limits evaluated using the images processed using the TLOCI method for speckle suppression \citep{Marois2014}, while the solid lines show the ones obtained using the images reduced using the PCA algorithm \citep{soummer2012, amara2012pynpoint}.}\label{f:limits}
\end{centering}
\end{figure*}

\subsection{IRDIS Long Slit Spectroscopy data}\label{sec:lss}
The LSS mode of IRDIS allows to obtain both medium (R$\sim$350) and low resolution spectra (R$\sim$ 50). For this study we only used the medium resolution spectrum, which not only allows the spectral classification for the sub-stellar companion, but also provides information on key diagnostics such as e.g., the K{\sc i} and Na{\sc i} lines that are identifiable with this resolving power.

The LSS data was analysed using the SILSS pipeline \citep{vigan2016b}, which has been developed specifically to analyse IRDIS LSS data. The pipeline combines the standard ESO pipeline with custom IDL routines to process the raw data into a final extracted spectrum for the companion. After creating the static calibrations (background, flat field, wavelength calibration), the pipeline calibrates the science data and corrects for the bad pixels. It also corrects for a known issue of the MRS data, which produces a variation of the PSF position with wavelength because of a slight tilt ($\sim$1 degree) of the grism in its mount. To correct for this effect, the pipeline measures the position of the off-axis PSF in the science data as a function of wavelength, and shifts the data in each spectral channel by the amount necessary to compensate for the chromatic shift. All individual frames are calibrated independently for the two IRDIS fields. No speckle subtraction has been applied given the negligible flux of the central star at the separation of the sub-stellar companion.

Extraction and wavelength calibration of the 1D spectrum for the star, the companion and the early-type standard have been performed using {\it IRAF}\footnote{IRAF is the Image Reduction and Analysis Facility, a general purpose software system for the reduction and analysis of astronomical data. IRAF is written and supported by National Optical Astronomy Observatories} tasks.
We have tested two different extraction procedures, one which uses a fixed window extraction of 6 pixels and another one which uses a window size which is function of $\lambda$ / D (where D is the telescope diameter). The latter results in windows of roughly 3 pixels at 0.9 $\mu$m and 7 pixels at 1.8 $\mu$m. As the spectra provided by the two extraction methods appear to be in good agreement, we decided to keep the standard extraction (without pixel weighting) with a fixed width of 6 pixels, as it ensures higher signal-to-noise ratios (SNR, especially in the blue part). The SNR is $\sim$ 15 at 1.3 $\mu$m.

As mentioned in Section~2, an early type star (The A3IV star \object{HD~77281}) was observed as part of the LSS observing sequence to obtain a more accurate wavelength solution. 

The spectrum of \object{HD~77281} was also used to correct the spectra of \object{HD~284149} for the contamination of telluric lines, using the IRAF task $telluric$. This routine allows to account for small difference in the line intensity as well as possible wavelength shifts between the spectra of the science target and the template star.

The contrast spectrum of the companion was obtained by dividing by the spectrum of the primary extracted in an aperture of the same width. Then, in order to perform a comparison with a library of spectra and theoretical models, we calculated the flux spectrum of the brown dwarf by multiplying the contrast spectrum by the flux spectrum of the primary. To calculate the flux spectrum of \object{HD~284149} we used the models by \cite{brott2005}\footnote{available for download at\\ \url{ftp://ftp.hs.uni-hamburg.de/pub/outgoing/phoenix/GAIA}}, with B$-$V values retrieved from Nomad \citep{zacharias04}, 2MASS \citep{skru06} and WISE photometric information \citep{wright2010} and adopted a reddening of E(B-V)=0.08 mag \citep[see e.g.,][]{bonavita2014}. We verified the reddening value from \cite{bonavita2014} using the IRSA tools available at \url{http://irsANSWER:ipac.caltech.edu/applications/DUST/} which yield a values of E(B-V) = 0.25-0.30. Considering that these values refer to total reddening within the Milky Way and considering the galactic position of \object{HD~284149}, we concluded that these values are fully compatible with the adopted ones.

The best spectral fit is obtained assuming $T_{\rm eff}$=6000K and log$g$=4.5 dex (with $g$ in cm~s$^{-2}$), in very good agreement with literature estimates in the ranges $T_{\rm eff}$=5876-6184 K
\citep{bailer2011}, $T_{\rm eff}$=5931K \citep{mcdonald2012}, and $T_{\rm eff}$=5970-6100 K \citep{bonavita2014}.

\section{Stellar Properties}
\label{sec:star}

A comprehensive analysis of various stellar properties and age indicators of the star was performed in \citet{bonavita2014}, yielding an estimate of $25^{+25}_{-10}$ Myr.
The availability of Gaia DR1 data sets \citep{gaia_dr1} as well as the recent revision of the ages of several young moving groups used as reference \citep{bell2015} calls for a reassessment of the stellar properties.
We also exploit photometric time series to refine the rotation period determination and better characterise the photometric variability of the star.

\subsection{Stellar properties from Gaia-DR1}
\label{sec:gaia}

\begin{figure*}
\begin{centering}
  \includegraphics[width=0.45\textwidth]{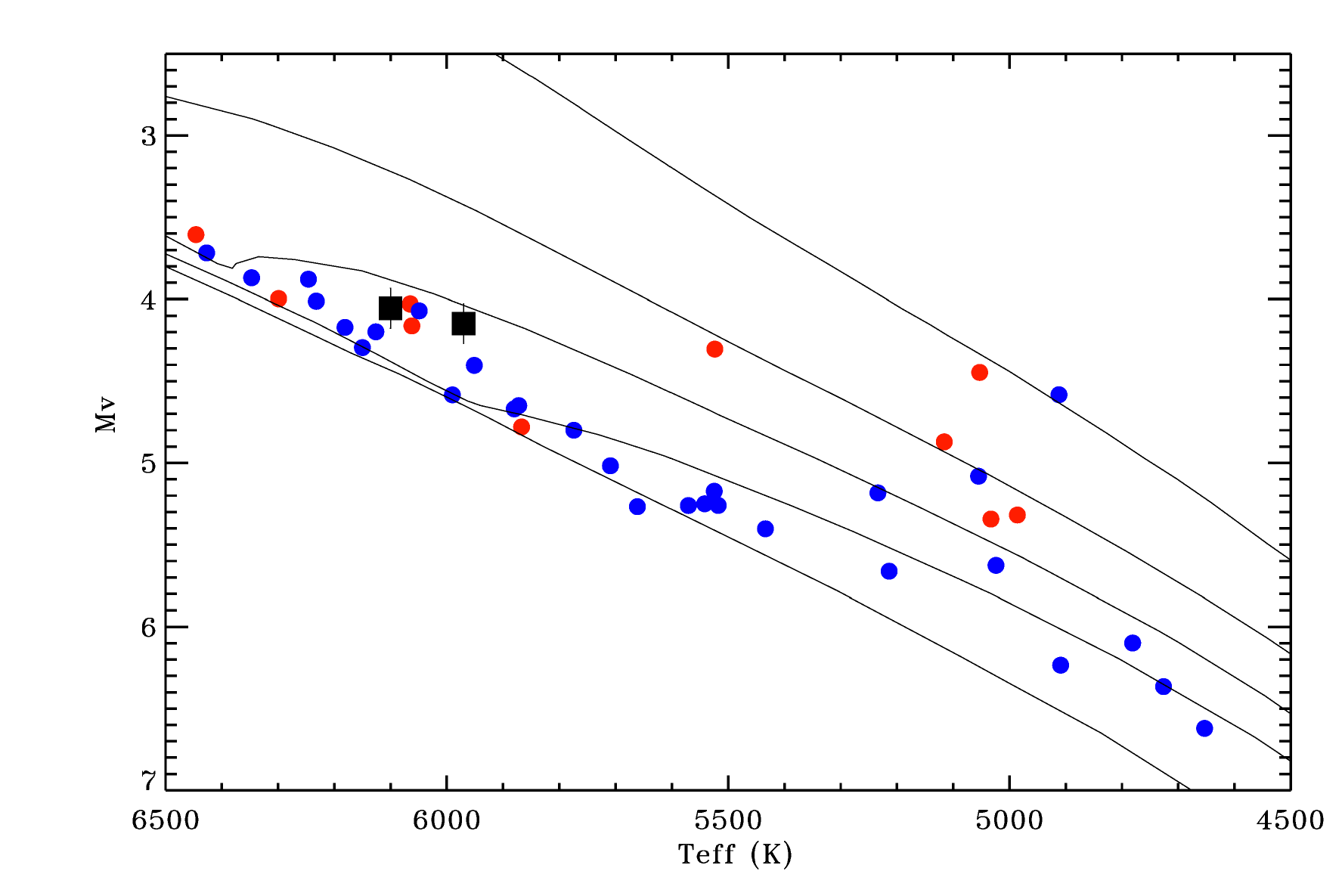}
  \includegraphics[width=0.45\textwidth]{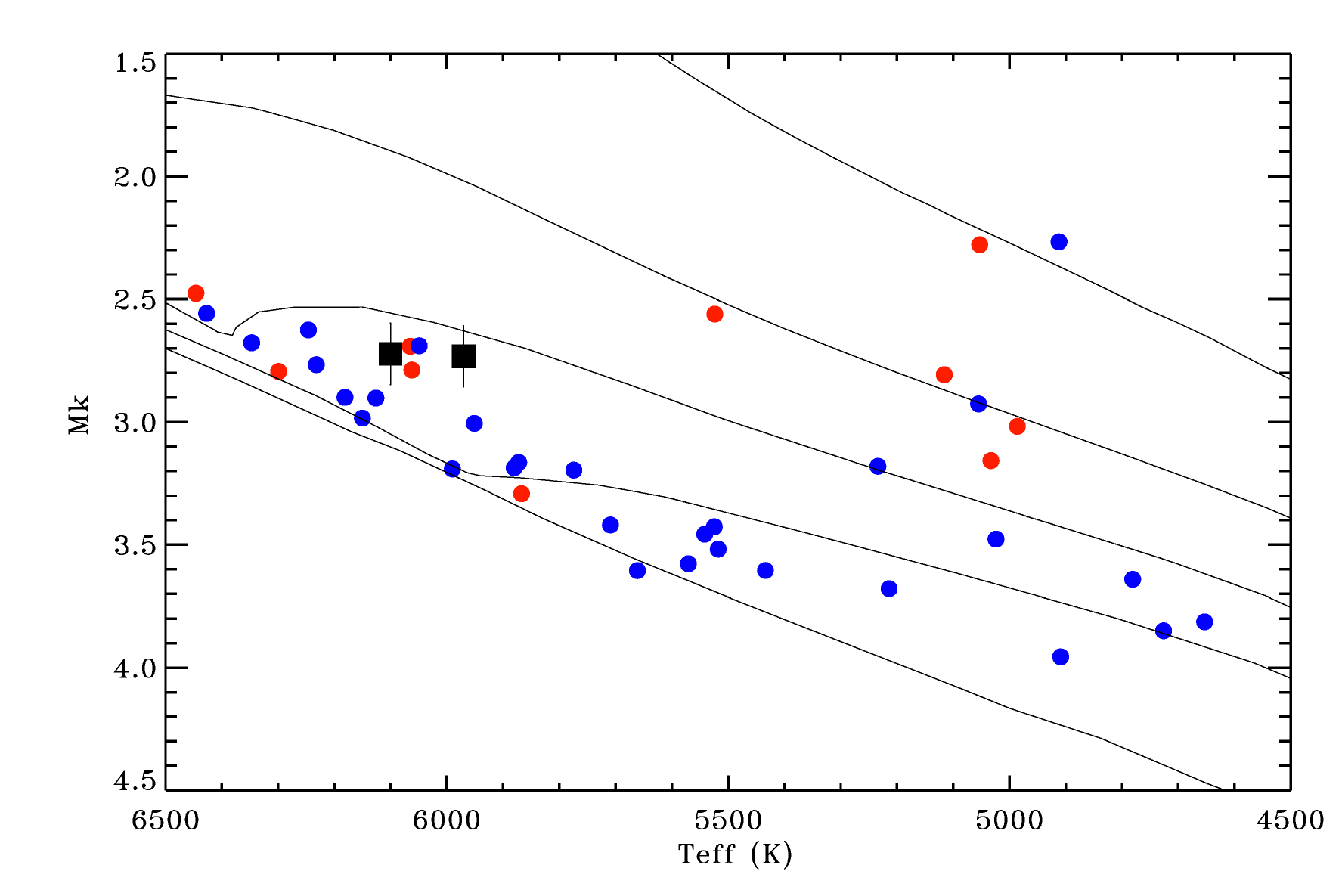}
\caption{Position of \object{HD~284149} (black squares) on the H-R diagram for the temperatures corresponding to F8 and G1 spectral classification \citep[see][for {\it V} and {\it Ks} band magnitudes (upper and lower panels, respectively]{bonavita2014}. Over plotted are the 5, 12, 20, 30, and 70 Myr isochrones of solar metallicity from \citet{bressan2012} and the members of the $\beta$ Pic~moving group (red circles) and Tuc-Hor association (blue circles) with available trigonometric parallax from the list of \citet{Pecaut13}.}
\label{fig:isoc}
\end{centering}
\end{figure*}

Gaia-DR1 yields a trigonometric parallax of 8.51$\pm$0.27 mas for \object{HD~284149}. Even after inclusion of a systematic error of 0.30 mas, as recommended by \cite{GDR1}, this represents a substantial improvement with respect to the value by \citet{vanleeuwen2007} (9.24$\pm$1.58~mas). The resulting value of the distance of \object{HD~284149} is then 117.50$\pm$5.5~pc.
The slightly larger intrinsic luminosity makes the star well detached from the ZAMS, allowing a reliable age estimate based on comparison with pre-main sequence evolutionary models, while previous attempts were inconclusive due to the error on parallax.
Figure~\ref{fig:isoc} represents an update of Figure~2b by \citet{bonavita2014}.

Ages younger than 15~Myr and older than 30~Myr appear unlikely.
On the other hand, plotting the members of $\beta$ Pic~moving group and Tuc-Hor association from \citet{Pecaut13} seems to suggest that, while the most probable locus of Tuc-Hor members is at fainter magnitudes, some individual members are in similar position to \object{HD~284149} in the  HR diagram. Nevertheless, an age as young as $\beta$ Pic moving group is favoured by this comparison. 
A dedicated analysis aimed at obtaining a more precise estimate of the age is presented in Section~3.3.

\subsection{Photometric analysis and rotation period}
\label{sec:photom}
We retrieved a photometric time series collected during the 2004 and 2006 observation seasons from the SuperWASP public archive \citep[Wide Angle Search for Planets][]{Butters10}. It consists of 3892 V-band measurements with an average photometric precision $\sigma$ = 0.006 mag. 
After the removal of outliers from the time series applying a moving boxcar filter with 3$\sigma$ threshold, we averaged consecutive data collected within 30 minutes, and finally we were left with 652 averaged magnitude values for the subsequent analysis.

We used the Lomb-Scargle periodogram analysis \citep[LS][]{Scargle82}, with the prescription of \cite{Horne1986}, on the SuperWASP time series to search for the rotation period of \object{HD~284149} (see Figure~\ref{periodogram}). From the computed stellar radius (R $\sim$ 1.36\,R$_\odot$, see below) and the measured projected rotational velocity, the rotation period is expected to be shorter than about 2 days. We therefore carried out our period search in the period range 0.1--10d.

In the left panel of Figure~\ref{periodogram}, we plot the LS periodogram as black solid line, whereas the red dotted line is the spectral window function. 
We detected  a number of highly significant power peaks with False Alarm Probability (FAP) $<$ 0.1\%.
The FAP is the probability that a power peak of that height simply arises from Gaussian noise in the data, and was estimated using a Monte-Carlo method, i.e., by generating 1000 artificial light curves obtained from the real one, keeping the date but permuting the magnitude values according to their uncertainty.\\
The most significant peak is at P = 1.051$\pm$0.005\,d (with a FAP of $\sim10^{-6}$), which we assume to represent the stellar rotation period. 
This value is similar, although formally significantly different, to the rotation period P = 1.073\,d previously measured by \cite{Grankin2007} and inferred from data collected at the Mt. Maidanak Observatory.
All the other significant power peaks at shorter periods in the LS periodogram are harmonics, arising from the one-day sampling interval imposed by the rotation of the Earth and the fixed longitude of the observation site. 
We note a second highly significant power peak at P = 1.074\,d, almost identical to the literature value. Such a period may arise from the presence of two spot groups at different average latitudes on a differentially rotating star. 
In this case, \object{HD~284149} would have a lower limit of the surface differential rotation of $\sim$2\%.
However, when we fold the light curve with the rotation period, as shown in the right panel of Figure~\ref{periodogram}, we note that in the 2006 observation season the light curve undergoes a rapid change of the phase of minimum ($\Delta\phi \simeq$ 0.25) and of the peak-to-peak amplitude (from $\Delta$V = 0.02\,mag to $\Delta$V = 0.04\,mag), suggesting a rapid reconfiguration of the active regions.
Indeed, it is interesting to consider the possible effect of active regions growth and decay (ARGD) on spurious variations of 
the photometric period. \cite{Dobson1990} and \cite{Donahue1992} have  proposed a method, the so-called pooled variance analysis, to estimate  the  time  scale  of  evolution  of  an  active  region.  This method is based on the analysis of the variance in photometric  time  series  over  different  time  scales.  
The variance profile of \object{HD~284149} is plotted in the left panel of Figure~\ref{fig:pool}.
We note that the variance monotonically increases and does not show any plateau at time scales longer than the rotation period, which is marked with a vertical dotted line in Figure~\ref{fig:pool}. This behaviour is characteristic of 'activity-dominated' stars \citep[see][]{messina2003}, that is of stars whose active regions are not stable, but evolve with the same time scale as the stellar rotation.
This would explain the large phase scatter, despite the high photometric precision, the presence of the two close peaks in the periodogram, the family of lower power peaks clustering around the two major peaks, and the difference with respect to the literature value. We conclude that secondary peaks are due to the ARGD rather than to surface differential rotation.

It is interesting to see that combining the available V-band photometry from Hipparcos (to which we applied a correction of -0.137\,mag to transform into the Johnson system), from ASAS \citep[All Sky Automated Survey, see ][]{ASAS}, \cite{Grankin2007} and SuperWASP (to which, to be consistent with ASAS data we applied a correction of -0.23~mag) \object{HD~284149} exhibits a long-term brightness variation which likely is related to a star spot cycle with an amplitude of $\Delta$V $\ga$ 0.2\,mag (see Figure~\ref{fig:pool}).

Using the brightest visual magnitude V = 9.55\,mag inferred from the ASAS time series, a reddening E(B$-$V) = 0.08~mag \citep{bonavita2014}, distance d = $117.50\pm5.5$~pc from Gaia-DR1, bolometric correction $BC_{\rm V} = -0.07\pm0.02$ \citep{Pecaut13}, we derive the luminosity $L = 2.20\pm0.45~L_{\odot}$.
Using the measured effective temperatures in the range T = 5970--6100~K \citep{bonavita2014}, we derive an average stellar radius $R = 1.36\pm0.33~R_{\odot}$.
Combining rotation period and projected rotational velocity $v\sin{i} = 27.0\pm1.9~kms^{-1}$ \citep{bonavita2014}, we  derive the inclination of the rotation axis $i = 25\pm5^{\circ}$.

\begin{figure*}
\includegraphics[width=100mm, height=180mm, angle=90]{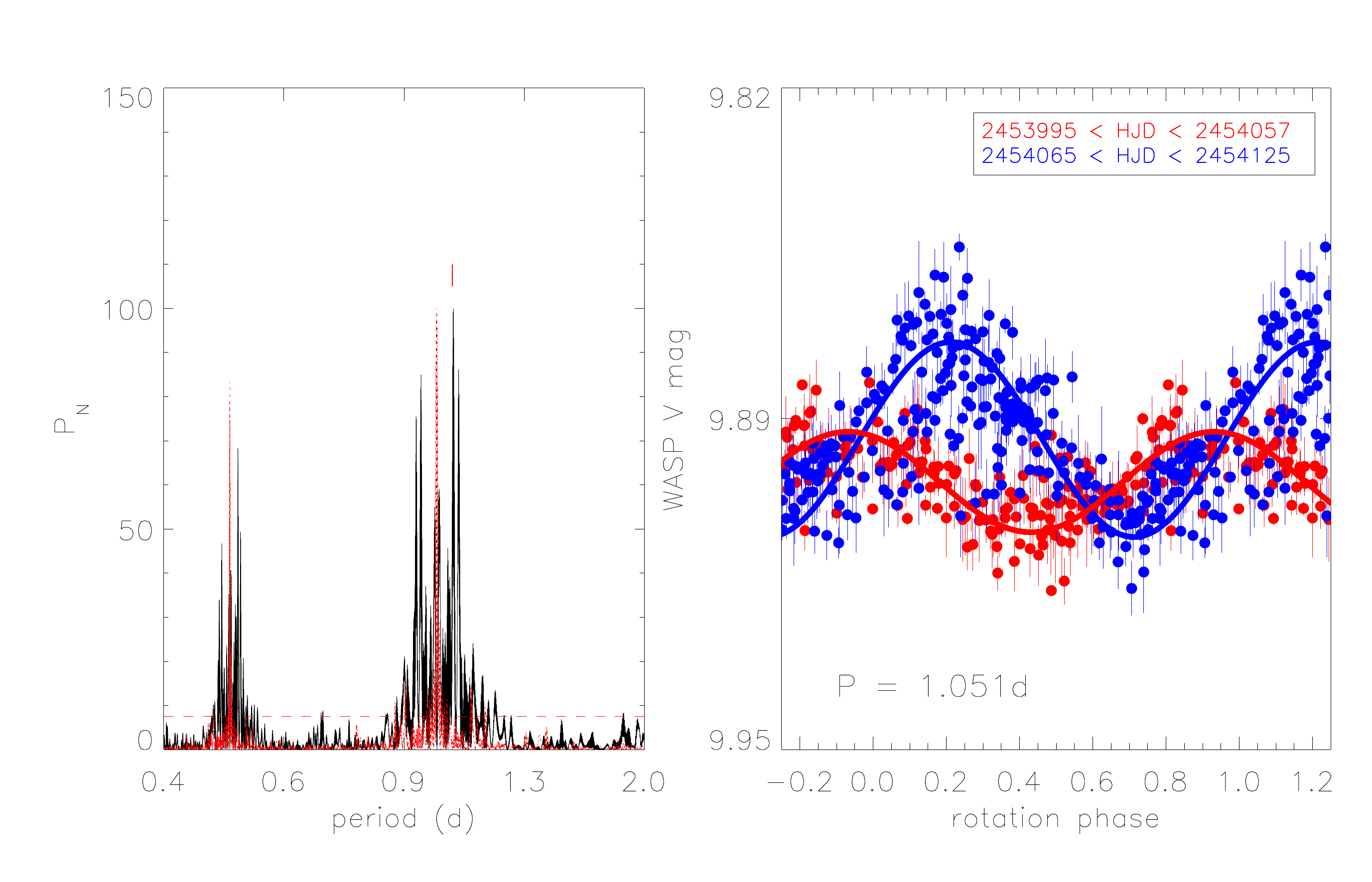}
\caption{Left panel: Lomb-Scargle periodogram of \object{HD~284149} of the SuperWASP V magnitude time series  collected in 2004 and 2006. The normalised power (P$_N$) (black solid line) is plotted versus period. The red dotted line represents the spectral window function, whereas the red horizontal dashed line indicates 
the power level corresponding to a FAP = 0.1\%. Right panel: SuperWASP V magnitudes of \object{HD~284149} collected in 2006 and  phased with the photometric rotation period $P = 1.051\pm0.005$~days. Solid lines are the sinusoidal fits to phased data collected in the date range 2453995 $<$ HJD $<$ 2454057 (red) and in 2454065 $<$ HJD $<$ 2454125 (blue). }
\label{periodogram}
\end{figure*}

\begin{figure*}[h]
\begin{centering}
\includegraphics[width=50mm, angle=90]{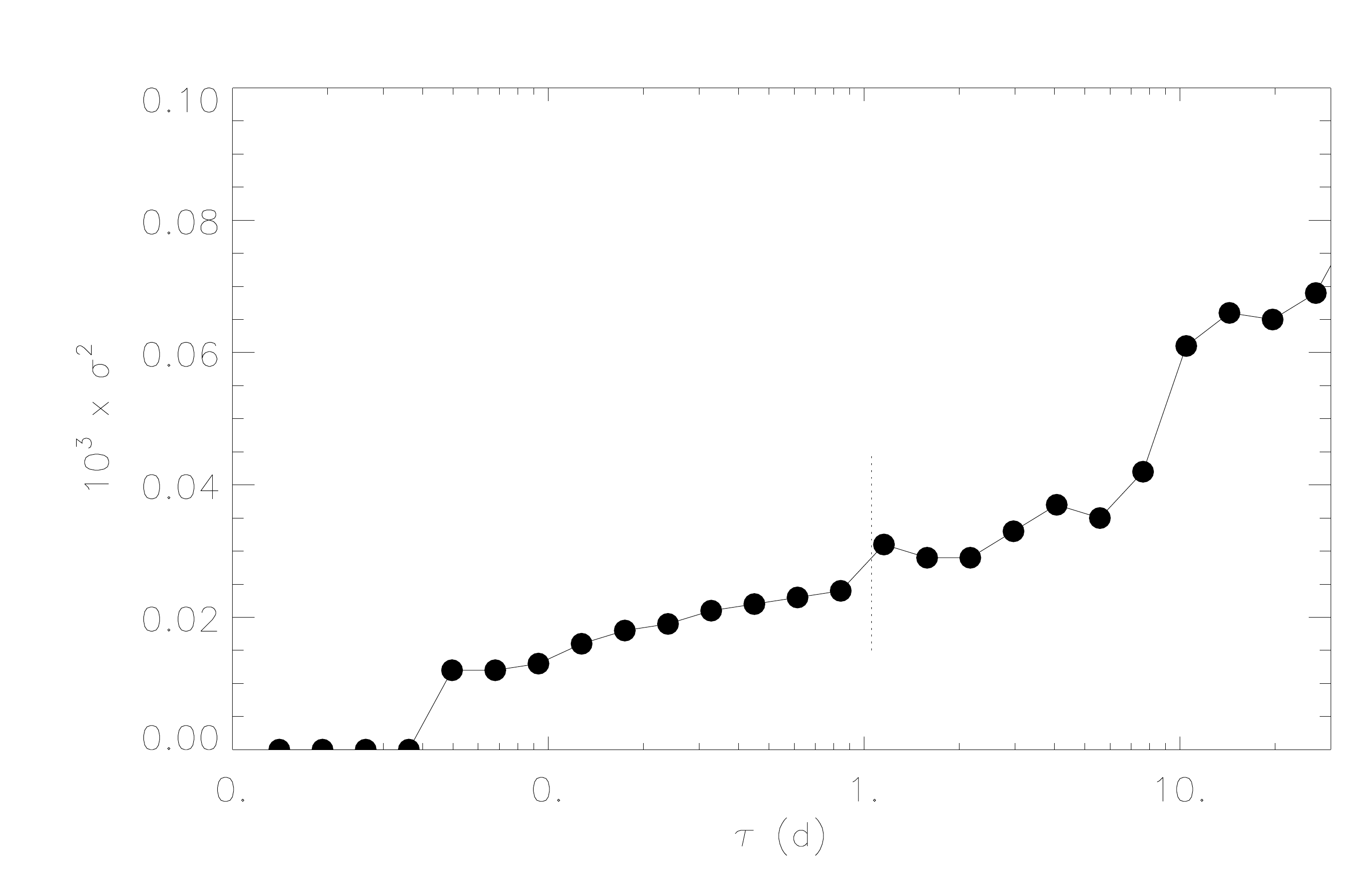}
\includegraphics[width=50mm,angle=90]{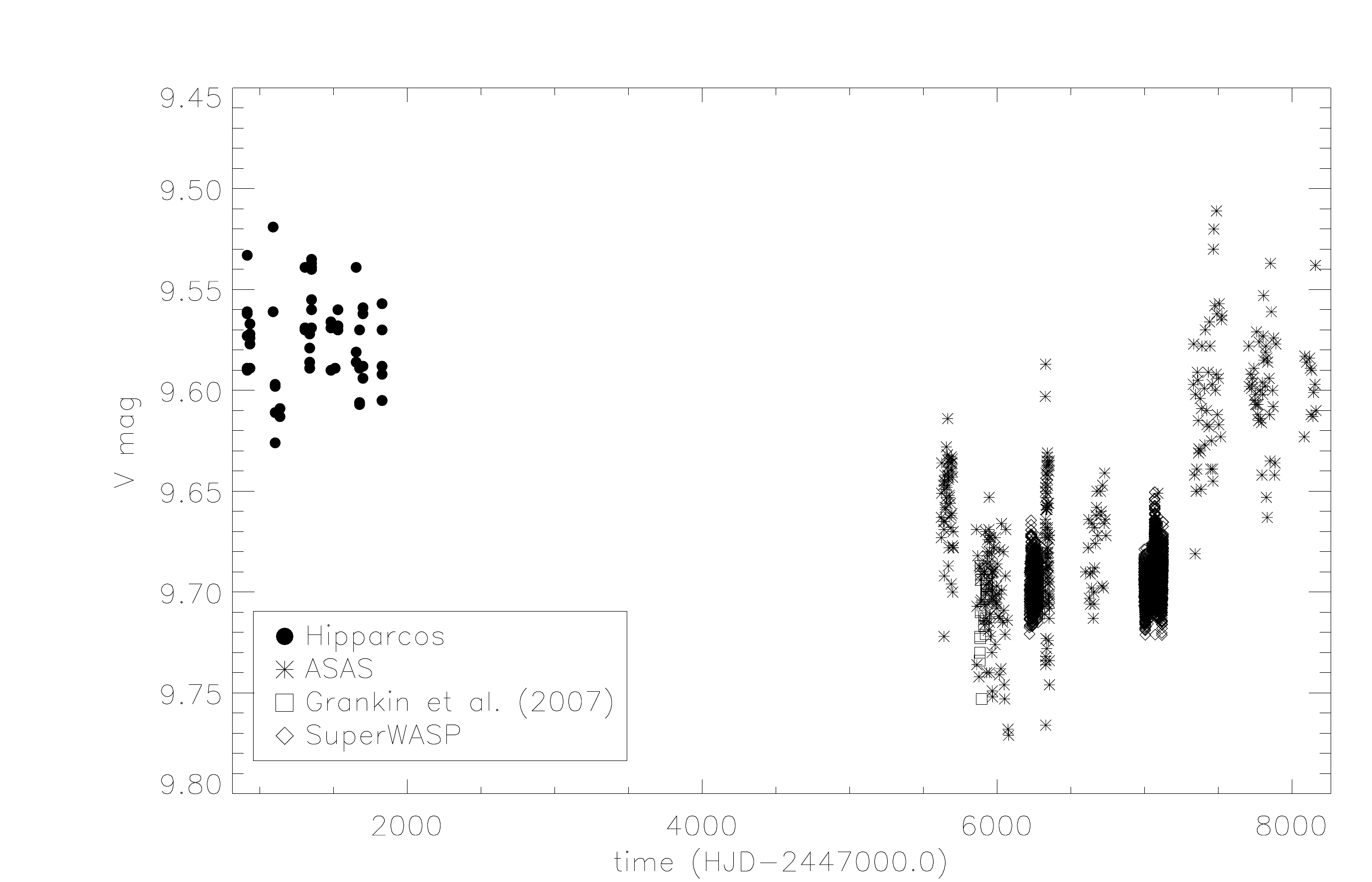}
\caption{Left Panel: Pooled variance profile of \object{HD~284149} computed on the SuperWASP V-mag time series. Right Panel: Long-term V-band magnitude time series of \object{HD~284149}.}
\label{fig:pool}
\end{centering}

\end{figure*}

\subsection{Stellar age}
\label{sec:age}
A comparison of the rotation period of \object{HD~284149} with the distribution of rotation periods of associations of known age allows us to constrain its age. 
We find that the rotation period P = 1.051\,d well fits into the period distribution of Tucana/Horologium, Carina, and Columba that have an age originally quoted to be 30 Myr but recently revised to 42-45 Myr \citep{bell2015}. On the other hand, \object{HD~284149} rotates slightly slower than most of the equal-mass association members of the 25$\pm$3 Myr $\beta$ Pic Associations \citep[e.g.][]{Messina16} with just a couple of confirmed members in the same region in a color-period diagram (Messina et al. 2017, submitted).
Finally \object{HD~284149} rotates significantly faster than the equal-mass members of Pleiades.\\
Therefore, on the basis of rotational properties, our target seems to have an age close to that of Tuc-Hor, but possibly as young as $\beta$ Pic moving group members.
The Lithium EW is compatible with both $\beta$ Pic moving group and Tuc-Hor moving group members \citep{bonavita2014} and clearly rules out ages as old as Pleiades or AB Dor MG.
The isochrone fitting \footnote{Figure~\ref{fig:isoc} is based on comparison with \citet{bressan2012} models.
The choice of adopted models does not affect significantly the results. Indeed we extended the comparison presented in \cite{desidera15}  to the recently published models by \citet{choi2016}, finding good agreement for the masses and ages of interest for this study.} indicates an age similar to that of $\beta$ Pic moving group members, although when plotting Tuc-Hor members on CMD a couple of individual late F - early G members are as bright as \object{HD~284149}. 
Therefore, we can conclude that \object{HD~284149} has an age between 20 and 45 Myr, with a most likely value of 35 Myr.

A finer age assessment is limited by the intrinsic spread of the various indicators at fixed age (as retrieved from moving group members). 
Improvements could arise from analysis of additional coeval stars coming with \object{HD~284149}.
{\cite{daemgen2015} found that this star, together with several other objects known as bona-fide Taurus members, has similar space position and kinematics to those of the Taurus star forming region, but also shows evidence of a distinctly older age. 
They proposed an age of about 20 Myr for what is now called the {\it Taurus-Ext Association} \citep[see Appendix A in][and references therein]{daemgen2015} from comparison of Lithium EW with that of the oldest Sco-Cen groups \citep{pecaut16}}.
Such proposed age for the group is at the young edge of our age range.

Finally, we mention that the low mass companion detected with SPHERE (see Sect. \ref{sec:close}) has negligible impact on the age indicators, due to its faintness with respect to the central star, apart from some minor effects on kinematics.

\section{Results}

\subsection{Discovery of a close low-mass stellar companion}
\label{sec:close}
\begin{figure}
\begin{centering}
\includegraphics[width=9cm]{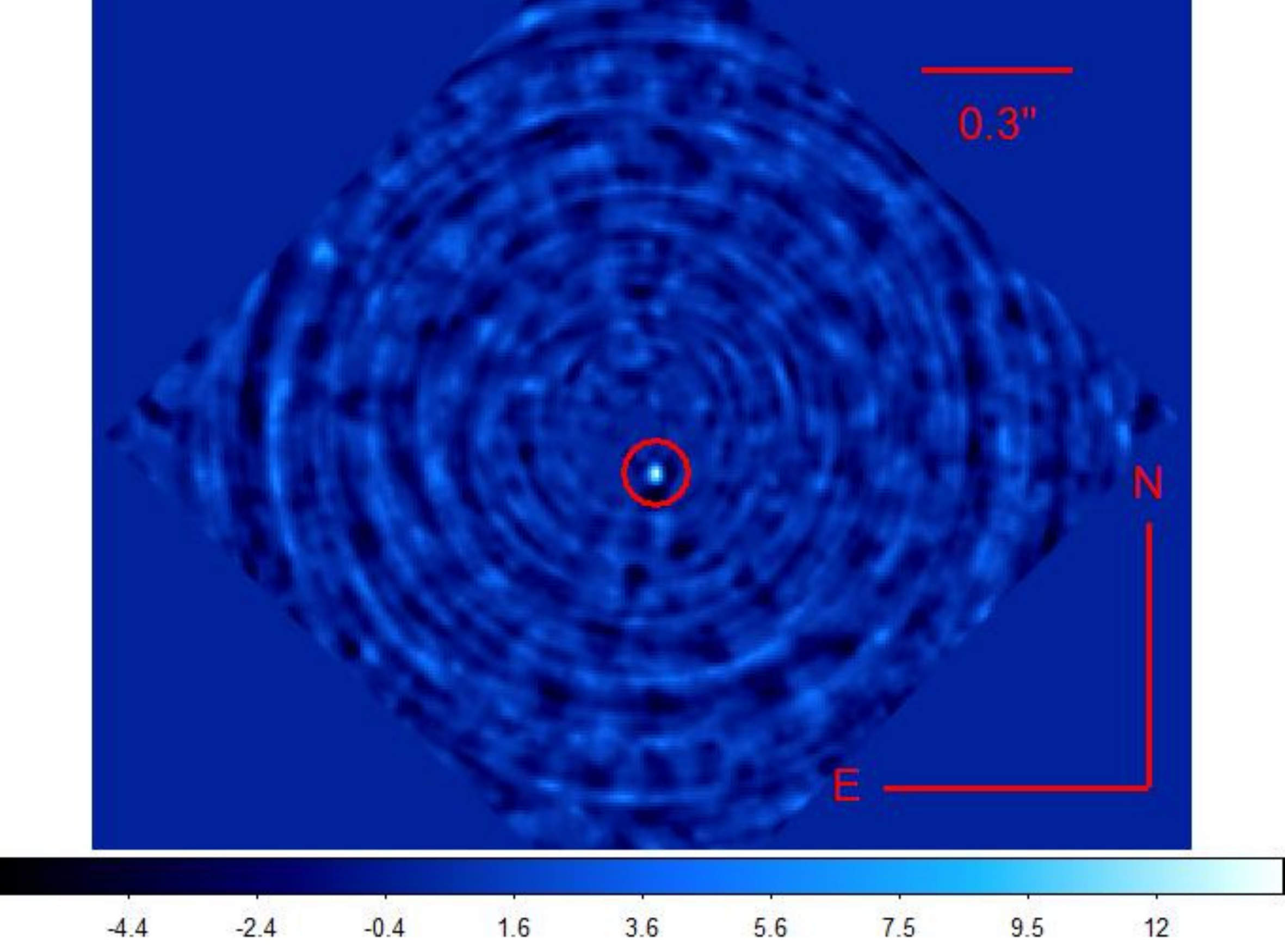}
\caption{Signal to noise map obtained for the non-coronagraphic IFS data 
of \object{HD~284149}. The position of the newly discovered low-mass stellar companion \object{HD~284149~B} is marked by a red circle. The peak of the SNR at the companion position is of $\sim$15.}\label{fig:SNmap}
\end{centering}
\end{figure}

The non-coronagraphic IFS data of \object{HD~284149} revealed a stellar companion at very small separation from the star.
The companion (hereafter \object{HD~284149B}~B) was visible but partially obscured by the coronagraph in the previous epoch, which then justified the choice for the non-coronagraphic mode for the following observations. 
\object{HD~284149~B} was successfully retrieved in both the IFS and the IRDIS non-coronagraphic images, both reduced using the PCA algorithm for speckle suppression. 
The right panel of Figure~\ref{fig:fig1} shows the processed IFS images, and the corresponding S/N map is shown in Figure\ref{fig:SNmap}. In both figures the position of \object{HD~284149~B} is marked by a red circle.  

\subsubsection{Astrometry and photometry}
\label{sec:411} 

\begin{table*}
    \caption{IFS and IRDIS astrometry of \object{HD~284149~B} from the non-coronagraphic data obtained in November 2015. 
    }
    \centering
    \begin{tabular}{cr@{\,$\pm$\,}lr@{\,$\pm$\,}lr@{\,$\pm$\,}lr@{\,$\pm$\,}lr@{\,$\pm$\,}lr@{\,$\pm$\,}}

    \hline\hline
    filter &
    \multicolumn{4}{c}{Separation} &
    \multicolumn{2}{c}{PA} &
    \multicolumn{2}{c}{$\Delta$ mag} &
    \multicolumn{2}{c}{Mass}\\
    
     & 
    \multicolumn{2}{c}{(mas)} &
    \multicolumn{2}{c}{(au)} &
    \multicolumn{2}{c}{($^{\circ}$)} &
    \multicolumn{2}{c}{} &
    \multicolumn{2}{c}{($M_{\odot}$)} \\
    \hline    
    {\it Y}  & 90.02 &  1.32 & 10.58 & 0.15 & 195.19 &  1.86  & 4.54  & 0.13  &  0.23 & 0.02 \\
    {\it J}  & 90.73 &  1.22 & 10.66 & 0.15 & 194.99 &  1.66  & 4.57  & 0.11  &  0.17 & 0.02 \\
    {\it H}  & 94.18 &  1.16 & 11.07 & 0.16 & 195.03 &  1.47  & 4.87  & 0.23  &  0.12 & 0.01 \\    
    {\it K1} & 80.16 & 18.57 &  9.42 & 0.66 & 187.53 & 67.03  & 4.07  & 0.24  &  0.19 & 0.02 \\
    {\it K2} & 81.72 & 18.27 &  9.60 & 0.66 & 193.22 & 34.00  & 3.79  & 0.21  &  0.20 & 0.02 \\ 
    \hline
    \multicolumn{8}{l}{ Adopted values} \\
             & 91.78 &  2.16 & 10.71 &  0.34 & 195.06 &  0.16  & \multicolumn{2}{c}{--}  &  0.16 &  0.04 \\
    \hline
    \end{tabular}
    \tablefoot{A true north position of $-1.7470\pm0.0048^{\circ}$ and pixel scale of $7.46\pm0.02~mas/pixel$ and $12.255\pm0.009~mas/pixel$, for IFS and IRDIS respectively \citep[see][for details]{2016SPIE.9908E..34M} were used. The value of the separation in AU was obtained using the value of the parallax of the star provided by the Gaia satellite \citep{GDR1}. The values of the masses are derived using the COND Models by \cite{baraffe03} and taking into account both the uncertainties on the photometry and on the age value. The uncertainties listed take into account all the possible sources of errors. Note that in the case of the astrometry, the dominant source of error is the uncertainty on the centering of the star, whether the error on the mass is dominated by the uncertainty on the age estimate.} \label{tab:photoastro_B}
\end{table*}

Table~\ref{tab:photoastro_B} shows the values of IRDIS and IFS astrometry and photometry for \object{HD~284149~B}.
The IFS photometry for \object{HD~284149~B} in the {\it Y}, {\it J} and {\it H} band was obtained considering the median contrast in the wavelength range between 0.95 and 1.15 $\mu$m for the {\it Y} band, between 1.15 and 1.35 $\mu$m for the {\it J} band and between 1.35 and 1.65 $\mu$m for the {\it H} band. 
The non-coronagraphic IRDIS images provided the contrast in the {\it K1} and {\it K2} bands instead. 
From these values, using the COND models by \cite{bate2003} and assuming an age of 35$^{+10}_{-15}$ Myr (see Section~3.3), we were able to estimate the mass of the companion for each spectral band (also reported in Table~\ref{tab:photoastro_B}). 

For each band we also obtained an independent value of the companion separation and position angle. 
The total uncertainty listed above takes into account all the possible sources of error but the final error bars are mainly dominated by the uncertainty on the centering of the star.
We derived a mass of $0.16\pm0.04~M_{\odot}$ and a separation of $91.8\pm2.2$~mas  for \object{HD~284149~B}, combining the measurements from the different bands (see Table~2 for details).

\begin{figure}
\begin{centering}
\includegraphics[width=0.45\textwidth]{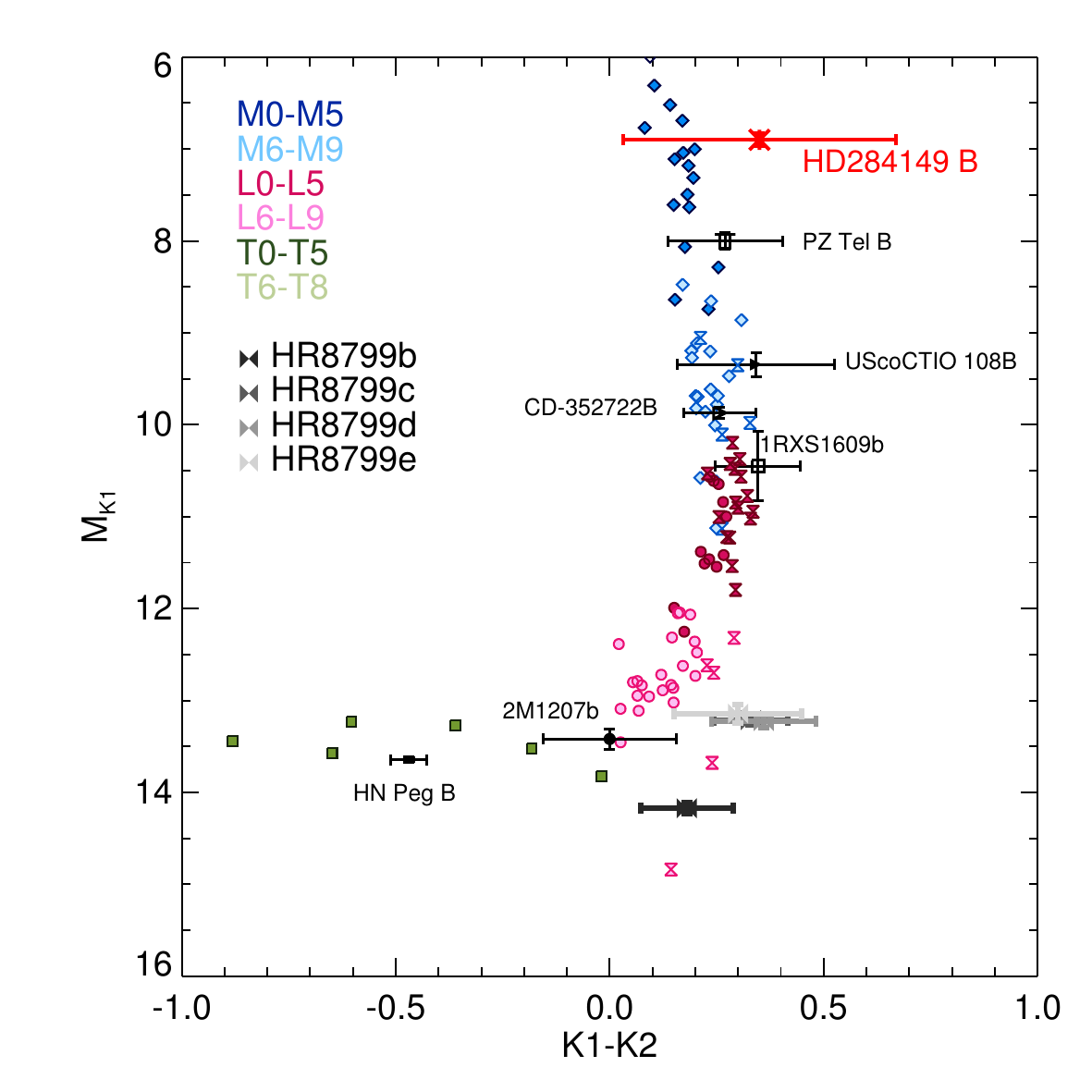}
\caption{Colour magnitude diagram showing the position of \object{HD 284149~B} (red bow tie) relative to that of M, L and T field dwarfs and of young known companions, based on the K1-K2 photometry from the non-coronographic IRDIS data (see text for details).}\label{f:cmd_B}
\end{centering}
\end{figure}

Figure~\ref{f:cmd_B} shows the location of \object{HD~284149~B} on a K-band based colour-magnitude diagrams (CMD).  
The positions of M, L and T field dwarfs and of young known companions are shown for comparison. 
Details on how the diagrams are generated are given in \cite{2016A&A...593A.119M} and \cite{2017A&A...603A..57S}. The field dwarf spectra used to generate the synthetic photometry are taken from the SpeXPrism library \citep{2014ASInC..11....7B}; \cite{2000ApJ...535..965L}, and from \cite{2015ApJ...804...92S}. We used for the most part distances reported in \cite{1980AJ.....85..454H}, \cite{2000AJ....120..447K}, \cite{2009ApJ...704..975J}, \cite{2012ApJ...752...56F}, and \cite{2013Sci...341.1492D}. We overlaid the photometry of young and/or dusty free-floating objects. Their absolute photometry was synthesised from their spectra \citep{2013ApJ...777L..20L, 2013ApJS..205....6M, allers2013, 2015ApJ...799..203G} and the corresponding parallaxes  \citep{2011ApJS..197...19K, 2012ApJ...752...56F, 2014A&A...568A...6Z, 2016ApJ...833...96L}.  We added the photometry of companions using the spectra and distances reported in \cite{2011ApJ...729..139W, 2015ApJ...804...96G, 2016ApJ...818L..12S, 2014MNRAS.445.3694D, 2015ApJ...802...61L, 2014ApJ...780L...4B, 2017AJ....154...10R, 2014A&A...562A.127B, 2010A&A...517A..76P, 2010ApJ...719..497L}.

The position of \object{HD~284149~B},  between the lower end of the M0-M5 sequence and the upper end of the M6-M9 sequence, seems to suggests a mid-M spectral type.

\subsubsection{Spectral characterisation}
\label{sec:412}
Exploiting the 39 IFS wavelengths we were able to extract a low resolution spectrum for the \object{HD~284149~B}. However, it is known that PCA tends to introduce biases on photometric data. Therefore we decided to use an alternative approach to extract the spectrum. We calculated a median data cube from the initial data set composed by all the calibrated data cubes after rotating each of them by the proper rotation angle  previously calculated in such a way that the whole data set is perfectly aligned. For each image of the median data cube  we then calculated the stellar profile by estimating the median of all the pixels at the same separation from the central  star which was then subtracted from the original image. On each wavelength frame of the resulting data cube we applied a three pixels radius aperture photometry. For each wavelength the resulting photometry was then normalised to the flux of the correspondent wavelength of the central star.
This approach also has the advantage of allowing to correct for the effect of the telluric lines as they affect both the primary and the secondary at the same way.

\begin{figure}
\begin{centering}
\includegraphics[width=9cm]{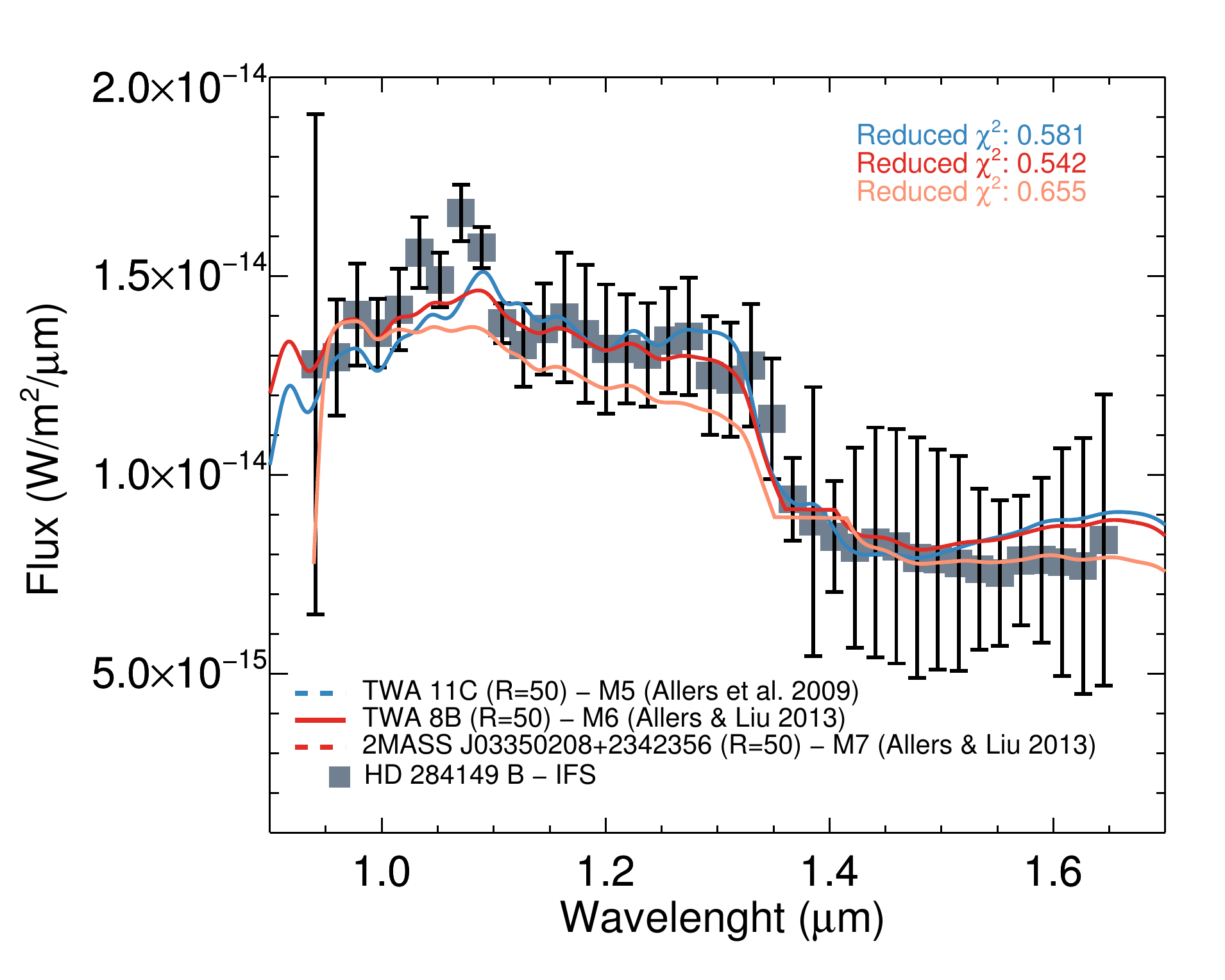}
\caption{ Comparison of the spectrum of \object{HD~284149~B} (grey squares) with the spectra of \object{TWA8~B} \citep{allers2009}, \object{TWA~11~C} and \object{2MASS~J03350208+2342356} \citep{allers2013}. The values of the reduced $\chi^2$ obtained for the fit are shown in the top right corner. \object{TWA8~B} (red curve) provides the best fit.}\label{f:spettroifs}
\end{centering}
\end{figure}

The extracted spectrum was then fitted with spectra of known objects from two different libraries: the Montreal Spectral Library\footnote{https://jgagneastro.wordpress.com/the-montreal-spectral-library} and the one from \citet{allers2013}. 
The best fit, as shown in Figure \ref{f:spettroifs}, was obtained with the spectrum of \object{TWA8~B} \citep{allers2009}, classified by \cite{allers2013} as a very low-gravity object (VL-G), with an assigned spectral type of M6. The spectra of \object{TWA~11~C} and \object{2MASS~J03350208+2342356}, classified as M5 and M7 respectively \citep{allers2013}, are shown for comparison. We therefore conclude that the spectra of \object{HD~284149~B} is compatible with a spectral type of M6$\pm$1 (No attempt was made to assign a gravity class to the object, given the limited resolution and signal to noise of the IFS spectra).
This is in good agreement with the estimate obtained using the IRDIS photometry (see Section~\ref{sec:411}). 

\subsubsection{Dynamical signatures of the close companion}
\label{sec:413}
The discovery of a stellar companion to \object{HD~284149} comes as confirmation to several dynamical signatures hinting to its existence.

A radial velocity difference of 3 km\,s$^{-1}$ between the early measurement by \citet{wichmann2000} and the recent sequence by \citet{nguyen12} was already noticed in \citet{bonavita2014}.
Assuming a mass of $\sim 0.167~M_{\odot}$ and a semi-major axis of 10.7 AU (equal to the projected separation), the expected RV semi-amplitude of the newly discovered companion is about 1.32 km\,s$^{-1}$ for an edge-on circular orbit.
We therefore conclude that the newly discovered companion is likely to be responsible for the observed radial velocity difference.
This opens the perspective for a determination of dynamical mass by coupling spectroscopic and imaging monitoring, which would be relevant given the young age of the system, as dynamical measurements of the mass are available only for few young binaries \citep[see e.g.][]{dupuy2016}.

An additional dynamical signature related to the presence of moderately close companions is the detection of significant differences in the proper motion of a star as measured at different epochs.
Such systems were first identified as $\Delta \mu$ binaries in the pioneering work by \citet{makarov2005}, who exploited the difference between the short-term proper motion measured by Hipparcos (epoch 1991, baseline 3.25 yr) and the long-term (a century timescale) proper motion by the Tycho-2 catalogue.
The release of Gaia-DR1 offers an additional opportunity to check for the detection of proper motion differences, providing proper motion measurements on a $\sim$24 yr baseline from the combination of Gaia-DR1 data with Hipparcos positions. 

\noindent \object{HD~284149} has no significant $\Delta \mu$ in \citet{makarov2005} but a comparison of its Gaia and Tycho-2 proper motions yields $\Delta \mu_{\alpha}=-3.32\pm1.01~mas/yr$ and $\Delta \mu_{\delta}=0.08\pm1.00~mas/yr$.
Significant proper motion difference is also seen between Gaia and UCAC4 measurements\footnote{As our results could be highly affected by systematic errors which could arise from an improper estimation of the proper motion error bars, we decided to consider only the value of $\Delta\mu$ obtained using Tycho-2 for our analysis.}, while Gaia and Hipparcos values do not differ significantly. 
Although \object{HD~284149} wasn't selected for observation because of its proper motion discrepancies, its newly detected companion then represents an ideal candidate for a dynamical characterisation that takes advantage from the combination of astrometry and imaging data. 

We therefore used the COPAINS (Code for Orbital Parametrization of Astrometrically Identified New Systems, Fontanive et al. private comm.) code to evaluate the characteristics of the possible companions compatible with the observed $\Delta \mu$.
The code uses Eq.~\ref{eq:deltamu}\footnote{In Eq.~\ref{eq:deltamu} $M_2$ is the mass of the secondary, $M_{Tot}$ is the total mass of the binary, $a$ is the semi-major axis in AU, $\Pi$ is the parallax of the system in mas, and $R_0$ takes into account the orbital phase so that $R_0=\left(\frac{1+e\cos E}{1-e\cos E}\right)^{1/2}$ where $e$ is the orbital eccentricity and $E$ is the eccentric anomaly}, derived by \citet{makarov2005}, to estimate the change in a star's proper motion induced by a given companion, for a range of possible masses and separations. 

\noindent A fine grid of mass and separation values is explored, and the expected $\Delta\mu$ is evaluated and compared with the observed one. 
In order to properly take into account the projection effects, for each point on the mass-separation grid the code considers $10^6$ possible orbital configurations, with eccentricities drawn from either a uniform or Gaussian \citep[see][and references therein for details]{bonavita2014} distribution. 

\begin{figure*}
\begin{centering}
\includegraphics[width=0.45\textwidth]{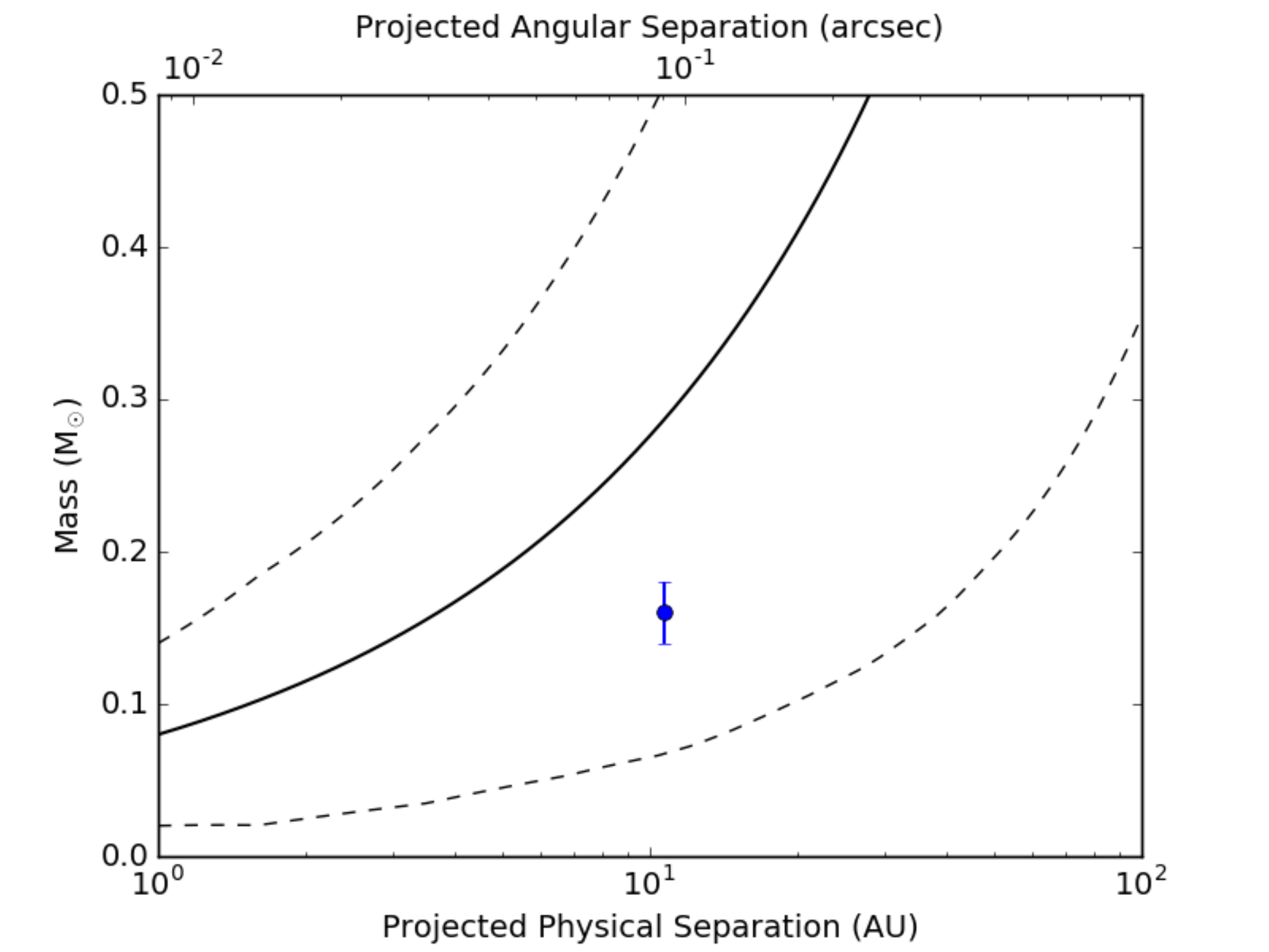}
\includegraphics[width=0.45\textwidth]{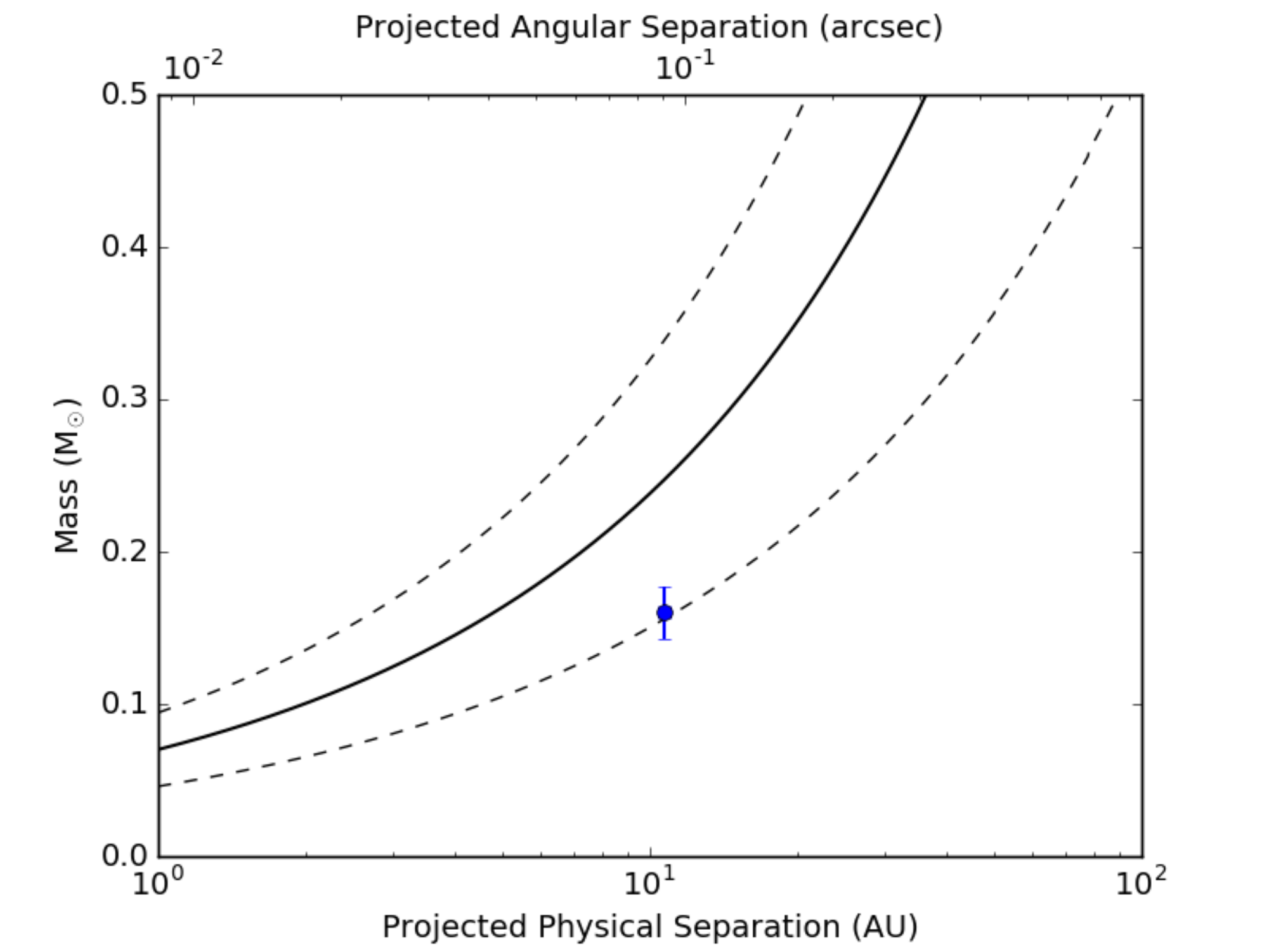}
\caption{Estimate of the mass and separation of the companions compatible with the observed $\Delta \mu$, assuming an orbit with an eccentricity randomly drawn from a flat (left panel) or Gaussian (right panel) distribution. The blue dot indicates the position of the detected companion.}\label{f:deltamu}
\end{centering}
\end{figure*}

\noindent Figure~\ref{f:deltamu} shows the results obtained for the two cases. 
In both panels the area enclosed by the two dashed lines shows the position on the mass-separation space of the companions that would cause a $\Delta\mu$ within one sigma from the observed one. Regardless of the assumption on the eccentricity distribution, the observed mass and separation of \object{HD~284149~B} (see Table~\ref{tab:photoastro_B}) seems to be compatible with the observed trend. 

\begin{figure*}
\begin{centering}
\includegraphics[width=0.45\textwidth]{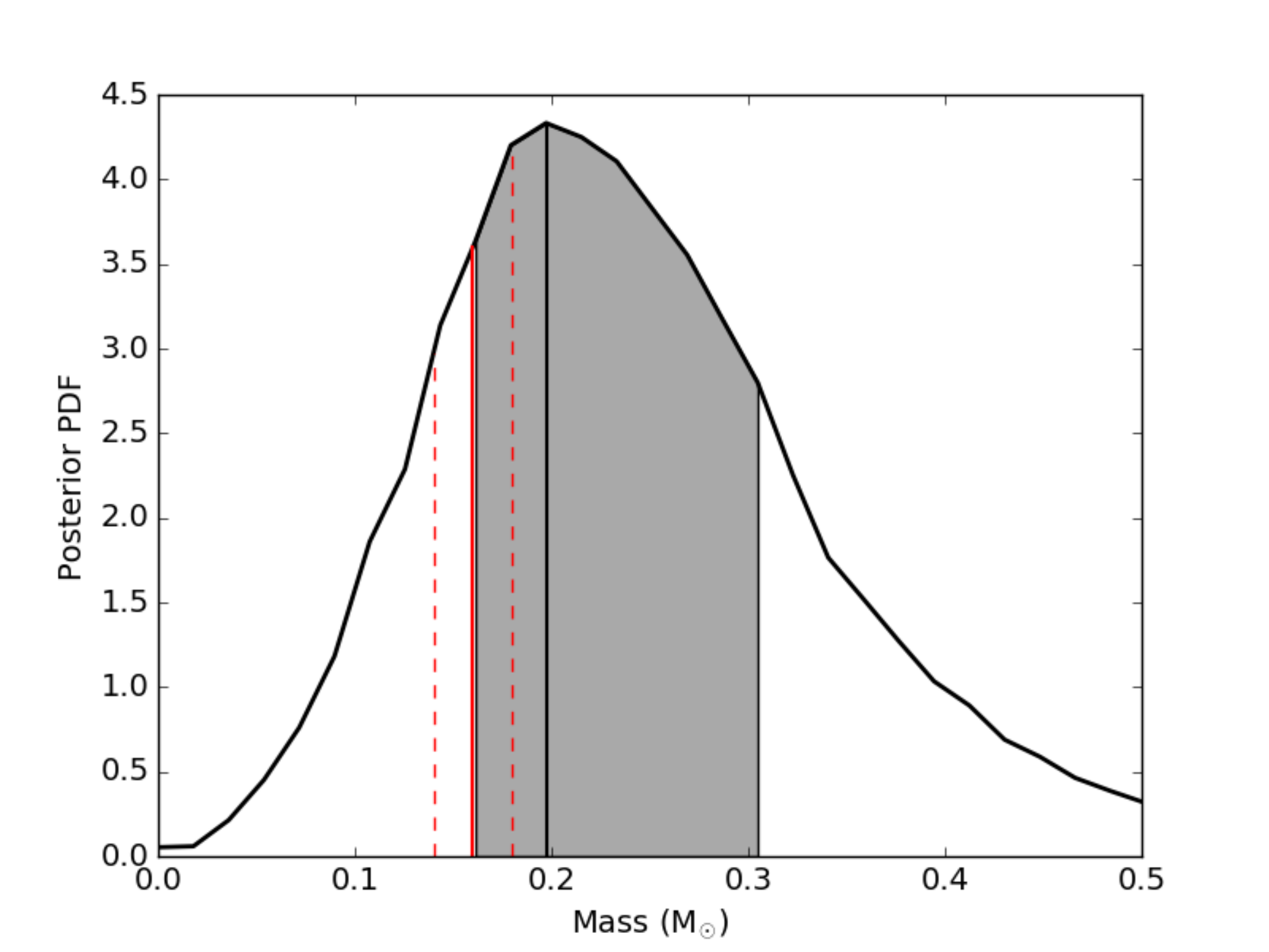}
\includegraphics[width=0.45\textwidth]{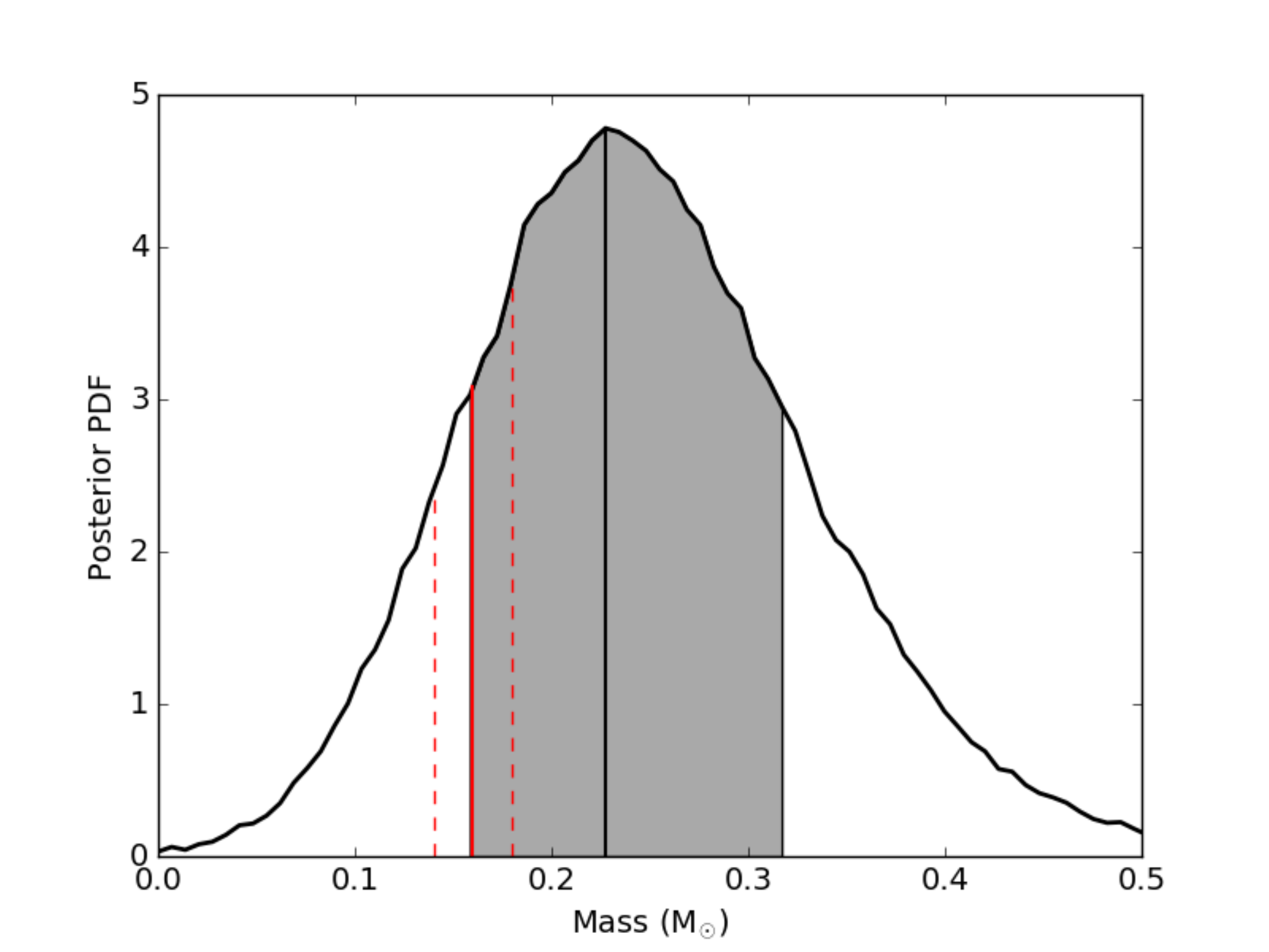}
\caption{Mass distribution of the companions compatible with the observed $\Delta\mu$, assuming a semi-major axis compatible with the observed projected separation of \object{HD~2841419}~B, and an eccentricity randomly drawn from a flat (left panel) or Gaussian (right panel) distribution. The black solid line shows the position of the most likely value and the shaded area highlights the region within a 1 sigma confidence level. The red solid line marks the value of the mass of \object{HD~2841419}~B (see Table~\ref{tab:photoastro_B})}\label{f:dmu_mass}
\end{centering}
\end{figure*}

\noindent Finally, the mass distribution for companions at the observed separation of \object{HD~284149~B} and compatible with the observed trend is shown in Figure~\ref{f:dmu_mass}. 
For the flat and Gaussian eccentricity priors, the posterior Mass distribution peaks at $0.20_{-0.04}^{+0.12}~M_{\odot}$ and $0.23_{+0.09}^{-0.07}~M_{\odot}$ respectively, and is therefore compatible with the value obtained from the SPHERE photometry ($0.16\pm0.04~M_{\odot}$ see Table~\ref{tab:photoastro_B}).

\begin{equation}
    \Delta\mu \leq \frac{2 \pi \Pi R_0 M_2}{\sqrt{aM_{Tot}}}
    \label{eq:deltamu}
\end{equation}


\subsection{Characterisation of the known brown dwarf companion}
\label{sec:4.2}

\subsubsection{IRDIS Astrometry}
\label{sec:421}

Precise relative astrometry for \object{HD~284149~b} was obtained from both the coronographic H2 and H3 images using the SHINE Specal pipeline (Galicher et al. private comm.).

For the non-coronagraphic images we used PSF fitting with the \emph{digiphot/allstar} routine in \emph{IRAF}. For each individual exposure in the image cubes (400 exposures per filter), we measure the relative pixel position between the two point sources using an 80\,pix-radius reference PSF composed from the unsaturated primary in all exposures per filter with the \emph{iraf/digiphot} task \emph{psf}. For efficiency, before applying \emph{allstar}, we cut out 160$\times$160 pixel regions around both the primary and the companion for every individual exposure (400 exposures total) and re-arrange these into two separate 20$\times$20 star grids per filter. The final astrometry was reconstructed by reverse-applying the offsets of this mapping process.
Measured pixel positions were transformed to separation and position angle by derotating with the sky rotation angle from the fits header and assuming a pixel scale of $12.255\pm0.009~mas/pix$ and a true north position of $-1.70\pm0.10\degree$ for the 2015-10-25 epoch and $-1.747\pm0.048\degree$ for the 2015-11-29 epoch \citep[see][]{maire2016}. 
The values obtained for both cases are reported in Table~\ref{tab:bdphot}.

The combined astrometry from the two epochs is sep=3669.14$\pm$0.91mas and PA=255.00$\pm$0.01$\degree$ taking into account all the random and systematic uncertainties. Adopting the new distance to the system from Gaia, we obtain a value of the projected separation of \object{HD~284149~b} of $431.2\pm 2.9$~AU.

\begin{figure}
\begin{centering}
\includegraphics[width=0.45\textwidth]{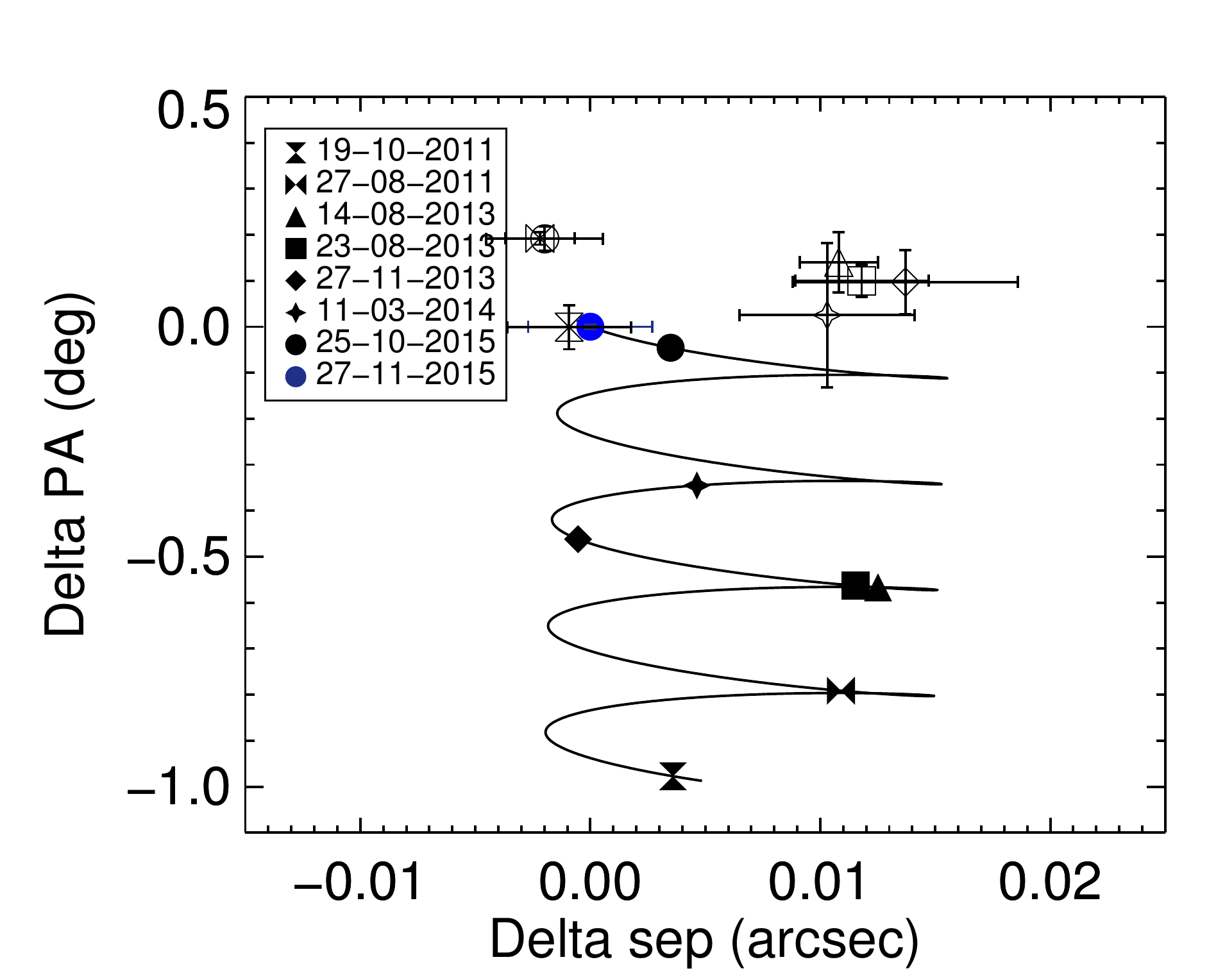}
\caption{Common proper motion analysis for \object{HD~284149~b}. The black solid line represents the motion of a background object. The filled symbols mark the positions of a background objects at the dates of the observations. The measured positions at the same dates are plotted with the corresponding open symbols. The blue filled circle marks the latest IRDIS\_EXT observation.}\label{f:cpm}
\end{centering}
\end{figure}

Figure~\ref{f:cpm} shows the updated common proper motion plot. 

\begin{table*}
    \caption{IRDIS Relative astrometry of \object{HD~284149~b}\label{tab:bdphot}}
    \centering
    \begin{tabular}{cr@{\,$\pm$\,}lr@{\,$\pm$\,}lr@{\,$\pm$\,}lr@{\,$\pm$\,}lr@{\,$\pm$\,}}
    \hline\hline
    filter&
    \multicolumn{4}{c}{sep} &
    \multicolumn{2}{c}{PA} \\
    & 
    \multicolumn{2}{l}{(mas)} &
    \multicolumn{2}{l}{(AU)} &    
    \multicolumn{2}{c}{($^{\circ}$)} \\

    \hline
    {\it H2}   & 3668.58 & 1.68                 & 431.06 & 5.75 & 255.01  & 0.01   \\ 
    {\it H3}   & 3669.03 & 1.33                 & 431.11 & 5.92 & 255.02  & 0.01   \\ 
    {\it K1}   & 3670.03 & $\pm$ 2.7 $\pm$ 0.01 & 431.23 & 5.89 & 254.82  & 0.01 $\pm$ 0.05   \\ 
    {\it K2}   & 3670.14 & $\pm$ 2.7 $\pm$ 0.01 & 431.24 & 5.88 & 254.83  & 0.01 $\pm$ 0.05  \\ 
        \hline
    \multicolumn{6}{l}{Adopted values} \\
               & 3669.14 & 0.92   & 431.16  & 2.93 & 255.01   & 0.01 \\

    \hline
    \end{tabular}
    \tablefoot{Reported numbers for {\it K1} and {\it K2} bands are mean values and their random uncertainties (standard error of the mean). When a second uncertainty is reported, it refers to the systematic uncertainty of the pixel scale and True North correction, respectively (see Sect.~\ref{sec:421}). The values of the separation in AU were obtained using the new parallax measurement from the Gaia mission \citep{GDR1}.
    }
\end{table*}

We did not attempt to constrain the orbital properties of the brown dwarf companion due to the insufficient time baseline of the
observations (4~yrs compared to an expected period of over 6000~yrs for a pole-on circular orbit) and because of the unknown orbit
of the close stellar companion, which could bias the derived orbital properties if not taken into account \citep{pearce2014}.

\subsubsection{IRDIS photometry}

Table~\ref{tab:bdphot} shows the values of the relative photometry of \object{HD~284149~b}. The values for the {\it H2} and {\it H3} bands were obtained using the coronagraphic data taken in October 2015, while we derived the {\it K1} and {\it K2} photometry from the non-coronagraphic observations. 
The values of the masses were obtained using the BT-Settl models by \cite{btsettl} and taking into account both the uncertainties on the photometry and on the age, the latter being once again the dominating source of error.
Combining these results we obtain for \object{HD~284149~b} a mass of $26\pm3~M_{Jup}$.

Finally, we used the IRDIS photometry to study the position of \object{HD~284149~b} in a color-magnitude diagram (CMD), compared with those of known MLTY field dwarfs, brown dwarfs and known directly imaged young companions (See Section~\ref{sec:411} for details on the construction of the plots).

\begin{figure*}
\begin{centering}
\includegraphics[width=0.45\textwidth]{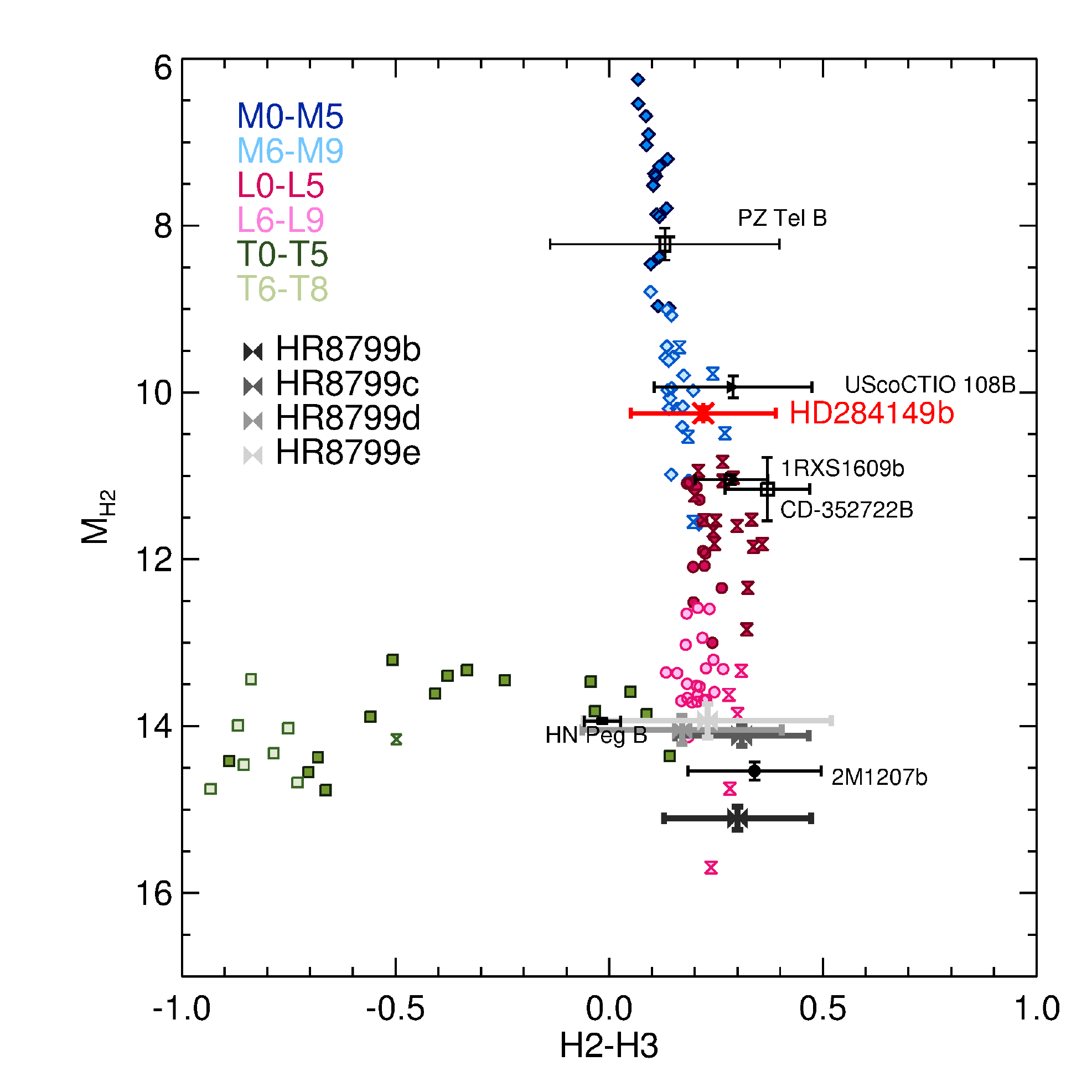}
\includegraphics[width=0.45\textwidth]{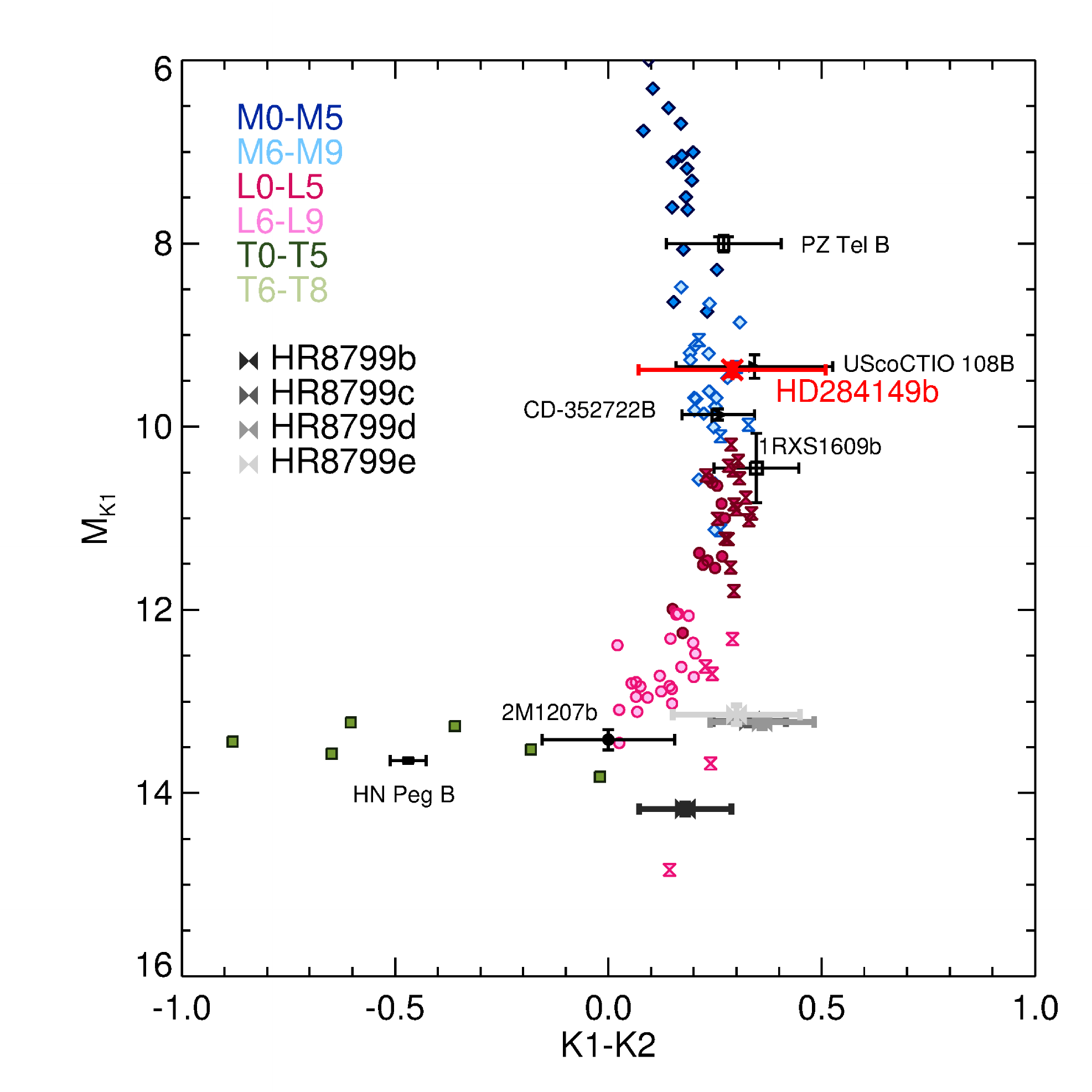}
\caption{Color–magnitude diagrams showing the position of \object{HD 284149~b} (red bow tie) relative to that of M, L and T field dwarfs and of young known companions, based on the H2-H3 photometry from the coronographic data (left panel) and the K1-K2 photometry from the non-coronographic data (right panel). See Section~\ref{sec:411} for all the appropriate references and details on the construction of the plots.}\label{f:cmd}
\end{centering}
\end{figure*}

As shown in Figure~\ref{f:cmd}, \object{HD~284149~b} (red bow tie) falls on the sequence of M6-M9 field dwarfs in both the {\it K} band and {\it H} band diagrams, and it is nicely bracketed by \object{UScoCTIO~108} \citep[an M9.5, see][for details]{bejar2008} and \object{CD--352722~B} \citep[an L1$\pm$1, see][]{wahhaj2011}.  

\begin{table}
    \caption{IRDIS Relative photometry of \object{HD~284149~b}\label{tab2}}
    \centering
    \begin{tabular}{cr@{\,$\pm$\,}lr@{\,$\pm$\,}lr@{\,$\pm$\,}}
    \hline\hline
    Filter&
    \multicolumn{2}{c}{$\Delta$ mag} &
    \multicolumn{2}{c}{Mass} \\ 
    &
    \multicolumn{2}{c}{} &
    \multicolumn{2}{c}{($M_{Jup}$)} \\
    \hline 
    {\it H2}   & 7.39 & 0.12 & 21.37 & 5.15 \\ 
    {\it H3}   & 7.17 & 0.12 & 20.11 & 3.76 \\ 
    {\it K1}   & 6.63 & 0.15 & 35.81 & 6.93 \\ 
    {\it K2}   & 6.34 & 0.16 & 34.94 & 7.09 \\ 
            \hline
    \multicolumn{4}{l}{Adopted values} \\
               & \multicolumn{2}{c}{ -- } & 26.28   &  2.93 \\

    \hline
    \end{tabular}
    \tablefoot{For the {\it K1} and {\it K2} bands, reported numbers are mean values and their random uncertainties (standard error of the mean). For all bands the values of the masses are derived using the BT-Settl Models by \cite{btsettl} assuming an age of 35~Myrs (see Section~\ref{sec:age}). Although the error on the mass is dominated by the uncertainty on the age, the listed values also take into account the contribute of the uncertainties on the photometry.
    }
\end{table}

\subsubsection{Spectral characterisation}
\label{sec:422}

The very high quality of the spectra obtained with the LSS mode of IRDIS (see Section~\ref{sec:lss}) allows us to put strong constraints on the spectral type of \object{HD~284149~b}, through the comparison with available libraries of spectra of similar objects. 

We first compared it with to the medium-resolution (R$\approx$2000, SXD mode) SpeX spectral library by \citet{allers2013}, which includes observed spectra for M, L, and T dwarfs (both young and old). In order to perform the comparison, those spectra were downgraded to our resolution of R$\approx$350. This was done by convolution with a Gaussian function of the appropriate full width half maximum.

\begin{figure*}[htbp]
\begin{centering}
\includegraphics[width=15.4cm]{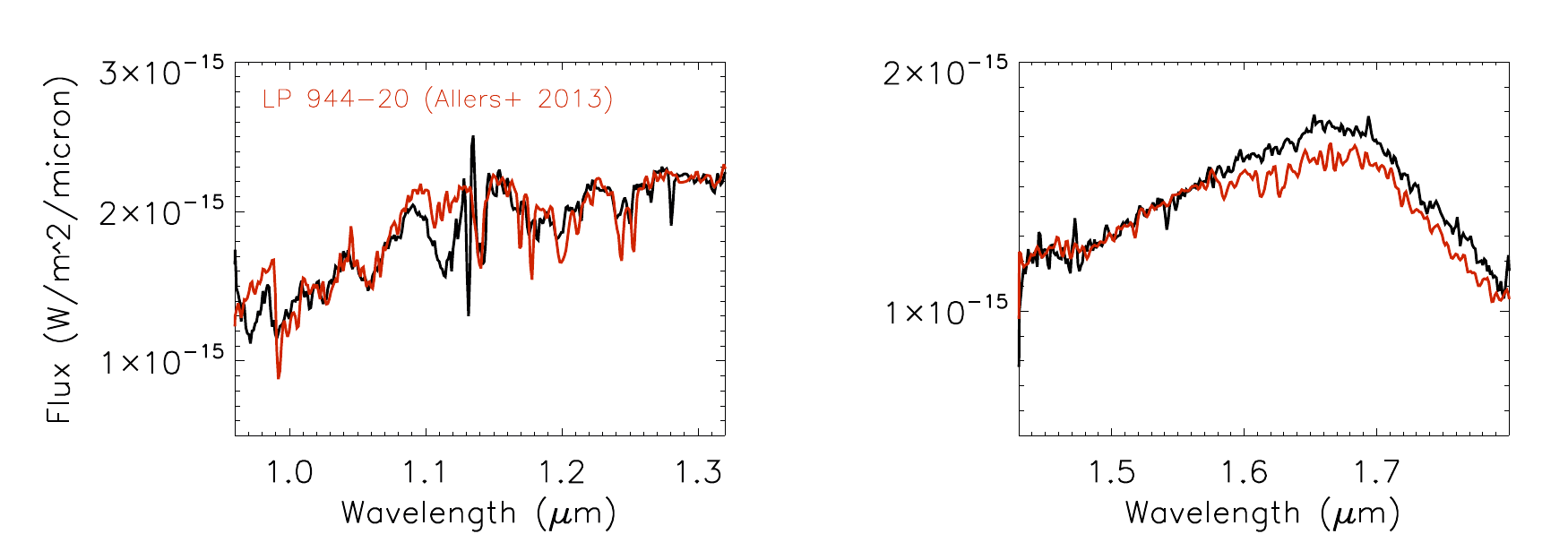}
\caption{Comparison of \object{HD~284149~b} with LP~944--20 from \citet{allers2013}}\label{f:allers}
\end{centering}
\end{figure*}

The result of the $\chi^2$ procedure that we have carried out in order to obtain the best fit is shown in Figure~\ref{f:allers}. 
The best fit is obtained using the spectra of \object{LP~944--20}, classified as M9{$\beta$ (\citealt{allers2013})}.
The plot clearly shows how, while the global shape of the pseudo-continuum is well represented by this spectral type, the gravity sensitive features (e.g., Na~{\sc i}, K~{\sc i}) of our target are significantly weaker. This suggests that \object{HD~284149~b} has a lower gravity, thus pointing to a younger age with respect to the template spectrum.

\begin{figure*}[htbp]
\begin{centering}
\includegraphics[width=15.4cm]{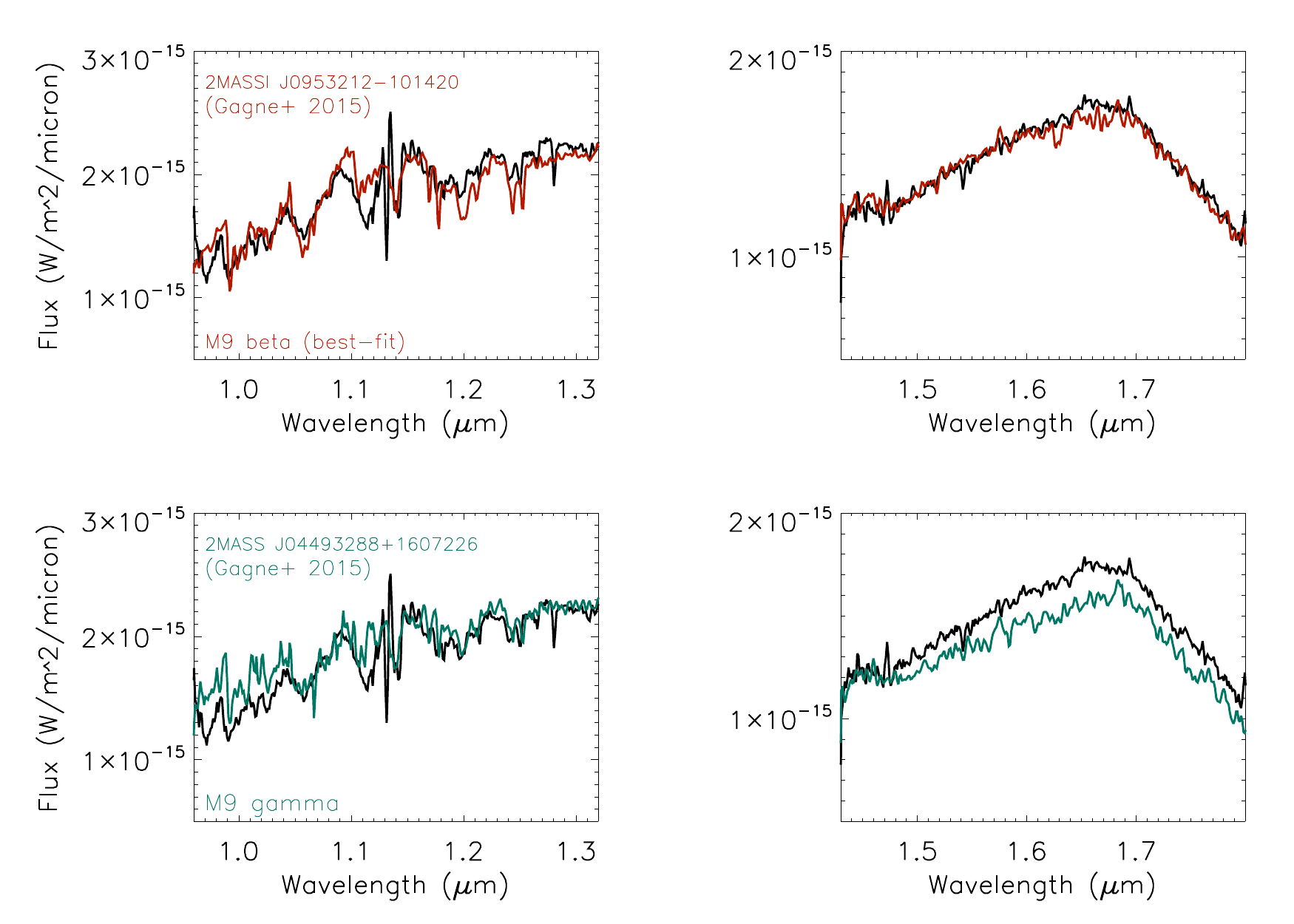}
\caption{Comparison of \object{HD~284149~b} with two M9-type brown dwarfs from the Montreal Spectral Library. While the global best fit is provided by the M9$\beta$, it is evident that individual spectral lines are too strong with respect to \object{HD~284149~b}, suggesting a lower gravity for this object (see the comparison with the M9$\gamma$ spectrum).}\label{f:montreal}
\end{centering}
\end{figure*}

\begin{figure*}[htbp]
\begin{centering}
\includegraphics[width=16cm]{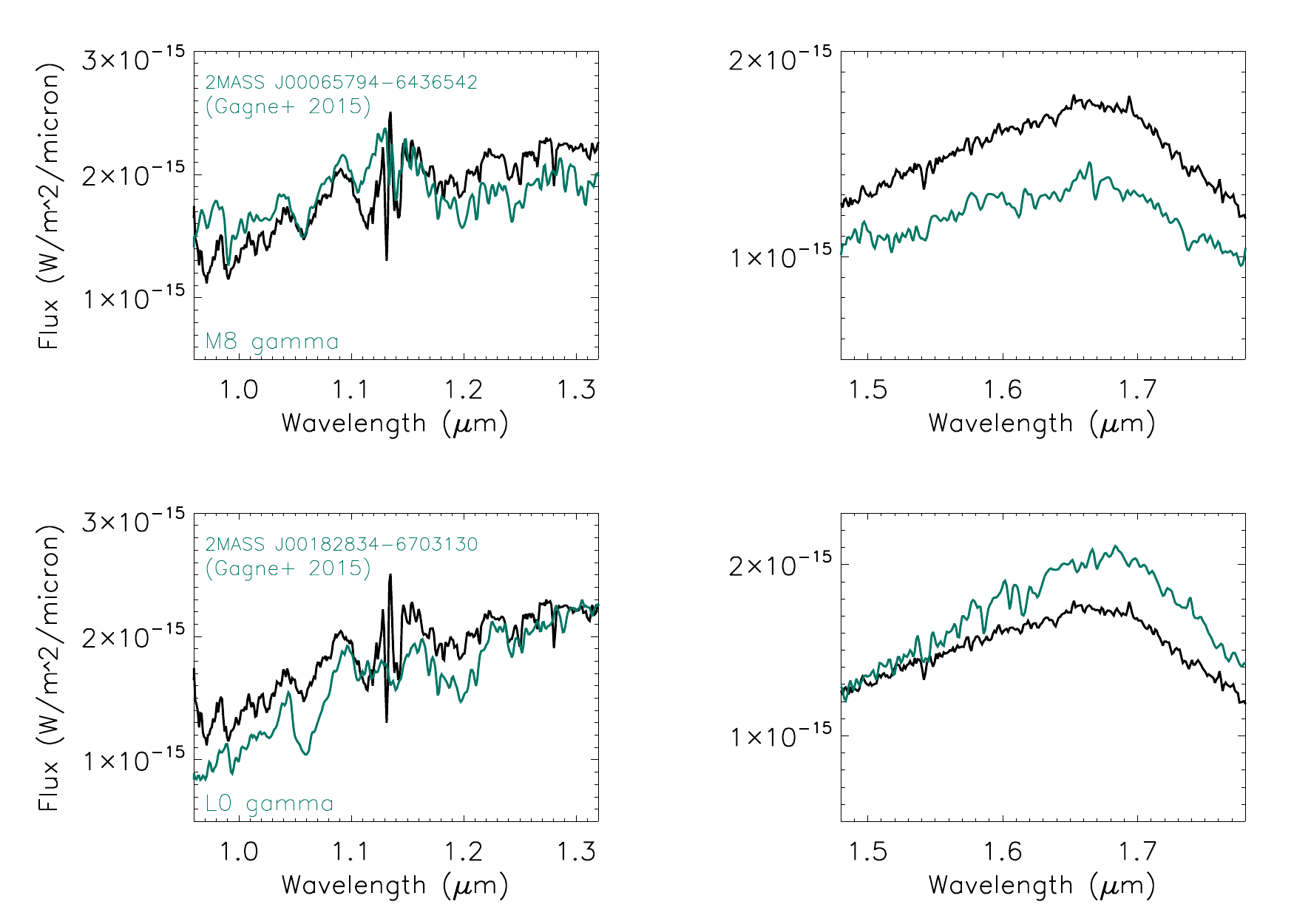}
\caption{Comparison of \object{HD~284149~b} with M8 and L0 young brown dwarfs from the Montreal Spectral Library (see text for discussion)}\label{f:montreal2}
\end{centering}
\end{figure*}

To further investigate this point, we also carried out a spectral comparison with template spectra for young brown dwarfs from the Montreal Spectral Library\footnote{available at \url{https://jgagneastro.wordpress.com/the-montreal-spectral-library/}, see e.g., \citet{gagne15}} (see Figure~\ref{f:montreal} and \ref{f:montreal2}). 

The best fitting spectra in this case is the one of \object{2MASS~J0953212$-$101420} (\citealt{gagne15}, top panels of Figure~\ref{f:montreal}), a M9 $\beta$. 
However, although the pseudo continuum is very well reproduced, again the spectral features are much weaker in our target. Conversely, the M9 $\gamma$ template provides a very good fit of the spectral lines (see lower panels of Figure~\ref{f:montreal}), despite the fact that \object{2MASS~J04493288+1607226} \citep{gagne15} is not a perfect match of the global continuum.

Thus, we can conclude that \object{HD~284149~b} has a spectral type of M9$\gamma$, the Greek letter pointing to a low surface gravity, hence to a young age \citep[see][]{kirkpatrick05}.

According to previous estimates obtained using NIR photometry, \object{HD~284149~b} was expected to have a spectral type between M8 and L1 \citep[see][]{bonavita2014}. 
As a final check we therefore compared our spectra with the M8 and L0/L1 spectra from the Montreal Spectral Library. Figure~\ref{f:montreal2} shows an example of such comparison, where only one example for each spectral type is plotted (blue solid line), together with our data (black solid line).
As clearly neither spectral type provide a good match of our data,  
we therefore conclude that the spectral type of \object{HD~284149~b} should be M9$\pm$0.5. 

It is noteworthy, in this context, that the gravity-sensitive spectral features of K~{\sc I} lines provide further support to our conclusion, as they also point towards a very low-gravity object (the Na~{\sc I} line at 1.14 $\mu$m could not be used because of the occurrence of cosmic rays on top of the feature). 
We measure EW(K~{\sc I})$_{1.169}$=2.70 \AA~ and EW(K~{\sc I})$_{1.177}$=4.72 \AA~, which provide a very strong indication of reduced gravity, confirmed by the position of \object{HD~284149~b} on diagnostic plots by \cite{allers2013} (see their Figure~23). 
As a further check, we have calculated the gravity score for the brown dwarf, following prescriptions given in \cite{allers2013}: by exploiting FeH bands, H-band continuum shape and K I lines we have obtained a four-digit score of 2212. Thus, our object can be classified as VL-G (i.e., very low gravity).

Note also that the very low-gravity we found for \object{HD~284149~b} allows an independent evidence on the system's age, as M9$\gamma$ objects are consistent with ages younger than $\sim$ 60 Myr (\citealt{martin17}).

As a further independent confirmation of our estimate we have also calculated the H$_2$O spectral index, as defined in \cite{allers2013}, which has been shown to be spectral-type sensitive and gravity insensitive \citep{allers2007, allers2013}. \object{HD~284149~b} is found to have an H-index of 1.137 which implies a spectral type of M9$\pm$1, thus in very good agreement with the value coming from the visual inspection. 

Finally, we have used the spectral type $vs$ $T_{\rm eff}$ 
relationship as given by \cite{filippazzo15} and found a value of $T_{\rm eff}$=2395$\pm$113 K. Our temperature estimate agrees very well with previous findings by \citet{bonavita2014}, who reported $T_{\rm eff}$=2337$^{+95}_{-182}$ K from the spectral type and the calibration by \cite{Pecaut13}.

\section{Discussion and conclusions}
This paper presents a detailed characterisation of the \object{HD~284149}~ABb system\footnote{Given the nature of the two companions, the IAU nomenclature would require naming that system HD 284149 (AB)C, where C refers to the brown dwarf. As the BD companion was already known as \object{HD~284149~b}, we nonetheless decided to keep the previous nomenclature to avoid future confusion.}. 
We were able to refine the estimate of the spectral type of the known sub-stellar companion \object{HD~284149~b} \citep{bonavita2014}, using high quality medium resolution spectra obtained with IRDIS in Long Slit Spectroscopy (LSS) Mode. Our results point toward an M9 spectral type and an effective temperature of 2300~K, with significant improvement with respect to the previous estimates. 

\noindent A reassessment of the stellar properties was carried out, also taking advantage of the availability of Gaia measurement of parallax and proper motion, resulting in a distance of 117.50$\pm$5.50 pc and a stellar age of 35 Myr. As a consequence we were able to refine the previous estimates of both separation and mass of \object{HD~284149~b} ($431.20\pm7.67~AU$ and $26\pm3~M_{Jup}$ respectively).

Finally, a close low-mass stellar companion (\object{HD~284149~B} $\sim$0.16$M_{\odot}$ at $\sim$0.1$^{\prime\prime}$) was resolved in the IRDIFS non-coronagraphic images. Such companion is compatible with the radial velocity difference pointed out by \citet{bonavita2014} as well as with the difference in proper motion between the Gaia and Tycho2 measurements. Therefore, there are good potential for dynamical mass determination for the \object{HD~284142}~ABb pair with future observations. 
\object{HD~284149}~ABb therefore adds to the short list of brown dwarf companions in circumbinary configuration \citep{bonavita2016}, supporting their conclusion that brown dwarfs in wide circumbinary orbits occur with a similar frequency with respect to around single stars.

\subsubsection*{Acknowledgements}
We thank the anonymous referee for extensive feedback that significantly improved the clarity of the paper.
This research made use of the Montreal Brown Dwarf and Exoplanet Spectral Library, maintained by Jonathan Gagn\'{e}.
This work has made use of data from the European Space Agency (ESA)
mission {\it Gaia} (\url{http://www.cosmos.esa.int/gaia}), processed by the {\it Gaia} Data Processing and Analysis Consortium (DPAC, \url{http://www.cosmos.esa.int/web/gaia/dpac/consortium}). Funding
for the DPAC has been provided by national institutions, in particular the institutions participating in the {\it Gaia} Multilateral Agreement.
SPHERE is an instrument designed and built by a consortium consisting of IPAG
(Grenoble,  France),  MPIA  (Heidelberg,  Germany),  LAM  (Marseille,  France),
LESIA  (Paris,  France),  Laboratoire  Lagrange  (Nice,  France),  INAF  -  Osservatorio di Padova (Italy), Observatoire Astronomique de l'Universit\'e de Gen\`eve (Switzerland), ETH Zurich
(Switzerland), NOVA (Netherlands), ONERA (France) and ASTRON (Netherlands) in collaboration with ESO. SPHERE was funded by ESO, with additional contributions  from  CNRS  (France),  MPIA  (Germany),  INAF  (Italy),  FINES (Switzerland) and NOVA (Netherlands). SPHERE also received funding from the
European Commission Sixth and Seventh Framework Programmes as part of the
Optical Infrared Coordination Network for Astronomy (OPTICON) under grant
number RII3-Ct-2004-001566 for FP6 (2004-2008), grant number 226604 for
FP7 (2009-2012) and grant number 312430 for FP7 (2013-2016). We acknowledge financial support from the Programme National de Planétologie (PNP) and the  Programme  National  de  Physique  Stellaire  (PNPS)  of  CNRS-INSU. 
Part of this work has been carried out within the frame of the National Centre for Competence in Research PlanetS supported by the Swiss National Science Foundation. S.D. acknowledges the financial support of the SNSF.
This work has also been supported by a grant from the French Labex OSUG@2020
(Investissements  d’avenir,  ANR10  LABX56).  
We thank the anonymous referee for extensive feedback that significantly improved the clarity of the paper.
We acknowledge support from the "Progetti Premiali" funding
scheme of the Italian Ministry of Education, University, and Research.
This  work  has  made  use  of  the SPHERE  Data  Centre,  jointly  operated  by  OSUG/IPAG  (Grenoble),  PYTHEAS/LAM/CeSAM (Marseille), OCA/Lagrange (Nice) and Observatoire de Paris/LESIA (Paris).

\bibliographystyle{aa}
\bibliography{main}

\begin{thebibliography}{119}
\expandafter\ifx\csname natexlab\endcsname\relax\def\natexlab#1{#1}\fi

\bibitem[{{Allard} {et~al.}(2012){Allard}, {Homeier}, {Freytag}, \&
  {Sharp}}]{btsettl}
{Allard}, F., {Homeier}, D., {Freytag}, B., \& {Sharp}, C.~M. 2012, in EAS
  Publications Series, Vol.~57, EAS Publications Series, ed. C.~{Reyl{\'e}},
  C.~{Charbonnel}, \& M.~{Schultheis}, 3--43

\bibitem[{{Allers} {et~al.}(2007){Allers}, {Jaffe}, {Luhman}, {Liu}, {Wilson},
  {Skrutskie}, {Nelson}, {Peterson}, {Smith}, \& {Cushing}}]{allers2007}
{Allers}, K.~N., {Jaffe}, D.~T., {Luhman}, K.~L., {et~al.} 2007, \apj, 657, 511

\bibitem[{{Allers} \& {Liu}(2013)}]{allers2013}
{Allers}, K.~N. \& {Liu}, M.~C. 2013, ApJ, 772, 79

\bibitem[{{Allers} {et~al.}(2009){Allers}, {Liu}, {Shkolnik}, {Cushing},
  {Dupuy}, {Mathews}, {Reid}, {Cruz}, \& {Vacca}}]{allers2009}
{Allers}, K.~N., {Liu}, M.~C., {Shkolnik}, E., {et~al.} 2009, ApJ, 697, 824

\bibitem[{Amara \& Quanz(2012)}]{amara2012pynpoint}
Amara, A. \& Quanz, S.~P. 2012, Monthly Notices of the Royal Astronomical
  Society, 427, 948

\bibitem[{{Antichi} {et~al.}(2009){Antichi}, {Dohlen}, {Gratton}, {Mesa},
  {Claudi}, {Giro}, {Boccaletti}, {Mouillet}, {Puget}, \&
  {Beuzit}}]{antichi2009}
{Antichi}, J., {Dohlen}, K., {Gratton}, R.~G., {et~al.} 2009, ApJ, 695, 1042

\bibitem[{{Artigau} {et~al.}(2015){Artigau}, {Gagn{\'e}}, {Faherty}, {Malo},
  {Naud}, {Doyon}, {Lafreni{\`e}re}, \& {Beletsky}}]{artigau2015}
{Artigau}, {\'E}., {Gagn{\'e}}, J., {Faherty}, J., {et~al.} 2015, ApJ, 806, 254

\bibitem[{{Bailer-Jones}(2011)}]{bailer2011}
{Bailer-Jones}, C.~A.~L. 2011, MNRAS, 411, 435

\bibitem[{{Bailey} {et~al.}(2014{\natexlab{a}}){Bailey}, {Meshkat}, {Reiter},
  {Morzinski}, {Males}, {Su}, {Hinz}, {Kenworthy}, {Stark}, {Mamajek},
  {Briguglio}, {Close}, {Follette}, {Puglisi}, {Rodigas}, {Weinberger}, \&
  {Xompero}}]{bailey2014}
{Bailey}, V., {Meshkat}, T., {Reiter}, M., {et~al.} 2014{\natexlab{a}}, ApJl,
  780, L4

\bibitem[{{Bailey} {et~al.}(2014{\natexlab{b}}){Bailey}, {Meshkat}, {Reiter},
  {Morzinski}, {Males}, {Su}, {Hinz}, {Kenworthy}, {Stark}, {Mamajek},
  {Briguglio}, {Close}, {Follette}, {Puglisi}, {Rodigas}, {Weinberger}, \&
  {Xompero}}]{2014ApJ...780L...4B}
{Bailey}, V., {Meshkat}, T., {Reiter}, M., {et~al.} 2014{\natexlab{b}}, ApJL,
  780, L4

\bibitem[{{Baraffe} {et~al.}(2003){Baraffe}, {Chabrier}, {Barman}, {Allard}, \&
  {Hauschildt}}]{baraffe03}
{Baraffe}, I., {Chabrier}, G., {Barman}, T.~S., {Allard}, F., \& {Hauschildt},
  P.~H. 2003, \aap, 402, 701

\bibitem[{{Bate} {et~al.}(2003){Bate}, {Bonnell}, \& {Bromm}}]{bate2003}
{Bate}, M.~R., {Bonnell}, I.~A., \& {Bromm}, V. 2003, MNRAS, 339, 577

\bibitem[{{B{\'e}jar} {et~al.}(2008){B{\'e}jar}, {Zapatero Osorio},
  {P{\'e}rez-Garrido}, {{\'A}lvarez}, {Mart{\'{\i}}n}, {Rebolo},
  {Vill{\'o}-P{\'e}rez}, \& {D{\'{\i}}az-S{\'a}nchez}}]{bejar2008}
{B{\'e}jar}, V.~J.~S., {Zapatero Osorio}, M.~R., {P{\'e}rez-Garrido}, A.,
  {et~al.} 2008, \apjl, 673, L185

\bibitem[{{Bell} {et~al.}(2015){Bell}, {Mamajek}, \& {Naylor}}]{bell2015}
{Bell}, C.~P.~M., {Mamajek}, E.~E., \& {Naylor}, T. 2015, MNRAS, 454, 593

\bibitem[{{Beuzit} {et~al.}(2008){Beuzit}, {Feldt}, {Dohlen}, {Mouillet},
  {Puget}, {Wildi}, {Abe}, {Antichi}, {Baruffolo}, {Baudoz}, {Boccaletti},
  {Carbillet}, {Charton}, {Claudi}, {Downing}, {Fabron}, {Feautrier},
  {Fedrigo}, {Fusco}, {Gach}, {Gratton}, {Henning}, {Hubin}, {Joos}, {Kasper},
  {Langlois}, {Lenzen}, {Moutou}, {Pavlov}, {Petit}, {Pragt}, {Rabou}, {Rigal},
  {Roelfsema}, {Rousset}, {Saisse}, {Schmid}, {Stadler}, {Thalmann}, {Turatto},
  {Udry}, {Vakili}, \& {Waters}}]{beuzit2008}
{Beuzit}, J.-L., {Feldt}, M., {Dohlen}, K., {et~al.} 2008, in SPIE, Vol. 7014,
  Ground-based and Airborne Instrumentation for Astronomy II, 701418

\bibitem[{{Biller} {et~al.}(2010){Biller}, {Liu}, {Wahhaj}, {Nielsen}, {Close},
  {Dupuy}, {Hayward}, {Burrows}, {Chun}, {Ftaclas}, {Clarke}, {Hartung},
  {Males}, {Reid}, {Shkolnik}, {Skemer}, {Tecza}, {Thatte}, {Alencar},
  {Artymowicz}, {Boss}, {de Gouveia Dal Pino}, {Gregorio-Hetem}, {Ida},
  {Kuchner}, {Lin}, \& {Toomey}}]{biller2010}
{Biller}, B.~A., {Liu}, M.~C., {Wahhaj}, Z., {et~al.} 2010, ApJl, 720, L82

\bibitem[{{Bonavita} {et~al.}(2014){Bonavita}, {Daemgen}, {Desidera},
  {Jayawardhana}, {Janson}, \& {Lafreni{\`e}re}}]{bonavita2014}
{Bonavita}, M., {Daemgen}, S., {Desidera}, S., {et~al.} 2014, ApJl, 791, L40

\bibitem[{{Bonavita} {et~al.}(2016){Bonavita}, {Desidera}, {Thalmann},
  {Janson}, {Vigan}, {Chauvin}, \& {Lannier}}]{bonavita2016}
{Bonavita}, M., {Desidera}, S., {Thalmann}, C., {et~al.} 2016, A\&A, 593, A38

\bibitem[{{Bonnefoy} {et~al.}(2014){Bonnefoy}, {Chauvin}, {Lagrange}, {Rojo},
  {Allard}, {Pinte}, {Dumas}, \& {Homeier}}]{2014A&A...562A.127B}
{Bonnefoy}, M., {Chauvin}, G., {Lagrange}, A.-M., {et~al.} 2014, A\&A, 562,
  A127

\bibitem[{{Bonnefoy} {et~al.}(2016){Bonnefoy}, {Zurlo}, {Baudino}, {Lucas},
  {Mesa}, {Maire}, {Vigan}, {Galicher}, {Homeier}, {Marocco}, {Gratton},
  {Chauvin}, {Allard}, {Desidera}, {Kasper}, {Moutou}, {Lagrange}, {Antichi},
  {Baruffolo}, {Baudrand}, {Beuzit}, {Boccaletti}, {Cantalloube}, {Carbillet},
  {Charton}, {Claudi}, {Costille}, {Dohlen}, {Dominik}, {Fantinel},
  {Feautrier}, {Feldt}, {Fusco}, {Gigan}, {Girard}, {Gluck}, {Gry}, {Henning},
  {Janson}, {Langlois}, {Madec}, {Magnard}, {Maurel}, {Mawet}, {Meyer},
  {Milli}, {Moeller-Nilsson}, {Mouillet}, {Pavlov}, {Perret}, {Pujet}, {Quanz},
  {Rochat}, {Rousset}, {Roux}, {Salasnich}, {Salter}, {Sauvage}, {Schmid},
  {Sevin}, {Soenke}, {Stadler}, {Turatto}, {Udry}, {Vakili}, {Wahhaj}, \&
  {Wildi}}]{bonnefoy2016}
{Bonnefoy}, M., {Zurlo}, A., {Baudino}, J.~L., {et~al.} 2016, A\&A, 587, A58

\bibitem[{{Bowler} {et~al.}(2017){Bowler}, {Liu}, {Mawet}, {Ngo}, {Malo},
  {Mace}, {McLane}, {Lu}, {Tristan}, {Hinkley}, {Hillenbrand}, {Shkolnik},
  {Benneke}, \& {Best}}]{bowler2017}
{Bowler}, B.~P., {Liu}, M.~C., {Mawet}, D., {et~al.} 2017, AJ, 153, 18

\bibitem[{{Bressan} {et~al.}(2012){Bressan}, {Marigo}, {Girardi}, {Salasnich},
  {Dal Cero}, {Rubele}, \& {Nanni}}]{bressan2012}
{Bressan}, A., {Marigo}, P., {Girardi}, L., {et~al.} 2012, MNRAS, 427, 127

\bibitem[{{Brott} \& {Hauschildt}(2005)}]{brott2005}
{Brott}, I. \& {Hauschildt}, P.~H. 2005, in ESA Special Publication, Vol. 576,
  The Three-Dimensional Universe with Gaia, ed. C.~{Turon}, K.~S. {O'Flaherty},
  \& M.~A.~C. {Perryman}, 565

\bibitem[{{Burgasser}(2014)}]{2014ASInC..11....7B}
{Burgasser}, A.~J. 2014, in Astronomical Society of India Conference Series,
  Vol.~11, Astronomical Society of India Conference Series

\bibitem[{{Butters} {et~al.}(2010){Butters}, {West}, {Anderson}, {Collier
  Cameron}, {Clarkson}, {Enoch}, {Haswell}, {Hellier}, {Horne}, {Joshi},
  {Kane}, {Lister}, {Maxted}, {Parley}, {Pollacco}, {Smalley}, {Street},
  {Todd}, {Wheatley}, \& {Wilson}}]{Butters10}
{Butters}, O.~W., {West}, R.~G., {Anderson}, D.~R., {et~al.} 2010, A\&A, 520,
  L10

\bibitem[{{Carson} {et~al.}(2013){Carson}, {Thalmann}, {Janson}, {Kozakis},
  {Bonnefoy}, {Biller}, {Schlieder}, {Currie}, {McElwain}, {Goto}, {Henning},
  {Brandner}, {Feldt}, {Kandori}, {Kuzuhara}, {Stevens}, {Wong}, {Gainey},
  {Fukagawa}, {Kuwada}, {Brandt}, {Kwon}, {Abe}, {Egner}, {Grady}, {Guyon},
  {Hashimoto}, {Hayano}, {Hayashi}, {Hayashi}, {Hodapp}, {Ishii}, {Iye},
  {Knapp}, {Kudo}, {Kusakabe}, {Matsuo}, {Miyama}, {Morino}, {Moro-Martin},
  {Nishimura}, {Pyo}, {Serabyn}, {Suto}, {Suzuki}, {Takami}, {Takato},
  {Terada}, {Tomono}, {Turner}, {Watanabe}, {Wisniewski}, {Yamada}, {Takami},
  {Usuda}, \& {Tamura}}]{carson2013}
{Carson}, J., {Thalmann}, C., {Janson}, M., {et~al.} 2013, ApJl, 763, L32

\bibitem[{{Chauvin} {et~al.}(2005{\natexlab{a}}){Chauvin}, {Lagrange}, {Dumas},
  {Zuckerman}, {Mouillet}, {Song}, {Beuzit}, \& {Lowrance}}]{chauvin2005a}
{Chauvin}, G., {Lagrange}, A.-M., {Dumas}, C., {et~al.} 2005{\natexlab{a}},
  A\&A, 438, L25

\bibitem[{{Chauvin} {et~al.}(2005{\natexlab{b}}){Chauvin}, {Lagrange},
  {Zuckerman}, {Dumas}, {Mouillet}, {Song}, {Beuzit}, {Lowrance}, \&
  {Bessell}}]{chauvin2005b}
{Chauvin}, G., {Lagrange}, A.-M., {Zuckerman}, B., {et~al.} 2005{\natexlab{b}},
  A\&A, 438, L29

\bibitem[{{Choi} {et~al.}(2016){Choi}, {Dotter}, {Conroy}, {Cantiello},
  {Paxton}, \& {Johnson}}]{choi2016}
{Choi}, J., {Dotter}, A., {Conroy}, C., {et~al.} 2016, ApJ, 823, 102

\bibitem[{{Claudi} {et~al.}(2008){Claudi}, {Turatto}, {Gratton}, {Antichi},
  {Bonavita}, {Bruno}, {Cascone}, {De Caprio}, {Desidera}, {Giro}, {Mesa},
  {Scuderi}, {Dohlen}, {Beuzit}, \& {Puget}}]{claudi2008}
{Claudi}, R.~U., {Turatto}, M., {Gratton}, R.~G., {et~al.} 2008, in SPIE, Vol.
  7014, Ground-based and Airborne Instrumentation for Astronomy II, 70143E

\bibitem[{{Daemgen} {et~al.}(2015){Daemgen}, {Bonavita}, {Jayawardhana},
  {Lafreni{\`e}re}, \& {Janson}}]{daemgen2015}
{Daemgen}, S., {Bonavita}, M., {Jayawardhana}, R., {Lafreni{\`e}re}, D., \&
  {Janson}, M. 2015, ApJ, 799, 155

\bibitem[{{De Rosa} {et~al.}(2014){De Rosa}, {Patience}, {Ward-Duong}, {Vigan},
  {Marois}, {Song}, {Macintosh}, {Graham}, {Doyon}, {Bessell}, {Lai},
  {McCarthy}, \& {Kulesa}}]{2014MNRAS.445.3694D}
{De Rosa}, R.~J., {Patience}, J., {Ward-Duong}, K., {et~al.} 2014, MNRAS, 445,
  3694

\bibitem[{{Delorme} {et~al.}(2013){Delorme}, {Gagn{\'e}}, {Girard}, {Lagrange},
  {Chauvin}, {Naud}, {Lafreni{\`e}re}, {Doyon}, {Riedel}, {Bonnefoy}, \&
  {Malo}}]{delorme2013}
{Delorme}, P., {Gagn{\'e}}, J., {Girard}, J.~H., {et~al.} 2013, A\&A, 553, L5

\bibitem[{{Desidera} {et~al.}(2015){Desidera}, {Covino}, {Messina}, {Carson},
  {Hagelberg}, {Schlieder}, {Biazzo}, {Alcal{\'a}}, {Chauvin}, {Vigan},
  {Beuzit}, {Bonavita}, {Bonnefoy}, {Delorme}, {D'Orazi}, {Esposito}, {Feldt},
  {Girardi}, {Gratton}, {Henning}, {Lagrange}, {Lanzafame}, {Launhardt},
  {Marmier}, {Melo}, {Meyer}, {Mouillet}, {Moutou}, {Segransan}, {Udry}, \&
  {Zaidi}}]{desidera15}
{Desidera}, S., {Covino}, E., {Messina}, S., {et~al.} 2015, A\&A, 573, A126

\bibitem[{{Dobson} {et~al.}(1990){Dobson}, {Donahue}, {Radick}, \&
  {Kadlec}}]{Dobson1990}
{Dobson}, A.~K., {Donahue}, R.~A., {Radick}, R.~R., \& {Kadlec}, K.~L. 1990, in
  Astronomical Society of the Pacific Conference Series, Vol.~9, Cool Stars,
  Stellar Systems, and the Sun, ed. G.~{Wallerstein}, 132--135

\bibitem[{{Dohlen} {et~al.}(2008){Dohlen}, {Langlois}, {Saisse}, {Hill},
  {Origne}, {Jacquet}, {Fabron}, {Blanc}, {Llored}, {Carle}, {Moutou}, {Vigan},
  {Boccaletti}, {Carbillet}, {Mouillet}, \& {Beuzit}}]{dohlen2008}
{Dohlen}, K., {Langlois}, M., {Saisse}, M., {et~al.} 2008, in SPIE, Vol. 7014,
  Ground-based and Airborne Instrumentation for Astronomy II, 70143L

\bibitem[{{Donahue} \& {Baliunas}(1992)}]{Donahue1992}
{Donahue}, R.~A. \& {Baliunas}, S.~L. 1992, ApJL, 393, L63

\bibitem[{{Dupuy} {et~al.}(2016){Dupuy}, {Forbrich}, {Rizzuto}, {Mann},
  {Aller}, {Liu}, {Kraus}, \& {Berger}}]{dupuy2016}
{Dupuy}, T.~J., {Forbrich}, J., {Rizzuto}, A., {et~al.} 2016, ApJ, 827, 23

\bibitem[{{Dupuy} \& {Kraus}(2013)}]{2013Sci...341.1492D}
{Dupuy}, T.~J. \& {Kraus}, A.~L. 2013, Science, 341, 1492

\bibitem[{{Faherty} {et~al.}(2012){Faherty}, {Burgasser}, {Walter}, {Van der
  Bliek}, {Shara}, {Cruz}, {West}, {Vrba}, \&
  {Anglada-Escud{\'e}}}]{2012ApJ...752...56F}
{Faherty}, J.~K., {Burgasser}, A.~J., {Walter}, F.~M., {et~al.} 2012, ApJ, 752,
  56

\bibitem[{{Filippazzo} {et~al.}(2015){Filippazzo}, {Rice}, {Faherty}, {Cruz},
  {Van Gordon}, \& {Looper}}]{filippazzo15}
{Filippazzo}, J.~C., {Rice}, E.~L., {Faherty}, J., {et~al.} 2015, \apj, 810,
  158

\bibitem[{{Forgan} {et~al.}(2015){Forgan}, {Parker}, \& {Rice}}]{forgan2015}
{Forgan}, D., {Parker}, R.~J., \& {Rice}, K. 2015, MNRAS, 447, 836

\bibitem[{{Forgan} \& {Rice}(2013)}]{forgan2013}
{Forgan}, D. \& {Rice}, K. 2013, MNRAS, 432, 3168

\bibitem[{{Gagn{\'e}} {et~al.}(2015){Gagn{\'e}}, {Faherty}, {Cruz},
  {Lafreni{\'e}re}, {Doyon}, {Malo}, {Burgasser}, {Naud}, {Artigau},
  {Bouchard}, {Gizis}, \& {Albert}}]{gagne15}
{Gagn{\'e}}, J., {Faherty}, J.~K., {Cruz}, K.~L., {et~al.} 2015, ApJs, 219, 33

\bibitem[{{Gaia Collaboration} {et~al.}(2016{\natexlab{a}}){Gaia
  Collaboration}, {Brown}, {Vallenari}, {Prusti}, {de Bruijne}, {Mignard},
  {Drimmel}, \& {co-authors}}]{gaia_dr1}
{Gaia Collaboration}, {Brown}, A.~G.~A., {Vallenari}, A., {et~al.}
  2016{\natexlab{a}}, ArXiv e-prints

\bibitem[{{Gaia Collaboration} {et~al.}(2016{\natexlab{b}}){Gaia
  Collaboration}, {Brown}, {Vallenari}, {Prusti}, {de Bruijne}, {Mignard},
  {Drimmel}, {Babusiaux}, {Bailer-Jones}, {Bastian}, \& et~al.}]{GDR1}
{Gaia Collaboration}, {Brown}, A.~G.~A., {Vallenari}, A., {et~al.}
  2016{\natexlab{b}}, A\&A, 595, A2

\bibitem[{{Gauza} {et~al.}(2015{\natexlab{a}}){Gauza}, {B{\'e}jar},
  {P{\'e}rez-Garrido}, {Rosa Zapatero Osorio}, {Lodieu}, {Rebolo}, {Pall{\'e}},
  \& {Nowak}}]{gauza2015}
{Gauza}, B., {B{\'e}jar}, V.~J.~S., {P{\'e}rez-Garrido}, A., {et~al.}
  2015{\natexlab{a}}, ApJ, 804, 96

\bibitem[{{Gauza} {et~al.}(2015{\natexlab{b}}){Gauza}, {B{\'e}jar},
  {P{\'e}rez-Garrido}, {Rosa Zapatero Osorio}, {Lodieu}, {Rebolo}, {Pall{\'e}},
  \& {Nowak}}]{2015ApJ...804...96G}
{Gauza}, B., {B{\'e}jar}, V.~J.~S., {P{\'e}rez-Garrido}, A., {et~al.}
  2015{\natexlab{b}}, ApJ, 804, 96

\bibitem[{{Gizis} {et~al.}(2015){Gizis}, {Allers}, {Liu}, {Harris}, {Faherty},
  {Burgasser}, \& {Kirkpatrick}}]{2015ApJ...799..203G}
{Gizis}, J.~E., {Allers}, K.~N., {Liu}, M.~C., {et~al.} 2015, ApJ, 799, 203

\bibitem[{{Grankin} {et~al.}(2007){Grankin}, {Artemenko}, \&
  {Melnikov}}]{Grankin2007}
{Grankin}, K.~N., {Artemenko}, S.~A., \& {Melnikov}, S.~Y. 2007, Information
  Bulletin on Variable Stars, 5752

\bibitem[{{Harrington} \& {Dahn}(1980)}]{1980AJ.....85..454H}
{Harrington}, R.~S. \& {Dahn}, C.~C. 1980, AJ, 85, 454

\bibitem[{{Horne} \& {Baliunas}(1986)}]{Horne1986}
{Horne}, J.~H. \& {Baliunas}, S.~L. 1986, ApJ, 302, 757

\bibitem[{{Janson} {et~al.}(2012){Janson}, {Bonavita}, {Klahr}, \&
  {Lafreni{\`e}re}}]{janson2012}
{Janson}, M., {Bonavita}, M., {Klahr}, H., \& {Lafreni{\`e}re}, D. 2012, ApJ,
  745, 4

\bibitem[{{Jenkins} {et~al.}(2009){Jenkins}, {Ramsey}, {Jones}, {Pavlenko},
  {Gallardo}, {Barnes}, \& {Pinfield}}]{2009ApJ...704..975J}
{Jenkins}, J.~S., {Ramsey}, L.~W., {Jones}, H.~R.~A., {et~al.} 2009, ApJ, 704,
  975

\bibitem[{{Kirkpatrick}(2005)}]{kirkpatrick05}
{Kirkpatrick}, J.~D. 2005, ARAA, 43, 195

\bibitem[{{Kirkpatrick} {et~al.}(2011){Kirkpatrick}, {Cushing}, {Gelino},
  {Griffith}, {Skrutskie}, {Marsh}, {Wright}, {Mainzer}, {Eisenhardt},
  {McLean}, {Thompson}, {Bauer}, {Benford}, {Bridge}, {Lake}, {Petty},
  {Stanford}, {Tsai}, {Bailey}, {Beichman}, {Bloom}, {Bochanski}, {Burgasser},
  {Capak}, {Cruz}, {Hinz}, {Kartaltepe}, {Knox}, {Manohar}, {Masters},
  {Morales-Calder{\'o}n}, {Prato}, {Rodigas}, {Salvato}, {Schurr}, {Scoville},
  {Simcoe}, {Stapelfeldt}, {Stern}, {Stock}, \& {Vacca}}]{2011ApJS..197...19K}
{Kirkpatrick}, J.~D., {Cushing}, M.~C., {Gelino}, C.~R., {et~al.} 2011, ApJS,
  197, 19

\bibitem[{{Kirkpatrick} {et~al.}(2000){Kirkpatrick}, {Reid}, {Liebert},
  {Gizis}, {Burgasser}, {Monet}, {Dahn}, {Nelson}, \&
  {Williams}}]{2000AJ....120..447K}
{Kirkpatrick}, J.~D., {Reid}, I.~N., {Liebert}, J., {et~al.} 2000, AJ, 120, 447

\bibitem[{{Kraus} {et~al.}(2017){Kraus}, {Herczeg}, {Rizzuto}, {Mann},
  {Slesnick}, {Carpenter}, {Hillenbrand}, \& {Mamajek}}]{kraus2017}
{Kraus}, A.~L., {Herczeg}, G.~J., {Rizzuto}, A.~C., {et~al.} 2017, ApJ, 838,
  150

\bibitem[{{Kraus} {et~al.}(2014){Kraus}, {Shkolnik}, {Allers}, \&
  {Liu}}]{kraus2014}
{Kraus}, A.~L., {Shkolnik}, E.~L., {Allers}, K.~N., \& {Liu}, M.~C. 2014, \aj,
  147, 146

\bibitem[{{Lachapelle} {et~al.}(2015){Lachapelle}, {Lafreni{\`e}re},
  {Gagn{\'e}}, {Jayawardhana}, {Janson}, {Helling}, \&
  {Witte}}]{2015ApJ...802...61L}
{Lachapelle}, F.-R., {Lafreni{\`e}re}, D., {Gagn{\'e}}, J., {et~al.} 2015, ApJ,
  802, 61

\bibitem[{{Lafreni{\`e}re} {et~al.}(2010){Lafreni{\`e}re}, {Jayawardhana}, \&
  {van Kerkwijk}}]{2010ApJ...719..497L}
{Lafreni{\`e}re}, D., {Jayawardhana}, R., \& {van Kerkwijk}, M.~H. 2010, ApJ,
  719, 497

\bibitem[{{Lagrange} {et~al.}(2010){Lagrange}, {Bonnefoy}, {Chauvin}, {Apai},
  {Ehrenreich}, {Boccaletti}, {Gratadour}, {Rouan}, {Mouillet}, {Lacour}, \&
  {Kasper}}]{lagrange2010}
{Lagrange}, A.-M., {Bonnefoy}, M., {Chauvin}, G., {et~al.} 2010, Science, 329,
  57

\bibitem[{{Langlois} {et~al.}(2013){Langlois}, {Vigan}, {Moutou}, {Sauvage},
  {Dohlen}, {Costille}, {Mouillet}, \& {Le Mignant}}]{2013aoel.confE..63L}
{Langlois}, M., {Vigan}, A., {Moutou}, C., {et~al.} 2013, in Proceedings of the
  Third AO4ELT Conference, ed. S.~{Esposito} \& L.~{Fini}, 63

\bibitem[{{Leggett} {et~al.}(2000){Leggett}, {Allard}, {Dahn}, {Hauschildt},
  {Kerr}, \& {Rayner}}]{2000ApJ...535..965L}
{Leggett}, S.~K., {Allard}, F., {Dahn}, C., {et~al.} 2000, ApJ, 535, 965

\bibitem[{{Liu} {et~al.}(2016){Liu}, {Dupuy}, \&
  {Allers}}]{2016ApJ...833...96L}
{Liu}, M.~C., {Dupuy}, T.~J., \& {Allers}, K.~N. 2016, ApJ, 833, 96

\bibitem[{{Liu} {et~al.}(2013){Liu}, {Magnier}, {Deacon}, {Allers}, {Dupuy},
  {Kotson}, {Aller}, {Burgett}, {Chambers}, {Draper}, {Hodapp}, {Jedicke},
  {Kaiser}, {Kudritzki}, {Metcalfe}, {Morgan}, {Price}, {Tonry}, \&
  {Wainscoat}}]{2013ApJ...777L..20L}
{Liu}, M.~C., {Magnier}, E.~A., {Deacon}, N.~R., {et~al.} 2013, ApJL, 777, L20

\bibitem[{{Luhman} {et~al.}(2017){Luhman}, {Mamajek}, {Shukla}, \&
  {Loutrel}}]{luhman2017}
{Luhman}, K.~L., {Mamajek}, E.~E., {Shukla}, S.~J., \& {Loutrel}, N.~P. 2017,
  AJ, 153, 46

\bibitem[{{Mace} {et~al.}(2013){Mace}, {Kirkpatrick}, {Cushing}, {Gelino},
  {Griffith}, {Skrutskie}, {Marsh}, {Wright}, {Eisenhardt}, {McLean},
  {Thompson}, {Mix}, {Bailey}, {Beichman}, {Bloom}, {Burgasser}, {Fortney},
  {Hinz}, {Knox}, {Lowrance}, {Marley}, {Morley}, {Rodigas}, {Saumon},
  {Sheppard}, \& {Stock}}]{2013ApJS..205....6M}
{Mace}, G.~N., {Kirkpatrick}, J.~D., {Cushing}, M.~C., {et~al.} 2013, ApJS,
  205, 6

\bibitem[{{Macintosh} {et~al.}(2015){Macintosh}, {Graham}, {Barman}, {De Rosa},
  {Konopacky}, {Marley}, {Marois}, {Nielsen}, {Pueyo}, {Rajan}, {Rameau},
  {Saumon}, {Wang}, {Patience}, {Ammons}, {Arriaga}, {Artigau}, {Beckwith},
  {Brewster}, {Bruzzone}, {Bulger}, {Burningham}, {Burrows}, {Chen}, {Chiang},
  {Chilcote}, {Dawson}, {Dong}, {Doyon}, {Draper}, {Duch{\^e}ne}, {Esposito},
  {Fabrycky}, {Fitzgerald}, {Follette}, {Fortney}, {Gerard}, {Goodsell},
  {Greenbaum}, {Hibon}, {Hinkley}, {Cotten}, {Hung}, {Ingraham},
  {Johnson-Groh}, {Kalas}, {Lafreniere}, {Larkin}, {Lee}, {Line}, {Long},
  {Maire}, {Marchis}, {Matthews}, {Max}, {Metchev}, {Millar-Blanchaer},
  {Mittal}, {Morley}, {Morzinski}, {Murray-Clay}, {Oppenheimer}, {Palmer},
  {Patel}, {Perrin}, {Poyneer}, {Rafikov}, {Rantakyr{\"o}}, {Rice}, {Rojo},
  {Rudy}, {Ruffio}, {Ruiz}, {Sadakuni}, {Saddlemyer}, {Salama}, {Savransky},
  {Schneider}, {Sivaramakrishnan}, {Song}, {Soummer}, {Thomas}, {Vasisht},
  {Wallace}, {Ward-Duong}, {Wiktorowicz}, {Wolff}, \& {Zuckerman}}]{51Eri}
{Macintosh}, B., {Graham}, J.~R., {Barman}, T., {et~al.} 2015, Science, 350, 64

\bibitem[{{Maire} {et~al.}(2016{\natexlab{a}}){Maire}, {Bonnefoy}, {Ginski},
  {Vigan}, {Messina}, {Mesa}, {Galicher}, {Gratton}, {Desidera}, {Kopytova},
  {Millward}, {Thalmann}, {Claudi}, {Ehrenreich}, {Zurlo}, {Chauvin},
  {Antichi}, {Baruffolo}, {Bazzon}, {Beuzit}, {Blanchard}, {Boccaletti}, {de
  Boer}, {Carle}, {Cascone}, {Costille}, {De Caprio}, {Delboulb{\'e}},
  {Dohlen}, {Dominik}, {Feldt}, {Fusco}, {Girard}, {Giro}, {Gisler}, {Gluck},
  {Gry}, {Henning}, {Hubin}, {Hugot}, {Jaquet}, {Kasper}, {Lagrange},
  {Langlois}, {Le Mignant}, {Llored}, {Madec}, {Martinez}, {Mawet}, {Milli},
  {M{\"o}ller-Nilsson}, {Mouillet}, {Moulin}, {Moutou}, {Orign{\'e}}, {Pavlov},
  {Petit}, {Pragt}, {Puget}, {Ramos}, {Rochat}, {Roelfsema}, {Salasnich},
  {Sauvage}, {Schmid}, {Turatto}, {Udry}, {Vakili}, {Wahhaj}, {Weber}, \&
  {Wildi}}]{maire2016}
{Maire}, A.-L., {Bonnefoy}, M., {Ginski}, C., {et~al.} 2016{\natexlab{a}},
  A\&A, 587, A56

\bibitem[{{Maire} {et~al.}(2016{\natexlab{b}}){Maire}, {Langlois}, {Dohlen},
  {Lagrange}, {Gratton}, {Chauvin}, {Desidera}, {Girard}, {Milli}, {Vigan},
  {Zins}, {Delorme}, {Beuzit}, {Claudi}, {Feldt}, {Mouillet}, {Puget},
  {Turatto}, \& {Wildi}}]{2016SPIE.9908E..34M}
{Maire}, A.-L., {Langlois}, M., {Dohlen}, K., {et~al.} 2016{\natexlab{b}}, in
  SPIE, Vol. 9908, Society of Photo-Optical Instrumentation Engineers (SPIE)
  Conference Series, 990834

\bibitem[{{Makarov} \& {Kaplan}(2005)}]{makarov2005}
{Makarov}, V.~V. \& {Kaplan}, G.~H. 2005, AJ, 129, 2420

\bibitem[{{Marois} {et~al.}(2014){Marois}, {Correia}, {Galicher}, {Ingraham},
  {Macintosh}, {Currie}, \& {De Rosa}}]{Marois2014}
{Marois}, C., {Correia}, C., {Galicher}, R., {et~al.} 2014, in SPIE, Vol. 9148,
  Adaptive Optics Systems IV, 91480U

\bibitem[{{Marois} {et~al.}(2006{\natexlab{a}}){Marois}, {Lafreni{\`e}re},
  {Doyon}, {Macintosh}, \& {Nadeau}}]{marois2006a}
{Marois}, C., {Lafreni{\`e}re}, D., {Doyon}, R., {Macintosh}, B., \& {Nadeau},
  D. 2006{\natexlab{a}}, ApJ, 641, 556

\bibitem[{{Marois} {et~al.}(2006{\natexlab{b}}){Marois}, {Lafreni{\`e}re},
  {Macintosh}, \& {Doyon}}]{2006ApJ...647..612M}
{Marois}, C., {Lafreni{\`e}re}, D., {Macintosh}, B., \& {Doyon}, R.
  2006{\natexlab{b}}, ApJ, 647, 612

\bibitem[{{Marois} {et~al.}(2008){Marois}, {Macintosh}, {Barman}, {Zuckerman},
  {Song}, {Patience}, {Lafreni{\`e}re}, \& {Doyon}}]{marois2008}
{Marois}, C., {Macintosh}, B., {Barman}, T., {et~al.} 2008, Science, 322, 1348

\bibitem[{{Marois} {et~al.}(2010){Marois}, {Zuckerman}, {Konopacky},
  {Macintosh}, \& {Barman}}]{marois2010}
{Marois}, C., {Zuckerman}, B., {Konopacky}, Q.~M., {Macintosh}, B., \&
  {Barman}, T. 2010, Nature, 468, 1080

\bibitem[{{Martin} {et~al.}(2017){Martin}, {Mace}, {McLean}, {Logsdon}, {Rice},
  {Kirkpatrick}, {Burgasser}, {McGovern}, \& {Prato}}]{martin17}
{Martin}, E.~C., {Mace}, G.~N., {McLean}, I.~S., {et~al.} 2017, \apj, 838, 73

\bibitem[{{McDonald} {et~al.}(2012){McDonald}, {Zijlstra}, \&
  {Boyer}}]{mcdonald2012}
{McDonald}, I., {Zijlstra}, A.~A., \& {Boyer}, M.~L. 2012, MNRAS, 427, 343

\bibitem[{{Meru} \& {Bate}(2010)}]{meru2010}
{Meru}, F. \& {Bate}, M.~R. 2010, MNRAS, 406, 2279

\bibitem[{{Mesa} {et~al.}(2015){Mesa}, {Gratton}, {Zurlo}, {Vigan}, {Claudi},
  {Alberi}, {Antichi}, {Baruffolo}, {Beuzit}, {Boccaletti}, {Bonnefoy},
  {Costille}, {Desidera}, {Dohlen}, {Fantinel}, {Feldt}, {Fusco}, {Giro},
  {Henning}, {Kasper}, {Langlois}, {Maire}, {Martinez}, {Moeller-Nilsson},
  {Mouillet}, {Moutou}, {Pavlov}, {Puget}, {Salasnich}, {Sauvage}, {Sissa},
  {Turatto}, {Udry}, {Vakili}, {Waters}, \& {Wildi}}]{mesa2015}
{Mesa}, D., {Gratton}, R., {Zurlo}, A., {et~al.} 2015, A\&A, 576, A121

\bibitem[{{Mesa} {et~al.}(2016{\natexlab{a}}){Mesa}, {Vigan}, {D'Orazi},
  {Ginski}, {Desidera}, {Bonnefoy}, {Gratton}, {Langlois}, {Marzari},
  {Messina}, {Antichi}, {Biller}, {Bonavita}, {Cascone}, {Chauvin}, {Claudi},
  {Curtis}, {Fantinel}, {Feldt}, {Garufi}, {Galicher}, {Henning}, {Incorvaia},
  {Lagrange}, {Millward}, {Perrot}, {Salasnich}, {Scuderi}, {Sissa}, {Wahhaj},
  \& {Zurlo}}]{mesa2016}
{Mesa}, D., {Vigan}, A., {D'Orazi}, V., {et~al.} 2016{\natexlab{a}}, ArXiv
  e-prints

\bibitem[{{Mesa} {et~al.}(2016{\natexlab{b}}){Mesa}, {Vigan}, {D'Orazi},
  {Ginski}, {Desidera}, {Bonnefoy}, {Gratton}, {Langlois}, {Marzari},
  {Messina}, {Antichi}, {Biller}, {Bonavita}, {Cascone}, {Chauvin}, {Claudi},
  {Curtis}, {Fantinel}, {Feldt}, {Garufi}, {Galicher}, {Henning}, {Incorvaia},
  {Lagrange}, {Millward}, {Perrot}, {Salasnich}, {Scuderi}, {Sissa}, {Wahhaj},
  \& {Zurlo}}]{2016A&A...593A.119M}
{Mesa}, D., {Vigan}, A., {D'Orazi}, V., {et~al.} 2016{\natexlab{b}}, A\&A, 593,
  A119

\bibitem[{{Messina} \& {Guinan}(2003)}]{messina2003}
{Messina}, S. \& {Guinan}, E.~F. 2003, A\&A, 409, 1017

\bibitem[{{Messina} {et~al.}(2016){Messina}, {Lanzafame}, {Feiden}, {Millward},
  {Desidera}, {Buccino}, {Curtis}, {Jofre'}, {Kehusmaa}, {Medhi}, {Monard}, \&
  {Petrucci}}]{Messina16}
{Messina}, S., {Lanzafame}, A.~C., {Feiden}, G.~A., {et~al.} 2016, ArXiv
  e-prints: 1607.06634

\bibitem[{{Naud} {et~al.}(2014){Naud}, {Artigau}, {Malo}, {Albert}, {Doyon},
  {Lafreni{\`e}re}, {Gagn{\'e}}, {Saumon}, {Morley}, {Allard}, {Homeier},
  {Beichman}, {Gelino}, \& {Boucher}}]{naud2014}
{Naud}, M.-E., {Artigau}, {\'E}., {Malo}, L., {et~al.} 2014, ApJ, 787, 5

\bibitem[{{Nguyen} {et~al.}(2012){Nguyen}, {Brandeker}, {van Kerkwijk}, \&
  {Jayawardhana}}]{nguyen12}
{Nguyen}, D.~C., {Brandeker}, A., {van Kerkwijk}, M.~H., \& {Jayawardhana}, R.
  2012, ApJ, 745, 119

\bibitem[{{Patience} {et~al.}(2010){Patience}, {King}, {de Rosa}, \&
  {Marois}}]{2010A&A...517A..76P}
{Patience}, J., {King}, R.~R., {de Rosa}, R.~J., \& {Marois}, C. 2010, A\&A,
  517, A76

\bibitem[{{Pavlov} {et~al.}(2008){Pavlov}, {M{\"o}ller-Nilsson}, {Feldt},
  {Henning}, {Beuzit}, \& {Mouillet}}]{pavlov08}
{Pavlov}, A., {M{\"o}ller-Nilsson}, O., {Feldt}, M., {et~al.} 2008, in SPIE,
  Vol. 7019, Advanced Software and Control for Astronomy II, 701939

\bibitem[{{Pearce} {et~al.}(2014){Pearce}, {Wyatt}, \& {Kennedy}}]{pearce2014}
{Pearce}, T.~D., {Wyatt}, M.~C., \& {Kennedy}, G.~M. 2014, MNRAS, 437, 2686

\bibitem[{{Pecaut} \& {Mamajek}(2013)}]{Pecaut13}
{Pecaut}, M.~J. \& {Mamajek}, E.~E. 2013, ApJs, 208, 9

\bibitem[{{Pecaut} \& {Mamajek}(2016)}]{pecaut16}
{Pecaut}, M.~J. \& {Mamajek}, E.~E. 2016, MNRAS, 461, 794

\bibitem[{{Pojmanski}(1997)}]{ASAS}
{Pojmanski}, G. 1997, ACA, 47, 467

\bibitem[{{Rajan} {et~al.}(2017){Rajan}, {Rameau}, {De Rosa}, {Marley},
  {Graham}, {Macintosh}, {Marois}, {Morley}, {Patience}, {Pueyo}, {Saumon},
  {Ward-Duong}, {Ammons}, {Arriaga}, {Bailey}, {Barman}, {Bulger}, {Burrows},
  {Chilcote}, {Cotten}, {Czekala}, {Doyon}, {Duch{\^e}ne}, {Esposito},
  {Fitzgerald}, {Follette}, {Fortney}, {Goodsell}, {Greenbaum}, {Hibon},
  {Hung}, {Ingraham}, {Johnson-Groh}, {Kalas}, {Konopacky}, {Lafreni{\`e}re},
  {Larkin}, {Maire}, {Marchis}, {Metchev}, {Millar-Blanchaer}, {Morzinski},
  {Nielsen}, {Oppenheimer}, {Palmer}, {Patel}, {Perrin}, {Poyneer},
  {Rantakyr{\"o}}, {Ruffio}, {Savransky}, {Schneider}, {Sivaramakrishnan},
  {Song}, {Soummer}, {Thomas}, {Vasisht}, {Wallace}, {Wang}, {Wiktorowicz}, \&
  {Wolff}}]{2017AJ....154...10R}
{Rajan}, A., {Rameau}, J., {De Rosa}, R.~J., {et~al.} 2017, AJ, 154, 10

\bibitem[{{Rameau} {et~al.}(2013){Rameau}, {Chauvin}, {Lagrange}, {Boccaletti},
  {Quanz}, {Bonnefoy}, {Girard}, {Delorme}, {Desidera}, {Klahr}, {Mordasini},
  {Dumas}, \& {Bonavita}}]{rameau2013}
{Rameau}, J., {Chauvin}, G., {Lagrange}, A.-M., {et~al.} 2013, ApJl, 772, L15

\bibitem[{{Samland} {et~al.}(2017){Samland}, {Molli{\`e}re}, {Bonnefoy},
  {Maire}, {Cantalloube}, {Cheetham}, {Mesa}, {Gratton}, {Biller}, {Wahhaj},
  {Bouwman}, {Brandner}, {Melnick}, {Carson}, {Janson}, {Henning}, {Homeier},
  {Mordasini}, {Langlois}, {Quanz}, {van Boekel}, {Zurlo}, {Schlieder},
  {Avenhaus}, {Beuzit}, {Boccaletti}, {Bonavita}, {Chauvin}, {Claudi}, {Cudel},
  {Desidera}, {Feldt}, {Fusco}, {Galicher}, {Kopytova}, {Lagrange}, {Le
  Coroller}, {Martinez}, {Moeller-Nilsson}, {Mouillet}, {Mugnier}, {Perrot},
  {Sevin}, {Sissa}, {Vigan}, \& {Weber}}]{2017A&A...603A..57S}
{Samland}, M., {Molli{\`e}re}, P., {Bonnefoy}, M., {et~al.} 2017, A\&A, 603,
  A57

\bibitem[{{Scargle}(1982)}]{Scargle82}
{Scargle}, J.~D. 1982, ApJ, 263, 835

\bibitem[{{Schneider} {et~al.}(2015){Schneider}, {Cushing}, {Kirkpatrick},
  {Gelino}, {Mace}, {Wright}, {Eisenhardt}, {Skrutskie}, {Griffith}, \&
  {Marsh}}]{2015ApJ...804...92S}
{Schneider}, A.~C., {Cushing}, M.~C., {Kirkpatrick}, J.~D., {et~al.} 2015, ApJ,
  804, 92

\bibitem[{{Sivaramakrishnan} \& {Oppenheimer}(2006)}]{2006ApJ...647..620S}
{Sivaramakrishnan}, A. \& {Oppenheimer}, B.~R. 2006, ApJ, 647, 620

\bibitem[{{Skrutskie} {et~al.}(2006){Skrutskie}, {Cutri}, {Stiening},
  {Weinberg}, {Schneider}, {Carpenter}, {Beichman}, {Capps}, {Chester},
  {Elias}, {Huchra}, {Liebert}, {Lonsdale}, {Monet}, {Price}, {Seitzer},
  {Jarrett}, {Kirkpatrick}, {Gizis}, {Howard}, {Evans}, {Fowler}, {Fullmer},
  {Hurt}, {Light}, {Kopan}, {Marsh}, {McCallon}, {Tam}, {Van Dyk}, \&
  {Wheelock}}]{skru06}
{Skrutskie}, M.~F., {Cutri}, R.~M., {Stiening}, R., {et~al.} 2006, AJ, 131,
  1163

\bibitem[{{Soummer} {et~al.}(2012){Soummer}, {Pueyo}, \&
  {Larkin}}]{soummer2012}
{Soummer}, R., {Pueyo}, L., \& {Larkin}, J. 2012, ApJL, 755, L28

\bibitem[{{Stone} {et~al.}(2016{\natexlab{a}}){Stone}, {Skemer}, {Kratter},
  {Dupuy}, {Close}, {Eisner}, {Fortney}, {Hinz}, {Males}, {Morley},
  {Morzinski}, \& {Ward-Duong}}]{stone2016}
{Stone}, J.~M., {Skemer}, A.~J., {Kratter}, K.~M., {et~al.} 2016{\natexlab{a}},
  ApJL, 818, L12

\bibitem[{{Stone} {et~al.}(2016{\natexlab{b}}){Stone}, {Skemer}, {Kratter},
  {Dupuy}, {Close}, {Eisner}, {Fortney}, {Hinz}, {Males}, {Morley},
  {Morzinski}, \& {Ward-Duong}}]{2016ApJ...818L..12S}
{Stone}, J.~M., {Skemer}, A.~J., {Kratter}, K.~M., {et~al.} 2016{\natexlab{b}},
  ApJL, 818, L12

\bibitem[{{Thalmann} {et~al.}(2008){Thalmann}, {Schmid}, {Boccaletti},
  {Mouillet}, {Dohlen}, {Roelfsema}, {Carbillet}, {Gisler}, {Beuzit}, {Feldt},
  {Gratton}, {Joos}, {Keller}, {Kragt}, {Pragt}, {Puget}, {Rigal}, {Snik},
  {Waters}, \& {Wildi}}]{Thalmann2008}
{Thalmann}, C., {Schmid}, H.~M., {Boccaletti}, A., {et~al.} 2008, in SPIE, Vol.
  7014, Ground-based and Airborne Instrumentation for Astronomy II, 70143F

\bibitem[{{van Leeuwen}(2007)}]{vanleeuwen2007}
{van Leeuwen}, F. 2007, A\&A, 474, 653

\bibitem[{{Vigan}(2016)}]{vigan2016b}
{Vigan}, A. 2016, {SILSS: SPHERE/IRDIS Long-Slit Spectroscopy pipeline},
  Astrophysics Source Code Library

\bibitem[{{Vigan} {et~al.}(2017){Vigan}, {Bonavita}, {Biller}, {Forgan},
  {Rice}, {Chauvin}, {Desidera}, {Meunier}, {Delorme}, {Schlieder}, {Bonnefoy},
  {Carson}, {Covino}, {Hagelberg}, {Henning}, {Janson}, {Lagrange}, {Quanz},
  {Zurlo}, {Beuzit}, {Boccaletti}, {Buenzli}, {Feldt}, {Girard}, {Gratton},
  {Kasper}, {Le Coroller}, {Mesa}, {Messina}, {Meyer}, {Montagnier},
  {Mordasini}, {Mouillet}, {Moutou}, {Reggiani}, {Segransan}, \&
  {Thalmann}}]{vigan2017}
{Vigan}, A., {Bonavita}, M., {Biller}, B., {et~al.} 2017, ArXiv e-prints

\bibitem[{{Vigan} {et~al.}(2016){Vigan}, {Bonnefoy}, {Ginski}, {Beust},
  {Galicher}, {Janson}, {Baudino}, {Buenzli}, {Hagelberg}, {D'Orazi},
  {Desidera}, {Maire}, {Gratton}, {Sauvage}, {Chauvin}, {Thalmann}, {Malo},
  {Salter}, {Zurlo}, {Antichi}, {Baruffolo}, {Baudoz}, {Blanchard},
  {Boccaletti}, {Beuzit}, {Carle}, {Claudi}, {Costille}, {Delboulb{\'e}},
  {Dohlen}, {Dominik}, {Feldt}, {Fusco}, {Gluck}, {Girard}, {Giro}, {Gry},
  {Henning}, {Hubin}, {Hugot}, {Jaquet}, {Kasper}, {Lagrange}, {Langlois}, {Le
  Mignant}, {Llored}, {Madec}, {Martinez}, {Mawet}, {Mesa}, {Milli},
  {Mouillet}, {Moulin}, {Moutou}, {Orign{\'e}}, {Pavlov}, {Perret}, {Petit},
  {Pragt}, {Puget}, {Rabou}, {Rochat}, {Roelfsema}, {Salasnich}, {Schmid},
  {Sevin}, {Siebenmorgen}, {Smette}, {Stadler}, {Suarez}, {Turatto}, {Udry},
  {Vakili}, {Wahhaj}, {Weber}, \& {Wildi}}]{vigan2016}
{Vigan}, A., {Bonnefoy}, M., {Ginski}, C., {et~al.} 2016, A\&A, 587, A55

\bibitem[{{Vigan} {et~al.}(2008){Vigan}, {Langlois}, {Moutou}, \&
  {Dohlen}}]{vigan2008}
{Vigan}, A., {Langlois}, M., {Moutou}, C., \& {Dohlen}, K. 2008, A\&A, 489,
  1345

\bibitem[{{Vigan} {et~al.}(2010){Vigan}, {Moutou}, {Langlois}, {Allard},
  {Boccaletti}, {Carbillet}, {Mouillet}, \& {Smith}}]{Vigan2010}
{Vigan}, A., {Moutou}, C., {Langlois}, M., {et~al.} 2010, MNRAS, 407, 71

\bibitem[{{Wagner} {et~al.}(2016){Wagner}, {Apai}, {Kasper}, {Kratter},
  {McClure}, {Robberto}, \& {Beuzit}}]{wagner2016}
{Wagner}, K., {Apai}, D., {Kasper}, M., {et~al.} 2016, Science, 353, 673

\bibitem[{{Wahhaj} {et~al.}(2011{\natexlab{a}}){Wahhaj}, {Liu}, {Biller},
  {Clarke}, {Nielsen}, {Close}, {Hayward}, {Mamajek}, {Cushing}, {Dupuy},
  {Tecza}, {Thatte}, {Chun}, {Ftaclas}, {Hartung}, {Reid}, {Shkolnik},
  {Alencar}, {Artymowicz}, {Boss}, {de Gouveia Dal Pino}, {Gregorio-Hetem},
  {Ida}, {Kuchner}, {Lin}, \& {Toomey}}]{2011ApJ...729..139W}
{Wahhaj}, Z., {Liu}, M.~C., {Biller}, B.~A., {et~al.} 2011{\natexlab{a}}, ApJ,
  729, 139

\bibitem[{{Wahhaj} {et~al.}(2011{\natexlab{b}}){Wahhaj}, {Liu}, {Biller},
  {Clarke}, {Nielsen}, {Close}, {Hayward}, {Mamajek}, {Cushing}, {Dupuy},
  {Tecza}, {Thatte}, {Chun}, {Ftaclas}, {Hartung}, {Reid}, {Shkolnik},
  {Alencar}, {Artymowicz}, {Boss}, {de Gouveia Dal Pino}, {Gregorio-Hetem},
  {Ida}, {Kuchner}, {Lin}, \& {Toomey}}]{wahhaj2011}
{Wahhaj}, Z., {Liu}, M.~C., {Biller}, B.~A., {et~al.} 2011{\natexlab{b}}, \apj,
  729, 139

\bibitem[{{Wichmann} {et~al.}(2000){Wichmann}, {Torres}, {Melo}, {Frink},
  {Allain}, {Bouvier}, {Krautter}, {Covino}, \& {Neuh{\"a}user}}]{wichmann2000}
{Wichmann}, R., {Torres}, G., {Melo}, C.~H.~F., {et~al.} 2000, A\&A, 359, 181

\bibitem[{{Wright} {et~al.}(2010){Wright}, {Eisenhardt}, {Mainzer}, {Ressler},
  {Cutri}, {Jarrett}, {Kirkpatrick}, {Padgett}, {McMillan}, {Skrutskie},
  {Stanford}, {Cohen}, {Walker}, {Mather}, {Leisawitz}, {Gautier}, {McLean},
  {Benford}, {Lonsdale}, {Blain}, {Mendez}, {Irace}, {Duval}, {Liu}, {Royer},
  {Heinrichsen}, {Howard}, {Shannon}, {Kendall}, {Walsh}, {Larsen}, {Cardon},
  {Schick}, {Schwalm}, {Abid}, {Fabinsky}, {Naes}, \& {Tsai}}]{wright2010}
{Wright}, E.~L., {Eisenhardt}, P.~R.~M., {Mainzer}, A.~K., {et~al.} 2010, AJ,
  140, 1868

\bibitem[{{Zacharias} {et~al.}(2004){Zacharias}, {Monet}, {Levine}, {Urban},
  {Gaume}, \& {Wycoff}}]{zacharias04}
{Zacharias}, N., {Monet}, D.~G., {Levine}, S.~E., {et~al.} 2004, in Bulletin of
  the American Astronomical Society, Vol.~36, American Astronomical Society
  Meeting Abstracts, 1418

\bibitem[{{Zapatero Osorio} {et~al.}(2014){Zapatero Osorio}, {B{\'e}jar},
  {Miles-P{\'a}ez}, {Pe{\~n}a Ram{\'{\i}}rez}, {Rebolo}, \&
  {Pall{\'e}}}]{2014A&A...568A...6Z}
{Zapatero Osorio}, M.~R., {B{\'e}jar}, V.~J.~S., {Miles-P{\'a}ez}, P.~A.,
  {et~al.} 2014, A\&A, 568, A6

\bibitem[{{Zurlo} {et~al.}(2016){Zurlo}, {Vigan}, {Galicher}, {Maire}, {Mesa},
  {Gratton}, {Chauvin}, {Kasper}, {Moutou}, {Bonnefoy}, {Desidera}, {Abe},
  {Apai}, {Baruffolo}, {Baudoz}, {Baudrand}, {Beuzit}, {Blancard},
  {Boccaletti}, {Cantalloube}, {Carle}, {Cascone}, {Charton}, {Claudi},
  {Costille}, {de Caprio}, {Dohlen}, {Dominik}, {Fantinel}, {Feautrier},
  {Feldt}, {Fusco}, {Gigan}, {Girard}, {Gisler}, {Gluck}, {Gry}, {Henning},
  {Hugot}, {Janson}, {Jaquet}, {Lagrange}, {Langlois}, {Llored}, {Madec},
  {Magnard}, {Martinez}, {Maurel}, {Mawet}, {Meyer}, {Milli},
  {Moeller-Nilsson}, {Mouillet}, {Orign{\'e}}, {Pavlov}, {Petit}, {Puget},
  {Quanz}, {Rabou}, {Ramos}, {Rousset}, {Roux}, {Salasnich}, {Salter},
  {Sauvage}, {Schmid}, {Soenke}, {Stadler}, {Suarez}, {Turatto}, {Udry},
  {Vakili}, {Wahhaj}, {Wildi}, \& {Antichi}}]{zurlo2016}
{Zurlo}, A., {Vigan}, A., {Galicher}, R., {et~al.} 2016, A\&A, 587, A57

\bibitem[{{Zurlo} {et~al.}(2014){Zurlo}, {Vigan}, {Mesa}, {Gratton}, {Moutou},
  {Langlois}, {Claudi}, {Pueyo}, {Boccaletti}, {Baruffolo}, {Beuzit},
  {Costille}, {Desidera}, {Dohlen}, {Feldt}, {Fusco}, {Henning}, {Kasper},
  {Martinez}, {Moeller-Nilsson}, {Mouillet}, {Pavlov}, {Puget}, {Sauvage},
  {Turatto}, {Udry}, {Vakili}, {Waters}, \& {Wildi}}]{zurlo2014}
{Zurlo}, A., {Vigan}, A., {Mesa}, D., {et~al.} 2014, A\&A, 572, A85

\end{thebibliography}

\end{document}